\documentclass[aps,prd,reprint,twocolumn,superscriptaddress,longbibliography,nofootinbib,floatfix,showpacs]{revtex4-1}
\usepackage{epsfig}
\usepackage{amsmath,amssymb,amsfonts}
\usepackage{hyperref}
\usepackage{mathrsfs}
\usepackage{bbm}
\usepackage{slashed}
\usepackage{graphicx}
\usepackage{verbatim}

\usepackage{bm}

\usepackage[usenames]{color}

\usepackage{tikz}
\usetikzlibrary{calc,decorations.markings,decorations.pathmorphing,positioning,shapes,snakes,arrows}

\definecolor{darkgreen}{rgb}{0.2,0.6,0}

\newcommand{\be}{\begin{equation}}
\newcommand{\ee}{\end{equation}}
\newcommand{\bw}{\begin{widetext}}
\newcommand{\ew}{\end{widetext}}
\newcommand{\bi}{\begin{itemize}}
\newcommand{\ei}{\end{itemize}}
\newcommand{\bea}{\begin{eqnarray}}
\newcommand{\eea}{\end{eqnarray}}
\newcommand{\bra}[1]{\langle\,#1\,|}          
\newcommand{\ket}[1]{|\,#1\,\rangle}          
\newcommand{\ud}{\mathrm{d}}

\newcommand{\LCm}{{\scriptscriptstyle -}} 
\newcommand{\LCp}{{\scriptscriptstyle +}}
\newcommand{\LCpm}{{\scriptscriptstyle \pm}}
\newcommand{\LCmp}{{\scriptscriptstyle \mp}}

\newcommand{\LCperp}{{\scriptscriptstyle \perp}}

\usepackage[T1]{fontenc} \usepackage[latin1]{inputenc}

\newcommand{\fsl}[1]{\slashed{#1}}

\begin{document}

\title{Single and double nonlinear Compton scattering}

\author{Victor Dinu} 
\email{dinu@barutu.fizica.unibuc.ro}
\affiliation{Department of Physics, University of Bucharest, P.O.~Box MG-11, M\u agurele 077125, Romania}

\author{Greger Torgrimsson}
\email{greger.torgrimsson@uni-jena.de}
\affiliation{Theoretisch-Physikalisches Institut, Abbe Center of Photonics,
Friedrich-Schiller-Universit\"at Jena, Max-Wien-Platz 1, D-07743 Jena, Germany}
\affiliation{Helmholtz Institute Jena, Fr\"obelstieg 3, D-07743 Jena, Germany}

\begin{abstract}
We study single, double and higher-order nonlinear Compton scattering where an electron interacts nonlinearly with a high-intensity laser and emits one, two or more photons. We study, in particular, how double Compton scattering is separated into one-step and two-step parts, where the latter is obtained from an incoherent product of two single-photon emissions. We include all contributions to double Compton scattering and show that the exchange term, which was not calculated in previous constant-crossed field studies, is in general on the same order of magnitude as the other one-step terms. Our approach reveals practically useful similarities between double Compton scattering and the trident process, which allows us to transfer some of our previous results for trident to double Compton scattering. We provide a new gluing approach for obtaining the dominant contribution to higher-order Compton scattering for long laser pulses. Unlike the standard gluing approach, our new approach does not require the intensity parameter $a_0$ to be much larger than one. 
For ``hard'' photons we obtain several saddle-point approximations for various field shapes.    
\end{abstract}
\maketitle

\section{Introduction}

In~\cite{Dinu:2017uoj} we studied the trident process~\cite{Baier,Ritus:1972nf,Hu:2010ye,Ilderton:2010wr,King:2013osa,King:2018ibi,Mackenroth:2018smh}, $e^\LCm\to 2e^\LCm+e^\LCp$, in plane-wave background fields, and derived compact expressions for the probability for arbitrary background field shapes. Here we will apply the same methods to another second-order process, namely double nonlinear Compton scattering~\cite{Morozov:1975uah,Lotstedt:2009zz,Loetstedt:2009zz,Seipt:2012tn,Mackenroth:2012rb,King:2014wfa}, where the incoming electron emits two photons, $e^\LCm\to e^\LCm+2\gamma$. This is also a process that one can separate into one-step and two-step parts, where the latter is obtained by incoherently gluing together the probabilities of two single-photon emissions. The two-step term is expected to be a good approximation of the total probability for sufficiently high intensities, or more precisely for $a_0=eE/(m\omega)\gg1$, where $E$ is the field strength and $\omega$ a typical/characteristic frequency of the (in general pulsed) background field. 
This two-step dominance is what makes it possible to use particle-in-cell (PIC) simulations to study complicated higher-order processes in high-intensity fields~\cite{RidgersCode,Gonoskov:2014mda,Osiris,Smilei}. This regime is also associated with the locally-constant-field (LCF) approximation, which entails further simplifications. There is now interest in going beyond or improving the standard LCF approximation~\cite{Ilderton:2018nws,DiPiazza:2018bfu}.  

In this paper we are interested in corrections to the two-step approximation. In particular, the one-step part can be separated into (what we call) direct\footnote{Note that we do not use ``direct'' as synonymous to the one-step term. By ``direct'' we mean instead the non-exchange part. The two-step term only has a direct part while the one-step term has both direct and exchange parts.} and exchange terms, where the latter comes from the cross term between the two terms in the amplitude which are related by exchanging the two emitted photons. A similar exchange term appears in the trident case, and in~\cite{Dinu:2017uoj} we showed that, while omitted in previous constant-crossed/LCF studies, it is in general on the same order of magnitude as the direct part of the one-step term. Here we make a similar investigation into the importance of the exchange term in double Compton scattering. That the exchange term can be important e.g. for $a_0\sim1$ was also found in~\cite{Seipt:2012tn}. 

For $a_0\sim1$ the one-step term is in general on the same order of magnitude as the two-step term. However, if the field is sufficiently long then the probability is again dominated by a term that can be expressed as an incoherent product of two single-photon emissions. If $a_0$ is not large one should of course not expect this two-step term to be the same as the LCF two-step term. While spin effects are usually neglected in PIC simulations, to obtain the complete two-step term in the LCF regime one has to sum the incoherent product over the spin of the intermediate electron~\cite{Morozov:1975uah,King:2014wfa}. In this paper we identify a term in double Compton scattering that dominates for sufficiently long pulses without assuming $a_0\gg1$ or any particular field shape, and then we show that this two-step term can be obtained from an appropriate sum of the incoherent product of two single-photon emissions. We do this for an arbitrary background field. For fields with linear polarization one can obtain the two-step term by summing over spin in essentially the same way as in the LCF regime~\cite{Morozov:1975uah,King:2014wfa}. However, for fields that do not have linear polarization things become more nontrivial, because in general one has to take into account the fact that there is a spin sum already on the amplitude level, which in general leads to a double spin sum on the probability level. We have found a simple prescription for obtaining the entire two-step term from the spin-dependent probability for single Compton scattering. This gluing approach is to the best of our knowledge new and seems promising for studying higher-order processes. We have checked that it gives the correct results for triple and quadruple Compton scattering, where the electron emits three and four photons.  

Calculating higher-order processes means performing higher-dimensional integrals. Numerical integration can quickly become challenging. In our approach we integrate analytically over the transverse components of the momenta, and then the longitudinal momentum spectrum is obtained by performing a number of lightfront-time ($x^\LCp$) integrals. The exponential part of these integrands can in general be expressed in terms of an ($x^\LCp$-dependent) effective mass, and the integrals can be performed with the saddle-point method. In fact, the integrals for double Compton scattering are very similar to the ones in the trident case~\cite{Dinu:2017uoj}, so we have for example been able to reuse saddle points we found in~\cite{Dinu:2017uoj} for double Compton scattering, and the new saddle-point results we provide here can also be translated to the trident case. For certain simple field shapes we can obtain simple analytical approximations, but the saddle-point method can also be useful even if one has to find the saddle points numerically, as it can offer a quick estimate and a check of exact numerical integrations. We show here that the saddle-point method can give a good approximation of even quite small and fast oscillations in the spectrum.      

In comparison with previous papers on double Compton scattering, note that our focus is on the longitudinal momentum spectra, which we obtain by performing all integrals over the transverse momenta. We have several reasons for this:
1) We can perform these integrals exactly analytically for arbitrary pulse shape. 
2) The total/integrated probabilities only depend on the longitudinal momentum of the initial particle, but not on its transverse momentum, so it is natural to consider how the initial longitudinal momentum is distributed among the final-state particles.
3) The longitudinal momentum spectra are Lorentz invariant, being expressed in terms of 4-vector products of the particles' momenta and the field's  wave vector, which makes them especially suitable for theoretical studies. 
4) Even after performing these integrals for the first-order processes the results are still general enough for the construction of gluing estimates, which would not have been the case if we had instead integrated over the longitudinal momenta (or summed over the spins). 
5) Higher orders in general depend on several momentum and spin variables, so by performing these integrals we reduce this to a more manageable number of parameters, while still being sure that we have not missed any important regions of phase space.  The previous points give motivation for reducing the number of parameters by integrating over the transverse rather than some other components of the momenta.
So, while different quantities might be more relevant for experiments, at least from an analytical/theoretical point of view it is natural to consider the longitudinal momentum spectrum integrated over the transverse momenta.

This paper is organized as follows. We focus first on double Compton scattering. In Sec.~\ref{Description of formalism} we give the necessary definitions. In Sec.~\ref{Exact analytical results section} we provide compact expressions for the exact probability for arbitrary field shapes. In Sec.~\ref{Two-step and one-step section} we separate the probability into one-step and two-step terms and compare with the incoherent product of two single-photon emissions. This comparison helped us to find a new gluing approach, which we confirm in Sec.~\ref{MultiphotonSection} for triple and quadruple Compton scattering. In Sec.~\ref{Saddle point approximation section} we derive simple analytical approximations for ``hard'' photons for various field shapes. In Sec.~\ref{saddlePointInterferenceSection} we apply the saddle-point method to fields with many oscillations and hence many contributing saddle points, which lead to interference effects in the momentum spectrum. We consider single Compton scattering and compare this saddle-point approximation with an exact numerical integration and find very good agreement. 
In Sec.~\ref{DCnumericalLCF} we consider double Compton scattering in the LCF approximation. We show, in particular, that the exchange term can continue to be on the same order as the direct part of the one-step term also for larger $\chi$.

\section{Definitions}\label{Description of formalism}

We use the same formalism and notation as in~\cite{Dinu:2017uoj}, which we briefly recall here for convenience. Lightfront coordinates are defined by $v^\LCpm=2v_\LCmp=v^0\pm v^3$ and $v^\LCperp=\{v^1,v^2\}$, and we use 
$\bar{x}=\{x^\LCm,x^\LCperp\}$ for coordinates and $\bar{p}=\{p_\LCm,p_\LCperp\}$ for momenta.
The plane-wave background field is given by $f_{\mu\nu}=k_\mu a'_\nu-k_\nu a'_\mu$, where $k_\mu=k_\LCp\delta_\mu^\LCp$ is a light-like wave vector and $a_\LCperp(\phi)$, with $\phi=kx$, is a polarization vector with an arbitrary dependence on lightfront time $x^\LCp$. We use units with $c=\hbar=1$ as well as $m_e=1$, and absorb the electron charge into the background field, i.e. $ea_\mu\to a_\mu$.

We have the same initial state as in~\cite{Dinu:2017uoj}, i.e an electron with momentum $p_\mu$ and spin $\sigma$,
\be
|\text{in}\rangle=\int\!\ud\tilde{p}\;f(p)b^\dagger(p\sigma)|0\rangle \;,
\ee
where
$\ud\tilde{p}=\theta(p_\LCm)\ud p_\LCm\ud^2p_\LCperp/(2p_\LCm(2\pi)^3)$
is the Lorentz-invariant momentum measure, $\theta(.)$ is the Heaviside step function, and $f(p)$ a sharply peaked wave packet\footnote{In this paper we do not consider effects of finite-sized wave packets. See~\cite{Angioi:2016vir} for such effects in photon emission by a single electron, and~\cite{Angioi:2017ygv} for two-electron wave packets and the difference in coherence compared to the classical prediction~\cite{Klepikov}.}~\cite{Ilderton:2012qe} (we also use $p$ for the position of this peak). The normalization of the initial state, $\langle\text{in}|\text{in}\rangle=1$, and of the mode operators,
$\{b(q,r),\bar{b}(q',r')\}=2p_\LCm\bar{\delta}(q-q')\delta_{rr'}$ where $\bar{\delta}(\dots)=(2\pi)^3\delta_{\LCm,\LCperp}(\dots)$, implies
\be
 \int\!\ud\tilde{p}\;|f|^2=1 \;.
\ee

We focus first on double Compton scattering, where the final state contains one electron with $p'_\mu$ and $\sigma'$ and two photons with momenta and polarization vectors $l_1^\mu,l_2^\mu$ and $\epsilon_1^\mu,\epsilon_2^\mu$. We use lightfront gauge, so in addition to $l\epsilon(l)=0$ we also have $k\epsilon=0$. 
The amplitude for two-photon emission, $M$, is defined via the evolution operator $U$ by
\be
\langle0|b(p'\sigma')\epsilon_1 a(l_1)\epsilon_2a(l_2)Ub^\dagger(p\sigma)|0\rangle
=:\bar{\delta}(p'+l_2+l_1-p)\frac{M}{k_\LCp} \;.
\ee
As in~\cite{Dinu:2017uoj}, in order to reduce the number of parameters on which the probability depends, we integrate analytically the probability over the Gaussian transverse momentum integrals~\cite{Dinu:2013hsd} and sum/average over spins and polarizations, 
\be\label{total-P}
\begin{split}
\mathbb{P}=&\frac{1}{4}\sum\limits_\text{spins}\int\!\ud\tilde{p}'\ud\tilde{l}_1\ud\tilde{l}_2\Big|\int\!\ud\tilde{p}\;f\frac{1}{k_\LCp}\bar{\delta}(p'+l_2+l_1-p)M\Big|^2 \\
=&\frac{1}{4}\sum\limits_\text{spins}\int\!\ud\tilde{l}_1\ud\tilde{l}_2\frac{\theta(kp')}{kpkp'}|M|^2 \;,
\end{split}
\ee
where the factor of $1/4$ is due to spin-averaging and the presence of identical particles, and $\bar{p}'=\bar{p}-\bar{l}_1-\bar{l}_2$.

We separate the amplitude into $M=M^{12}+M^{21}$, where $M^{21}$ is obtained from $M^{12}$ by replacing $l_1\leftrightarrow l_2$ and $\epsilon_1\leftrightarrow \epsilon_2$, which on the probability level gives
$|M|^2=|M^{12}|^2+|M^{21}|^2+2\text{Re }\bar{M}^{21}M^{12}$. We refer to the first two terms as the direct part and the cross term as the exchange part, i.e. 
\be\label{PdirDef}
\mathbb{P}_{\rm dir}=\frac{1}{4}\sum\limits_\text{spins}\int\!\ud\tilde{l}_1\ud\tilde{l}_2\frac{\theta(kp')}{kpkp'}|M^{12}|^2+(1\leftrightarrow2) \;,
\ee
where $(1\leftrightarrow2)$ is obtained from the first term by replacing $l_1\leftrightarrow l_2$ and $\epsilon_1\leftrightarrow\epsilon_2$, and
\be\label{PexDef}
\mathbb{P}_{\rm ex}=\frac{1}{2}\sum\limits_\text{spins}\int\!\ud\tilde{l}_1\ud\tilde{l}_2\frac{\theta(kp')}{kpkp'}\text{Re }\bar{M}^{21}M^{12} \;.
\ee

\begin{figure}
\centering
\includegraphics[width=\linewidth]{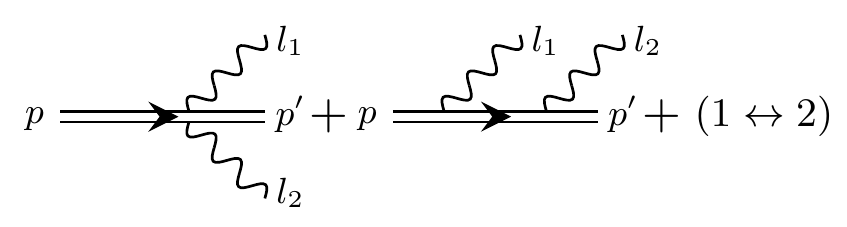}
\caption{This diagram shows the separation of the amplitude for double Compton scattering. The first and the second terms represent $M_1^{12}$ and $M_2^{12}$, respectively. All particles, including the intermediate electron in the second diagram, are on shell.}
\label{FeynmanDiagram}
\end{figure}

We have relegated the calculation of the amplitude to the appendix as it only involves standard methods. The important thing to note is that the amplitude contains two terms, $M^{12}=M_1^{12}+M_2^{12}$, where $M_1^{12}$ has one $x^\LCp$ integral and $M_2^{12}$ has two. These terms are illustrated in Fig.~\ref{FeynmanDiagram}.
As in~\cite{Dinu:2017uoj}, this leads to a separation of the direct and the exchange part of the probability into three terms with different numbers of $x^\LCp$ integrals,
\be\label{PdirectDef}
\begin{split}
\{\mathbb{P}_{\rm dir}^{11},&\mathbb{P}_{\rm dir}^{12},\mathbb{P}_{\rm dir}^{22}\}:=\frac{1}{4}\sum\limits_\text{spins}\int\!\ud\tilde{l}_1\ud\tilde{l}_2
\frac{\theta(kp')}{kpkp'}\\
&\{|M_1^{12}|^2,2\text{Re}\bar{M}_1^{12}M_2^{12},|M_2^{12}|^2\}+(1\leftrightarrow2) \;,
\end{split}
\ee
\be\label{PexchangeDef}
\begin{split}
\{\mathbb{P}_{\rm ex}^{11},&\mathbb{P}_{\rm ex}^{12},\mathbb{P}_{\rm ex}^{22}\}:=\frac{1}{2}\sum\limits_\text{spins}\int\!\ud\tilde{l}_1\ud\tilde{l}_2\frac{\theta(kp')}{kpkp'} \\
&\text{Re}\{\bar{M}_1^{21}M_1^{12},\bar{M}_1^{21}M_2^{12}+(1\leftrightarrow2),\bar{M}_2^{21}M_2^{12}\} \;.
\end{split}
\ee

We perform the Gaussian integrals over the transverse components of the photon momenta $l_{1\LCperp}$ and $l_{2\LCperp}$, and define a longitudinal momentum spectrum $\mathbb{P}(q)$ as
\be
\mathbb{P}=\int_0^1\ud q_1\ud q_2\theta(s_2)\mathbb{P}(q) \;,
\ee
where $q_i=kl_i/kp$ and $s_2=kp'/kp=1-q_1-q_2$. We also define $b_0=kp$, $s_1=1-q_1$, $s_{\bar{1}}=1-q_2$ and $s_0=1$.

\section{Exact analytical results}\label{Exact analytical results section}

\begin{figure*}
	\centering
	\includegraphics[width=\textwidth]{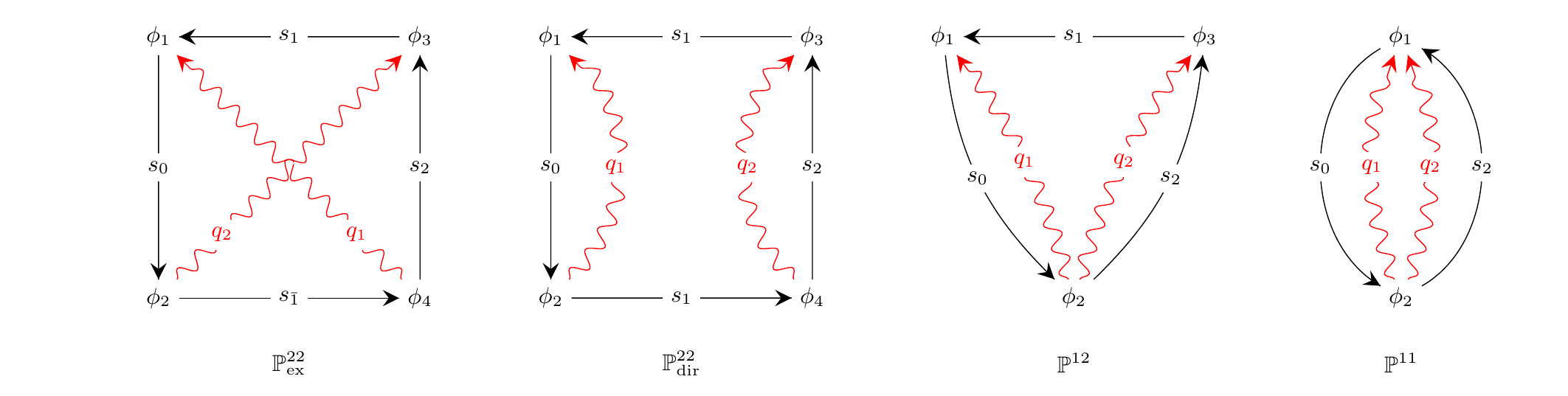}
	\caption{These diagrams illustrate the different contributions to the probability, with black, solid lines representing electrons and red, wavy lines for photons.}
	\label{ProbabilityDiagrams}
\end{figure*}

The different contributions are illustrated in Fig.~\ref{ProbabilityDiagrams}.
For the direct part of the simplest term we find
\be\label{P11dir}
\mathbb{P}_{\rm dir}^{11}(q)=\frac{\alpha^2s_2}{8\pi^2}\left[\frac{1}{s_1^2}+\frac{1}{s_{\bar{1}}^2}\right]\int\frac{-\ud\phi_{12}}{(\theta_{21}+i\epsilon)^2}\exp\left\{\frac{ir_{20}\Theta_{21}}{2b_0}\right\} \;,
\ee
where $r_{ij}=(1/s_i)-(1/s_j)$, $\ud\phi_{12}=\ud\phi_1\ud\phi_2$, $\theta_{ij}=\phi_i-\phi_j$,
$\Theta_{ij}:=\theta_{ij}M_{ij}^2$, and $M$ is an effective mass given by~\cite{Kibble:1975vz} 
\be
M_{ij}^2:=\langle\pi\rangle_{ij}^2=1+\langle{\bf a}^2\rangle_{ij}-\langle{\bf a}\rangle_{ij}^2 \;,
\ee
where the lightfront-time average is
\be
\langle F\rangle_{ij}:=\frac{1}{\theta_{ij}}\int_{\phi_j}^{\phi_i}\!\ud\phi\; F(\phi) \;,
\ee
and where the Lorentz momentum is given by
\be\label{LorentzMomentum}
\pi_\mu(\phi)=p_\mu-a_\mu+\frac{2ap-a^2}{2kp}k_\mu \;.
\ee
The exchange part $\mathbb{P}_{\rm ex}^{11}(\epsilon_1,\epsilon_2)$ depends nontrivially on the polarization vectors, but after summing over polarization vectors we find $\mathbb{P}_{\rm ex}^{11}(q)=0$, in contrast to the trident case~\cite{Dinu:2017uoj} where the corresponding term is nonzero.  
For the terms with three $x^\LCp$ integrals we find
\be\label{P12dir}
\begin{split}
\mathbb{P}_{\rm dir}^{12}(q)=\text{Re}&\frac{i\alpha^2}{8\pi^2b_0}\int\frac{\ud\phi_{123}
\theta(\theta_{31})(q_1q_2-s_2D_{12})}{s_1^3(\theta_{21}+i\epsilon)(\theta_{23}+i\epsilon)} \\
&\exp\left\{\frac{i}{2b_0}[r_{21}\Theta_{23}+r_{10}\Theta_{21}]\right\}+(1\leftrightarrow2) 
\end{split}
\ee
and
\be
\begin{split}
\mathbb{P}_{\rm ex}^{12}(q)=\text{Re}&\frac{-i\alpha^2}{8\pi^2b_0}\int\frac{\ud\phi_{123}\theta(\theta_{31})D_{12}}{s_{\bar{1}}(\theta_{21}+i\epsilon)(\theta_{23}+i\epsilon)} \\
&\exp\left\{\frac{i}{2b_0}[r_{21}\Theta_{23}+r_{10}\Theta_{21}]\right\}+(1\leftrightarrow2) \;,
\end{split}
\ee
where $D_{12}={\bf \Delta}_{12}\!\cdot\!{\bf \Delta}_{32}$ and
\be\label{DeltaDefinition}
{\bf\Delta}_{ij}:={\bf a}(\phi_i)-\langle{\bf a}\rangle_{ij} \;.
\ee
The $i\epsilon$ factors initially make the transverse momentum integrals converge and at this stage provide a prescription for how to avoid the singularities in the $\phi$ integrals. This is equivalent to a shift in the $\phi$-integration contours. From now on we leave these $i\epsilon$ factors implicit, they can be reinstated by replacing $\phi_{1,3}\to\phi_{1,3}-i\epsilon/2$ and $\phi_{2,4}\to\phi_{2,4}+i\epsilon/2$.
For the direct term with four $x^\LCp$ integrals we find
\be\label{P22dir}
\begin{split}
\mathbb{P}_{\rm dir}^{22}(q)=&-\frac{\alpha^2}{8\pi^2b_0^2}\int\ud\phi_{1234}\frac{\theta(\theta_{31})\theta(\theta_{42})}{s_1^2\theta_{21}\theta_{43}}e^{\frac{i}{2b_0}[r_{21}\Theta_{43}+r_{10}\Theta_{21}]} \\
&\bigg\{Q_{21}^{10}Q_{43}^{21} 
-\frac{q_1q_2}{4s_1^2}\bigg[({\bf w}_2-{\bf w}_1)\!\cdot\!({\bf w}_4-{\bf w}_3) \\
&+\frac{(1+s_1)(s_1+s_2)}{s_2}W_{1234}\bigg]\bigg\} 
+(1\leftrightarrow2) \;,
\end{split}
\ee
where 
\be\label{Qijdef}
Q_{ij}^{kl}=\frac{\kappa_{kl}}{2}\left(\frac{2ib_0}{r_{kl}\theta_{ij}}+{\bf\Delta}_{ij}\!\cdot\!{\bf\Delta}_{ji}+1\right)-1 \;,
\ee
\be\label{Wijdef}
\begin{split}
W_{ijkl}:=&(\mathbf{w}_{i}\!\times\!\mathbf{w}_{j})\!\cdot\!(\mathbf{w}_{k}\!\times\!\mathbf{w}_{l}) \\
=&(\mathbf{w}_{i}\!\cdot\!\mathbf{w}_{k})(\mathbf{w}_{j}\!\cdot\!\mathbf{w}_{l})-(\mathbf{w}_{i}\!\cdot\!\mathbf{w}_{l})(\mathbf{w}_{j}\!\cdot\!\mathbf{w}_{k}) \;,
\end{split}
\ee
$\kappa_{ij}=(s_i/s_j)+(s_j/s_i)$,
and where ${\bf w}_1={\bf\Delta}_{12}$, ${\bf w}_2={\bf\Delta}_{21}$, ${\bf w}_3={\bf\Delta}_{34}$ and ${\bf w}_4={\bf\Delta}_{43}$. 
For linear polarization we have $W_{ijkl}=0$. In contrast to the trident case, here we have a dot product between the two steps even for linear polarization.
Finally, the last term is given by
\be\label{P22ex}
\begin{split}
\mathbb{P}_{\rm ex}^{22}(q)=&\text{Re}\frac{\alpha^2}{16\pi^2b_0^2}\int\ud\phi_{1234}\frac{\theta(\theta_{42})\theta(\theta_{31})}{s_1s_{\bar{1}}s_2d_0} \\
&\exp\bigg\{\frac{i}{2b_0}\frac{q_1q_2}{s_1s_{\bar{1}}s_2d_0}\bigg(\theta_{23}\theta_{41}\left[\frac{\Theta_{41}}{q_2}+\frac{\Theta_{23}}{q_1}\right] \\
+&\theta_{21}\theta_{43}\left[\frac{\Theta_{43}}{s_2}-\Theta_{21}\right]+\theta_{31}\theta_{42}\left[\frac{\Theta_{31}}{s_1}-\frac{\Theta_{42}}{s_{\bar{1}}}\right]\bigg)\bigg\} \\
&\left\{F_0+f_0-\frac{2ib_0}{d_0}(f_1+z_1)+\left[\frac{2b_0}{d_0}\right]^2z_2\right\} \;,
\end{split}
\ee
where
\be
d_0=-\frac{\theta_{42}\theta_{31}}{s_1s_{\bar{1}}}+\frac{\theta_{21}\theta_{43}}{s_2} \;,
\ee
\be
\begin{split}
F_0=&-(\kappa_{02}+\kappa_{1{\bar{1}}})({\bf d}_1\!\cdot\!{\bf d}_4)({\bf d}_2\!\cdot\!{\bf d}_3) \\
&-(\kappa_{02}-\kappa_{1{\bar{1}}})({\bf d}_1\!\times\!{\bf d}_4)\!\cdot\!({\bf d}_2\!\times\!{\bf d}_3) \;,
\end{split}
\ee
\be
\begin{split}
f_0=-\frac{1}{s_1s_{\bar{1}}s_2}[&(s_1q_2{\bf d}_1-s_{\bar{1}}q_1{\bf d}_2)\!\cdot\!(s_{\bar{1}}q_2{\bf d}_4-s_1q_1{\bf d}_3)\\
&+(q_1{\bf d}_2+q_2s_2{\bf d}_4)\!\cdot\!(q_2{\bf d}_1+q_1s_2{\bf d}_3)] \;,
\end{split}
\ee
\be
\begin{split}
f_1=&\kappa_{02}\left(\theta_{21}{\bf d}_1\!\cdot\!{\bf d}_2-\frac{\theta_{43}}{s_2}{\bf d}_3\!\cdot\!{\bf d}_4\right)+ \\
&\kappa_{1{\bar{1}}}\left(\frac{\theta_{31}}{s_1}{\bf d}_1\!\cdot\!{\bf d}_3-\frac{\theta_{42}}{s_{\bar{1}}}{\bf d}_2\!\cdot\!{\bf d}_4\right)+ \\
&(\kappa_{02}+\kappa_{1{\bar{1}}})\left(\frac{\theta_{41}}{q_2}{\bf d}_1\!\cdot\!{\bf d}_4+\frac{\theta_{23}}{q_1}{\bf d}_2\!\cdot\!{\bf d}_3\right) \;,
\end{split}
\ee
\be
\begin{split}
z_1=&-\frac{q_1^2}{s_1q_2}\left(3-\frac{s_{\bar{1}}s_2}{s_1}\right)\phi_1+\frac{q_2^2}{s_{\bar{1}}q_1}\left(3-\frac{s_1s_2}{s_{\bar{1}}}\right)\phi_2 \\
&-\frac{q_2^2}{s_1s_2q_1}\left(3-\frac{s_{\bar{1}}}{s_1s_2}\right)\phi_3+\frac{q_1^2}{s_{\bar{1}}s_2q_2}\left(3-\frac{s_1}{s_{\bar{1}}s_2}\right)\phi_4
\end{split}
\ee
and
\be
z_2=-\kappa_{02}\frac{\theta_{43}\theta_{21}}{s_2}-\kappa_{1{\bar{1}}}\frac{\theta_{31}\theta_{42}}{s_1s_{\bar{1}}}+(\kappa_{02}+\kappa_{1{\bar{1}}})\frac{\theta_{23}\theta_{41}}{q_1q_2} \;.
\ee
The field enters the prefactor via
\be
\begin{split}
{\bf d}_1=&\frac{q_2}{s_{\bar{1}}s_2d_0}\left[-\frac{\theta_{43}\theta_{21}}{s_0}{\bf\Delta}_{12}+\frac{\theta_{42}\theta_{31}}{s_1}{\bf\Delta}_{13}+\frac{\theta_{23}\theta_{41}}{q_2}{\bf\Delta}_{14}\right]
\\
{\bf d}_2=&\frac{q_1}{s_1s_2d_0}\left[-\frac{\theta_{43}\theta_{21}}{s_0}{\bf\Delta}_{21}+\frac{\theta_{41}\theta_{23}}{q_1}{\bf\Delta}_{23}+\frac{\theta_{31}\theta_{42}}{s_{\bar{1}}}{\bf\Delta}_{24}\right]
\\	
{\bf d}_3=&\frac{q_1}{s_0s_{\bar{1}}d_0}\left[-\frac{\theta_{42}\theta_{31}}{s_1}{\bf\Delta}_{31}+\frac{\theta_{41}\theta_{23}}{q_1}{\bf\Delta}_{32}+\frac{\theta_{21}\theta_{43}}{s_2}{\bf\Delta}_{34}\right]
\\
{\bf d}_4=&\frac{q_2}{s_0s_1d_0}\left[\frac{\theta_{23}\theta_{41}}{q_2}{\bf\Delta}_{41}-\frac{\theta_{31}\theta_{42}}{s_{\bar{1}}}{\bf\Delta}_{42}+\frac{\theta_{21}\theta_{43}}{s_2}{\bf\Delta}_{43}\right] \;.
\end{split}
\ee
These expressions for $\mathbb{P}_{\rm ex}^{22}$ look remarkably similar to the corresponding ones in the trident case~\cite{Dinu:2017uoj}. In fact, one can show that~\eqref{P22ex} can be obtained from Eq.~(26) in~\cite{Dinu:2017uoj} by replacing $\phi_3-i\epsilon\leftrightarrow\phi_4+i\epsilon$ everywhere except in the step functions, taking into account that $s_3^{\rm trident}=-s_2=1-s_1-s_{\bar{1}}$ and $s_2^{\rm trident}=s_{\bar{1}}$, and multiplying the prefactor by an overall factor of $-1$, which seems natural since we now have identical bosons instead of fermions. One can also show that the individual terms in the prefactor can be obtained in the same way. In particular\footnote{Note that the expressions for ${\bf d}_i$ given here are simpler than the equivalent ones given in~\cite{Dinu:2017uoj}. There are only three instead of four terms, and ${\bf d}_i$ only involves ${\bf\Delta}_{ij}$ with $j\ne i$.}, ${\bf d}_3^{\rm trident}\to {\bf d}_4^{\rm here}$ and ${\bf d}_4^{\rm trident}\to {\bf d}_3^{\rm here}$. This means that $\mathbb{P}_{\rm ex}^{22}$ in double Compton scattering has the same symmetries as in the trident case and can be calculated in a similar way.  

\section{Two-step and one-step terms}\label{Two-step and one-step section}

In this section we compare~\eqref{P22dir} with the product of two single-photon emissions. To treat the electron spin we use the following representation of the Dirac matrices
\begin{align}
	\gamma^0=&
	\begin{pmatrix}
		0&0&1&0 \\
		0&0&0&1 \\
		1&0&0&0 \\
		0&1&0&0
	\end{pmatrix}
	&
	\gamma^1=&
	\begin{pmatrix}
		0&0&0&1 \\
		0&0&1&0 \\
		0&-1&0&0 \\
		-1&0&0&0
	\end{pmatrix}
	\nonumber\\
	\gamma^2=&
	\begin{pmatrix}
		0&0&0&-i \\
		0&0&i&0 \\
		0&i&0&0 \\
		-i&0&0&0
	\end{pmatrix}
	&
	\gamma^3=&
	\begin{pmatrix}
		0&0&1&0 \\
		0&0&0&-1 \\
		-1&0&0&0 \\
		0&1&0&0
	\end{pmatrix} 
\;
\end{align}
and the following spinor basis (cf.~\cite{Kogut:1969xa})
\be
u_{\scriptscriptstyle\uparrow}=\frac{1}{\sqrt{2p_\LCm}}\begin{pmatrix}1\\0\\2p_\LCm\\-p_1-ip_2 \end{pmatrix} \qquad
u_{\scriptscriptstyle\downarrow}=\frac{1}{\sqrt{2p_\LCm}}\begin{pmatrix}p_1-ip_2\\2p_\LCm\\0\\1 \end{pmatrix} \;.
\ee
This spinor basis is particularly convenient for the quantities that we calculate here.
An arbitrary spinor can be expressed as a linear combination of these,
\be
u=\cos\left(\frac{\rho}{2}\right)u_{\scriptscriptstyle\uparrow}+\sin\left(\frac{\rho}{2}\right)e^{i\lambda}u_{\scriptscriptstyle\downarrow} \;.
\ee
Instead of $\rho$ and $\lambda$ we express the spin dependence in terms of the components of the unit vector ${\bf n}$ that points in the average spin direction for ${\bf p}=0$, i.e. 
\be\label{initialSpinDirection}
{\bf n}:=\frac{1}{2}u^\dagger{\bf\Sigma}u({\bf p}=0)=\{\cos\lambda\sin\rho,\sin\lambda\sin\rho,\cos\rho\} \;,
\ee
where 
${\bf\Sigma}=i\{\gamma^2\gamma^3,\gamma^3\gamma^1,\gamma^1\gamma^2\}$.

Now, the probability of single-photon emission, summed over photon polarization and transverse momenta, is given by
\be\label{PCn0n1}
\mathbb{P}_C=\langle\mathbb{P}\rangle+{\bf n}_0\!\cdot\!{\bf P}_0+{\bf P}_1\!\cdot\!{\bf n}_1+{\bf n}_0\!\cdot\!{\bf P}_{01}\!\cdot\!{\bf n}_1 \;,
\ee 
where ${\bf n}_0$ and ${\bf n}_1$ are the spin vectors of the initial and final electron, respectively. 
The first term $\langle\mathbb{P}\rangle$ gives the probability averaged\footnote{So, $2\langle\mathbb{P}\rangle$ gives the probability summed rather than averaged over the final electron's spin.} over initial and final spins, 
\be\label{glueAve}
\langle\mathbb{P}\rangle=\frac{i\alpha}{4\pi b_0s_0^2}\int\!\frac{\ud^2\phi}{\theta_{21}}Q_{21}^{10}e^{\frac{ir_{10}}{2b_0}\Theta_{21}} \;.
\ee
The remaining terms give the spin dependence,
\be\label{glueP0}
{\bf P}_0=\frac{i\alpha}{4\pi b_0s_0^2}\frac{q_1}{s_0}\int\frac{\ud^2\phi}{\theta}\left[{\bf1}+\left[1+\frac{s_0}{s_1}\right]\hat{\bf k}\;{\bf X}\right]\!\cdot\!{\bf V}e^{\frac{ir_{10}}{2b_0}\Theta} \;,
\ee
\be\label{glueP1}
{\bf P}_1=\frac{i\alpha}{4\pi b_0s_0^2}\frac{q_1}{s_1}\int\frac{\ud^2\phi}{\theta}{\bf V}\!\cdot\!\left[{\bf1}+\left[1+\frac{s_1}{s_0}\right]{\bf X}\;\hat{\bf k}\right]e^{\frac{ir_{10}}{2b_0}\Theta} \;,
\ee
and
\be\label{glueP01}
\begin{split}
{\bf P}_{01}=\frac{i\alpha}{4\pi b_0s_0^2}&\int\frac{\ud^2\phi}{\theta}\left[\frac{q_1}{s_1}\hat{\bf k}\;{\bf X}-\frac{q_1}{s_0}{\bf X}\;\hat{\bf k}-\frac{q_1^2}{2s_0s_1}\hat{\bf k}\;\hat{\bf k}\right. \\
+&\left.\left[\frac{2ib_0}{r_{10}\theta}+D_1\right]\left[{\bf 1}+\frac{q_1^2}{2s_0s_1}\hat{\bf k}\;\hat{\bf k}\right]\right]e^{\frac{ir_{10}}{2b_0}\Theta} \;,
\end{split}
\ee
where $\hat{\bf k}=\{0,0,1\}$, $\hat{\bf k}\;{\bf X}\!\cdot\!{\bf V}=\hat{\bf k}({\bf X}\!\cdot\!{\bf V})$ etc., $D_1={\bf\Delta}_{12}\!\cdot\!{\bf\Delta}_{21}$ and
\be\label{XVdef}
{\bf X}=\frac{1}{2}({\bf w}_2+{\bf w}_1)
\qquad
{\bf V}=\frac{1}{2}{\bm\sigma}_2\!\cdot\!({\bf w}_2-{\bf w}_1) \;,
\ee
where the Pauli matrix is given as usual by
\be
{\bm\sigma}_2=\begin{pmatrix}0&-i\\i&0\end{pmatrix} \;.
\ee
Note that ${\bf n}_1$ gives the average spin direction for ${\bf p}_1=0$ and we have integrated over $p_{1\LCperp}$ with ${\bf n}_1$ fixed. Regardless of whether or not this is the most directly relevant quantity for spin-sensitive experiments, we show below that~\eqref{PCn0n1} can be very useful for studying multi-photon emission.
For a detailed investigation of spin effects in nonlinear Compton scattering see~\cite{Seipt:2018adi}.    

In evaluating these expressions we can put $s_0=1$. One reason for keeping $s_0$ explicit is that it helps us to glue together two single-photon emissions, which one might expect to be done according to 
\be
\frac{1}{4}\sum_{{\bf n}_0,{\bf n}_1,{\bf n}_2}\mathbb{P}_C(s_0\to1)\mathbb{P}_C(s_0\to s_1,s_1\to s_2)+(1\leftrightarrow2) \;,
\ee
where one factor of $1/2$ comes from averaging over the spin of the initial electron and another factor of $1/2$ comes from the symmetrization. 
We can write this as
\be\label{glueTwoAve}
\begin{split}
\frac{2^2}{2}\Big\langle&[\langle\mathbb{P}\rangle+{\bf n}_0\!\cdot\!{\bf P}_0+{\bf P}_1\!\cdot\!{\bf n}_1+{\bf n}_0\!\cdot\!{\bf P}_{01}\!\cdot\!{\bf n}_1] \\
&[\langle\mathbb{P}\rangle+{\bf n}_1\!\cdot\!{\bf P}_0+{\bf P}_1\!\cdot\!{\bf n}_2+{\bf n}_1\!\cdot\!{\bf P}_{01}\!\cdot\!{\bf n}_2]\Big\rangle \\
=&2(\langle\mathbb{P}\rangle\langle\mathbb{P}\rangle+{\bf P}_1\!\cdot\!\langle{\bf n}_1{\bf n}_1\rangle\!\cdot\!{\bf P}_0) \;,
\end{split}
\ee
where the factor of $2^2$ is due to the replacement of the sum of two spins with their average for ${\bf n}_1$ and ${\bf n}_2$, and we have omitted the arguments of the probability terms (the second factor in each term is obtained by making the appropriate replacements in~\eqref{glueAve}, \eqref{glueP0}, \eqref{glueP1} and~\eqref{glueP01}). 
It is easy to show that the $\langle\mathbb{P}\rangle\langle\mathbb{P}\rangle$ term gives the $QQ$-term in~\eqref{P22dir}. The remaining terms are more subtle. We first note that these terms can be expressed as
\be\label{VXXVterm}
\begin{split}
&-\frac{1}{4}\left[({\bf w}_2-{\bf w}_1)\!\cdot\!({\bf w}_4-{\bf w}_3)+\frac{(1+s_1)(s_1+s_2)}{s_2}W_{1234}\right] \\
&={\bf V}_1\!\cdot\!\left[{\bf 1}+\left(1+\frac{s_1}{s_0}\right)\left(1+\frac{s_1}{s_2}\right){\bf X}_1\;{\bf X}_2\right]\!\cdot\!{\bf V}_2 \\
&={\bf V}_1\!\cdot\!\left[{\bf 1}+\left(1+\frac{s_1}{s_0}\right){\bf X}_1\;\hat{\bf k}\right]\!\cdot\!\left[{\bf 1}+\left(1+\frac{s_1}{s_2}\right)\hat{\bf k}\;{\bf X}_2\right]\!\cdot\!{\bf V}_2
\;,
\end{split}
\ee
where ${\bf V}_1$ and ${\bf X}_1$ are given by~\eqref{XVdef}, and ${\bf V}_2$ and ${\bf X}_2$ are obtained by replacing $\phi_2\to\phi_4$, $\phi_1\to\phi_3$ in~\eqref{XVdef}.   
This should be compared with the corresponding term in ${\bf P}_1\!\cdot\!\langle{\bf n}_1{\bf n}_1\rangle\!\cdot\!{\bf P}_0$, i.e.
\be\label{VnnVterm}
\begin{split}
{\bf V}_1\cdot\!&\left[{\bf 1}+\left(1+\frac{s_1}{s_0}\right){\bf X}_1\;\hat{\bf k}\right]\!\cdot\!\langle{\bf n}_1{\bf n}_1\rangle \\
&\!\cdot\!\left[{\bf 1}+\left(1+\frac{s_1}{s_2}\right)\hat{\bf k}\;{\bf X}_2\right]\!\cdot\!{\bf V}_2 \;.
\end{split}
\ee
The gluing approach works if~\eqref{VnnVterm} gives~\eqref{VXXVterm} after summing over ${\bf n}_1$. 
In~\eqref{glueTwoAve} we have only used $\langle1\rangle=1$ and $\langle{\bf n}\rangle=0$.
For linear polarization with ${\bf a}\propto{\bf e}_1$ we have ${\bf X}\!\cdot\!{\bf V}=0$, and then we can simply sum over ${\bf n}_1=\pm{\bf e}_2$. For arbitrary polarization we cannot in general obtain~\eqref{VXXVterm} from~\eqref{VnnVterm} unless we let ${\bf n}_1$ depend on both $\phi_1$ and $\phi_2$ (or $\phi_3$ and $\phi_4$). For arbitrary polarization in the LCF regime we have
\be
{\bm w}_1\approx-\frac{\theta_{21}}{2}{\bm a}'(\sigma_{21}) \qquad {\bm w}_2\approx\frac{\theta_{21}}{2}{\bm a}'(\sigma_{21}) \;,
\ee
where $\sigma_{ij}=(\phi_i+\phi_j)/2$,
so then we can neglect the ${\bf X}$ terms and obtain~\eqref{VXXVterm} by choosing the spin direction to be perpendicular to the locally constant field and $\hat{\bf k}$, i.e. either ${\bf n}_1=\pm\hat{\bf k}\times\hat{\bf a}(\sigma_{21})$ or ${\bf n}_1=\pm\hat{\bf k}\times\hat{\bf a}(\sigma_{43})$, where $\hat{\bf a}={\bf a}/|{\bf a}|$. In the LCF regime and for linear polarization our gluing approach reduces to the one in~\cite{Morozov:1975uah,King:2014wfa}, and then we have the same one-step/two-step separation as in~\cite{Morozov:1975uah,King:2014wfa}.  

The reason that the naive gluing approach does not always work is because we actually have a sum over the spin of the intermediate electron already on the amplitude level, so, instead of having on the probability level just one sum over ${\bf n}_1$, one should have one sum for the amplitude and a second sum for its complex conjugate, 
\be
\mathbb{P}=\sum_{{\bf n}_1,{\bf n}'_1}\dots u({\bf n}_1,p_1)\bar{u}({\bf n}_1,p_1)\dots u({\bf n}'_1,p_1)\bar{u}({\bf n}_1',p_1)\dots
\ee
where the sum is over $\pm{\bf n}$ (or $\rho$ and $\rho+\pi$) for some ${\bf n}$.
While the momentum $p_1$ is the same in the amplitude and its complex conjugate, the spins ${\bf n}_1$ and ${\bf n}'_1$ need not be the same. Let 
\be
\mathbb{P}_{\rm same}=\sum_{{\bf n}_1={\bf n}'_1}\dots \qquad \mathbb{P}_{\rm diff}=\sum_{{\bf n}_1\ne{\bf n}'_1}\dots
\ee
Compared to~\eqref{glueTwoAve}, one can show that 
\be
\mathbb{P}_{\rm same}=2(\langle\mathbb{P}\rangle\langle\mathbb{P}\rangle+{\bf P}_1\!\cdot\!{\bf n}_1{\bf n}_1\!\cdot\!{\bf P}_0)
\ee
and
\be
\mathbb{P}_{\rm diff}=-2({\bf P}_1\!\times\!{\bf n}_1)\!\cdot\!({\bf n}_1\!\times\!{\bf P}_0) \;.
\ee
These clearly depend on the spin directions $\pm{\bf n}_1$ one chooses to sum over, but their sum is independent of ${\bf n}_1$,
\be
\mathbb{P}_{\rm same}+\mathbb{P}_{\rm diff}=2(\langle\mathbb{P}\rangle\langle\mathbb{P}\rangle+{\bf P}_1\!\cdot\!{\bf P}_0)
\ee
As we saw above, for linear polarization or in the LCF regime we can choose ${\bf n}_1$ such that $\mathbb{P}_{\rm diff}$ vanishes, but in general we need to include this term. Fortunately, our results suggests a simple cure for the naive gluing approach: Include factors of $2$ in the overall prefactor as if we only had one sum over ${\bf n}_1$ as above, and then simplify using  $\langle1\rangle=1$, $\langle{\bf n}_1\rangle=0$ and importantly $\langle{\bf n}_1{\bf n}_1\rangle={\bf 1}$, where the last ingredient is motivated by the contribution from ${\bf n}_1\ne{\bf n}'_1$. We show in the next section that this simple procedure also works for triple and quadruple nonlinear Compton scattering. Note that this improved gluing procedure gives us the dominant term for sufficiently long pulses, for any polarization and field shape, and we can in particular go beyond the usual LCF regime (where gluing first order, albeit spin-averaged, processes is a basic component of PIC codes for $a_0\gg1$) and consider $a_0\sim1$.

In the gluing approach one also has to make sure that the second step happens after the first, which can be done by including a step function $\theta(\sigma_{43}-\sigma_{21})$.
In~\eqref{P22dir} we have two step functions, which we deal with in the same way as in~\cite{Dinu:2017uoj}, i.e. we write $\mathbb{P}_{\rm dir}^{22}=\mathbb{P}_{\rm dir}^{22\to2}+\mathbb{P}_{\rm dir}^{22\to1}$ where $\mathbb{P}_{\rm dir}^{22\to2}$ and $\mathbb{P}_{\rm dir}^{22\to1}$ are obtained, respectively, from the first and second term in
\be\label{thetaSeparate}
\begin{split}
\theta(\theta_{42})\theta(\theta_{31})=&\theta(\sigma_{43}-\sigma_{21})\bigg\{1-\\
&\theta\left(\frac{|\theta_{43}-\theta_{21}|}{2}-[\sigma_{43}-\sigma_{21}]\right)\bigg\} \;.
\end{split}
\ee  
It is $\mathbb{P}_{\rm two}:=\mathbb{P}_{\rm dir}^{22\to2}$ (rather than $\mathbb{P}_{\rm dir}^{22}$) which we refer to as the two-step term. Although it can be obtained from the above gluing approach, we can obtain it without reference to the gluing approach by selecting one part of the exact/total probability. This part scales quadratically in the volume/pulse length and dominates for sufficiently long pulses.

\section{Multiphoton emission}\label{MultiphotonSection}

\subsection{Triple Compton}

\begin{figure}
	\centering
	\includegraphics[width=0.8\linewidth]{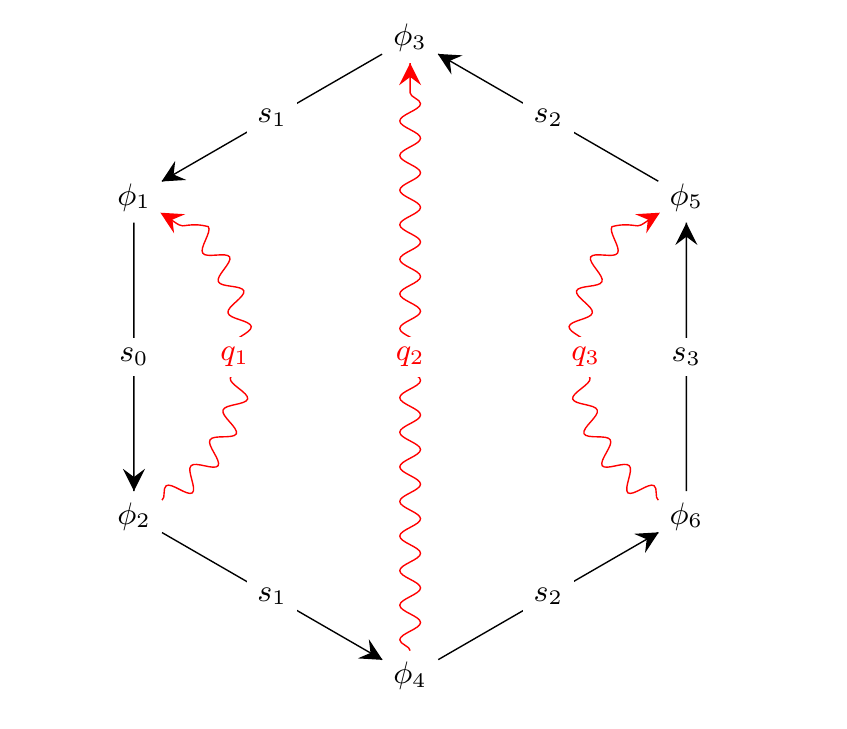}
	\caption{This diagram illustrates $\mathbb{P}_{\rm dir}^{33}$ for triple Compton scattering.}
	\label{hexagonDiagram}
\end{figure}

In this section we calculate the three-step part of triple nonlinear Compton scattering, i.e. the part of the probability of three-photon emission that dominates for long pulses, illustrated in Fig.~\ref{hexagonDiagram}. 
The emission of three photons by an electron colliding with a single photon has been studied in~\cite{Lotstedt:2012zz}, but to the best of our knowledge nonlinear triple Compton scattering has not been studied in the regime we are interested in here.
This is in principle a straightforward generalization of our results for the two-step part of double Compton scattering, except that it takes more time to simplify the prefactor. 
After some simplification we find
\be\label{P33dir}
\begin{split}
\mathbb{P}_{\rm dir}^{33}=&\frac{-i\alpha^3}{48\pi^3b_0^3s_1^2s_2^2}\int\ud^6\phi\frac{\theta(\theta_{64})\theta(\theta_{42})\theta(\theta_{53})\theta(\theta_{31})}{\theta_{65}\theta_{43}\theta_{21}} \\
&e^{\frac{i}{2b_0}\left[r_{32}\Theta_{65}+r_{21}\Theta_{43}+r_{10}\Theta_{21}\right]}
\bigg\{Q_{21}^{10}Q_{43}^{21}Q_{65}^{32} \\
+&Q_{21}^{10}\frac{q_2q_3}{s_2^2}{\bf V}_2\!\cdot\!\left[{\bf 1}+\left[1+\frac{s_2}{s_1}\right]\left[1+\frac{s_2}{s_3}\right]{\bf X}_2{\bf X}_3\right]\!\cdot\!{\bf V}_3 \\ 
+&\frac{q_1q_2}{s_1^2}{\bf V}_1\!\cdot\!\left[{\bf 1}+\left[1+\frac{s_1}{s_0}\right]\left[1+\frac{s_1}{s_2}\right]{\bf X}_1{\bf X}_2\right]\!\cdot\!{\bf V}_2Q_{65}^{32} \\
+&\frac{q_1q_3}{s_1s_2}{\bf V}_1\!\cdot\!\left[
-\frac{q_2^2}{2s_1s_2}\left[1+\frac{s_1}{s_0}\right]\left[1+\frac{s_2}{s_3}\right]{\bf X}_1{\bf X}_3 \right. \\
+&\left[\frac{2ib_0}{r_{21}\theta_{43}}+D_2\right]\left[{\bf 1}+\frac{\kappa_{21}}{2}\left[1+\frac{s_1}{s_0}\right]\left[1+\frac{s_2}{s_3}\right]{\bf X}_1{\bf X}_3\right] \\
+&\left.\frac{q_2}{s_2}\left[1+\frac{s_1}{s_0}\right]{\bf X}_1{\bf X}_2-\frac{q_2}{s_1}\left[1+\frac{s_2}{s_3}\right]{\bf X}_2{\bf X}_3\right]\!\cdot\!{\bf V}_3\bigg\} \\
&+\text{permutations}\;,
\end{split}
\ee 
where $s_1=1-q_1$, $s_2=1-q_1-q_2$, $s_3=1-q_1-q_2-q_3>0$,
$D_2={\bf\Delta}_{34}\cdot{\bf\Delta}_{43}$,
${\bf V}_3$ and ${\bf X}_3$ are obtained by replacing $\phi_2\to\phi_6$, $\phi_1\to\phi_5$ in~\eqref{XVdef},  
and ``permutation'' is an instruction to sum over all permutations of the emitted photons. 
Note that the exponential part is a simple generalization from single and double Compton scattering.
Compare this with the result of the gluing approach described in the previous section, which in this case gives
\be
\begin{split}
	\frac{2^3}{3!}\Big\langle
	&[\langle\mathbb{P}\rangle+{\bf n}_0\!\cdot\!{\bf P}_0+{\bf P}_1\!\cdot\!{\bf n}_1+{\bf n}_0\!\cdot\!{\bf P}_{01}\!\cdot\!{\bf n}_1] \\
	&[\langle\mathbb{P}\rangle+{\bf n}_1\!\cdot\!{\bf P}_0+{\bf P}_1\!\cdot\!{\bf n}_2+{\bf n}_1\!\cdot\!{\bf P}_{01}\!\cdot\!{\bf n}_2] \\
	&[\langle\mathbb{P}\rangle+{\bf n}_2\!\cdot\!{\bf P}_0+{\bf P}_1\!\cdot\!{\bf n}_3+{\bf n}_2\!\cdot\!{\bf P}_{01}\!\cdot\!{\bf n}_3]\Big\rangle \\
	&+\text{permutations} \\
	=&\frac{4}{3}\Big(\langle\mathbb{P}\rangle\langle\mathbb{P}\rangle\langle\mathbb{P}\rangle \\
	&+\langle\mathbb{P}\rangle{\bf P}_1\!\cdot\!\langle{\bf n}_2{\bf n}_2\rangle\!\cdot\!{\bf P}_0
	+{\bf P}_1\!\cdot\!\langle{\bf n}_1{\bf n}_1\rangle\!\cdot\!{\bf P}_0\langle\mathbb{P}\rangle \\
	&+{\bf P}_1\!\cdot\!\langle{\bf n}_1{\bf n}_1\rangle\!\cdot\!{\bf P}_{01}\!\cdot\!\langle{\bf n}_2{\bf n}_2\rangle\!\cdot\!{\bf P}_0
	\Big) \\
	&+\text{permutations} \;,
\end{split}
\ee
where the arguments are again suppressed. 
The factor of $2^3$ comes from the (initial) assumption that we are summing over two spin states for ${\bf n}_1$, ${\bf n}_2$ and ${\bf n}_3$,
and for linear polarization ${\bf a}\propto{\bf e}_1$ we can obtain~\eqref{P33dir} by summing over ${\bf n}_1=\pm{\bf e}_2$ and ${\bf n}_2=\pm{\bf e}_2$.
For arbitrary polarization we can obtain~\eqref{P33dir} from the following procedure: We write an overall factor of $2^N/N!$ and replace all sums with $\langle...\rangle$, and then we simplify with $\langle1\rangle=1$, $\langle{\bf n}\rangle=0$ and $\langle{\bf n}{\bf n}\rangle={\bf 1}$. Note again that it is the replacement $\langle{\bf n}{\bf n}\rangle={\bf 1}$ that allows us to obtain all terms in the general case. 

We already have a factorization into the different steps (with appropriate spin/polarization sums) before performing the transverse momentum integrals.
Because the momenta are related via momentum conservation, one might have thought that performing the transverse momentum integrals could have led to a non-factorized result. To understand why we still have factorization, note first that after integrating single Compton scattering over the transverse momenta of the final particles, the results~\eqref{glueAve}, \eqref{glueP0}, \eqref{glueP1} and~\eqref{glueP01} do not depend on the initial transverse momentum. Similarly, after performing the integrals over the transverse momenta of the final electron and the photon emitted from the last vertex, this step becomes independent on the other transverse momenta and hence factorizes, and then the same thing happens for the second step.

In analogy to~\eqref{thetaSeparate}, we define the three-step $\mathbb{P}_{\rm three}$ by replacing the product of step functions in~\eqref{P33dir} according to
\be
\theta(\theta_{64})\theta(\theta_{42})\theta(\theta_{53})\theta(\theta_{31})\to\theta(\sigma_{65}-\sigma_{43})\theta(\sigma_{43}-\sigma_{21}) \;.
\ee  

\subsection{Quadruple Compton}

\begin{figure}
	\centering
	\includegraphics[width=0.8\linewidth]{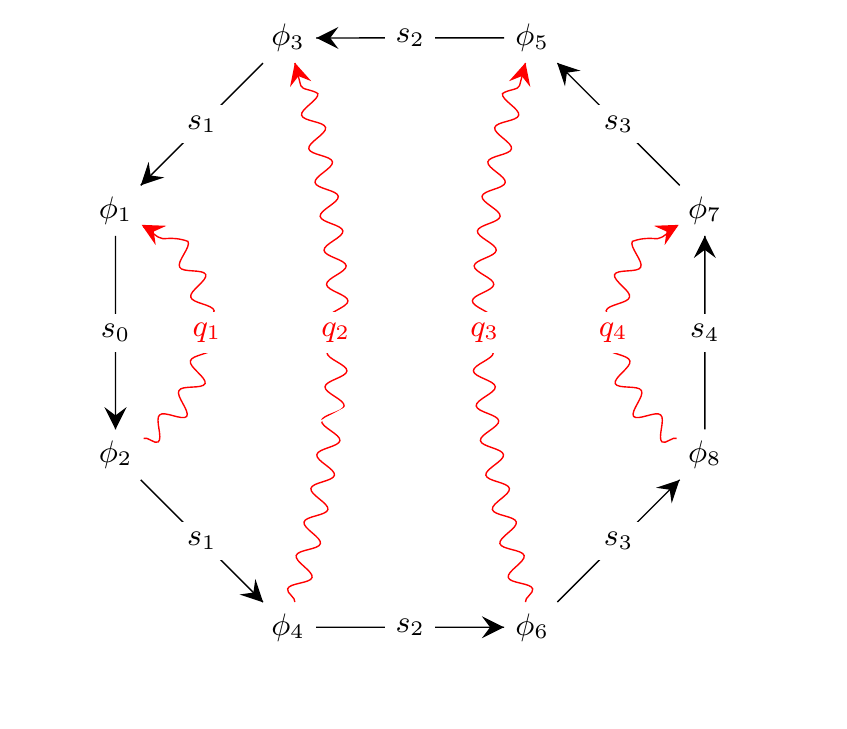}
	\caption{This diagram illustrates $\mathbb{P}_{\rm dir}^{44}$ for quadruple Compton scattering.}
	\label{quadrupleDiagram}
\end{figure}
We have also checked that the above gluing procedure gives the correct result for quadruple nonlinear Compton scattering, i.e. the emission of four photons, which is illustrated in Fig.~\ref{quadrupleDiagram}. Here our gluing approach is not only useful for interpreting the expressions, it is also very useful for simplifying the complicated prefactor. We have checked that the result can be expressed neatly and compactly as
\be
\begin{split}
	\frac{2^4}{4!}\Big\langle
	&[\langle\mathbb{P}\rangle+{\bf n}_0\!\cdot\!{\bf P}_0+{\bf P}_1\!\cdot\!{\bf n}_1+{\bf n}_0\!\cdot\!{\bf P}_{01}\!\cdot\!{\bf n}_1] \\
	&[\langle\mathbb{P}\rangle+{\bf n}_1\!\cdot\!{\bf P}_0+{\bf P}_1\!\cdot\!{\bf n}_2+{\bf n}_1\!\cdot\!{\bf P}_{01}\!\cdot\!{\bf n}_2] \\
	&[\langle\mathbb{P}\rangle+{\bf n}_2\!\cdot\!{\bf P}_0+{\bf P}_1\!\cdot\!{\bf n}_3+{\bf n}_2\!\cdot\!{\bf P}_{01}\!\cdot\!{\bf n}_3] \\
	&[\langle\mathbb{P}\rangle+{\bf n}_3\!\cdot\!{\bf P}_0+{\bf P}_1\!\cdot\!{\bf n}_4+{\bf n}_3\!\cdot\!{\bf P}_{01}\!\cdot\!{\bf n}_4]\Big\rangle \\
	&+\text{permutations} \;,
\end{split}	  
\ee
where $\langle1\rangle=1$, $\langle{\bf n}\rangle=0$ and $\langle{\bf n}{\bf n}\rangle={\bf 1}$ for each ${\bf n}_i$.
Even with the help of an advanced symbolic-calculation program such as Mathematica, obtaining or confirming this result by a direct calculation can take some time. Instead of calculating the prefactor from the trace of a long expression, we replaced all factors of $\slashed{p}_i+1$ (which would appear in the trace) by sums of $u\bar{u}$ expressed with a particular spinor representation. Note again that we only obtain all terms by replacing $\langle{\bf n}{\bf n}\rangle={\bf 1}$ to account for the terms that would be missing if one replaces the double sums over the spins of the intermediate states, i.e. ${\bf n}_1$, ${\bf n}_2$ and ${\bf n}_3$ in this case, with single sums as explained above for double Compton scattering.    

Although we have not yet proved that this gluing procedure works at arbitrarily high orders, the fact that it does work for double, triple and quadruple Compton scattering suggests that we have a method for obtaining the exact $N$-step part for $N$-Compton scattering for arbitrary $N$, where the $N$-step dominates for sufficiently long pulses.
We plan to further study this gluing approach and to generalize it to other higher-order processes involving more than one fermion, like the trident process.

\section{Saddle-point approximation}\label{Saddle point approximation section}

In this section we obtain saddle-point approximations, which help us to understand the structure and relative importance of the various terms. We can expect these approximations to be good for $\chi\ll1$ as long as $q_1$ and $q_2$ are not too small, so we are in particular outside the infrared region and do not have to worry about IR divergences. We also have to assume that $a_0$ is not too small.
The calculations are very similar to the ones in~\cite{Dinu:2017uoj}, except that this time, in order to avoid IR divergences, we do not integrate over the longitudinal momenta. We consider linearly polarized fields, $a(\phi)=a_0 f(\phi)$. In this section we focus on the dominant contribution from a single saddle point located around a single field maximum.  

\subsection{Locally constant fields}\label{LCFsaddleApprox1}

We consider first the LCF regime where we can expand the probability in $1/a_0\ll1$. For the one-step terms we find
\be
\mathbb{P}_{\rm dir}^{11}=\frac{\alpha^2}{16\pi^\frac{3}{2}}\frac{s_2}{\sqrt{r_{20}}}\left[\frac{1}{s_1^2}+\frac{1}{s_{\bar{1}}^2}\right]\int\!\frac{\ud\phi}{b_0}\chi^\frac{3}{2}\exp\left\{-\frac{2r_{20}}{3\chi}\right\} \;,
\ee
where $\chi(\phi)=a_0f'(\phi)b_0$,
\be
\begin{split}
	\mathbb{P}_{\rm dir}^{12}=&\frac{\alpha^2}{48\pi^\frac{3}{2}}\frac{4q_1q_2+s_2}{s_1^3\sqrt{r_{20}}}\left[\frac{1}{r_{10}}-\frac{1}{r_{21}}\right]\int\!\frac{\ud\phi}{b_0}\chi^\frac{3}{2}e^{-\frac{2r_{20}}{3\chi}} \\
	&+(1\leftrightarrow2) \;,
\end{split}
\ee
\be
\begin{split}
	\mathbb{P}_{\rm ex}^{12}=&\frac{\alpha^2}{48\pi^\frac{3}{2}}\frac{1}{s_{\bar{1}}\sqrt{r_{20}}}\left[\frac{1}{r_{10}}-\frac{1}{r_{21}}\right]\int\!\frac{\ud\phi}{b_0}\chi^\frac{3}{2}e^{-\frac{2r_{20}}{3\chi}} \\
	&+(1\leftrightarrow2) \;,
\end{split}
\ee
\be\label{dironeleadlcfsaddle}
\mathbb{P}^{22\to1}=-\frac{\alpha^2}{4\pi^\frac{3}{2}}\sqrt{r_{20}}\left[\frac{q_1}{q_2}+\frac{q_2}{q_1}+\frac{s_1s_{\bar{1}}}{q_1q_2}\right]\int\!\frac{\ud\phi}{b_0}\chi^\frac{1}{2}e^{-\frac{2r_{20}}{3\chi}}
\ee
and
\be
\mathbb{P}_{\rm ex}^{22}=-\mathbb{P}^{22\to1} \;,
\ee
and for the two-step term we find
\be\label{twosteplcfsaddle}
\begin{split}
	\mathbb{P}^{22\to2}=&\frac{\alpha^2}{8\pi}\sqrt{\frac{q_1q_2}{s_2}}\frac{1}{s_1}\left[\frac{q_1}{q_2}+\frac{q_2}{q_1}+\frac{s_1s_{\bar{1}}}{q_1q_2}\right] \\
	&\int\!\frac{\ud\sigma_1}{b_0}\int_{\sigma_1}\!\frac{\ud\sigma_2}{b_0}\sqrt{\chi(\sigma_{2})\chi(\sigma_{1})}e^{-\frac{2r_{10}}{3\chi(\sigma_1)}-\frac{2r_{21}}{3\chi(\sigma_2)}} \\
	&+(1\leftrightarrow2)\;.
\end{split}
\ee
For a constant field for which $\chi(\phi)$ is zero outside an interval of length $\Delta\phi$, we simply have $\int\ud\phi\to\Delta\phi$ and $\int\ud\sigma_1\ud\sigma_2\theta(\sigma_2-\sigma_1)\to\Delta\phi^2/2$.  

For a pulsed field we can also perform the remaining $\phi$ integral with the saddle-point approximation. Let us for simplicity assume one dominant field maximum with $f'(0)=1$, $f''(0)=0$ and $f^{(3)}(0)=-\zeta$. By performing the above $\phi$-integrals with the saddle-point method we find that the results are obtained from the corresponding constant field results by replacing $\chi\to\chi_0$, 
\be
\Delta\phi\to\sqrt{\frac{3\pi\chi_0}{\zeta r_{20}}}
\ee 
for the one-step terms, so for example
\be
\mathbb{P}^{22\to1}=-\frac{\sqrt{3}\alpha^2 a_0}{4\pi\sqrt{\zeta}}\left[\frac{q_1}{q_2}+\frac{q_2}{q_1}+\frac{s_1s_{\bar{1}}}{q_1q_2}\right]e^{-\frac{2r_{20}}{3\chi}} \;,
\ee
and
\be
\frac{\Delta\phi^2}{2}\left(\frac{1}{s_1}+\frac{1}{s_{\bar{1}}}\right)\to\frac{3\pi\chi_0}{2\zeta}\left(\frac{1}{s_1\sqrt{r_{21}r_{10}}}+\frac{1}{s_{\bar{1}}\sqrt{r_{2{\bar{1}}}r_{{\bar{1}}0}}}\right)
\ee
for the two-step term, which simplifies to
\be\label{P22to2LCFpulse}
\mathbb{P}^{22\to2}=\frac{3\alpha^2a_0^2}{8\zeta}\left[\frac{q_1}{q_2}+\frac{q_2}{q_1}+\frac{s_1s_{\bar{1}}}{q_1q_2}\right]e^{-\frac{2r_{20}}{3\chi}} \;.
\ee

We see a few things that are similar to the trident case: All terms have the same exponential, and $\mathbb{P}^{11}$ and $\mathbb{P}^{12}$ are smaller than $\mathbb{P}^{22\to1}$ by a factor of $\chi$. We also see that the exchange terms are on the same order of magnitude as the direct part of the one-step term. In fact, here $\mathbb{P}_{\rm ex}^{22}$ cancels $\mathbb{P}^{22\to1}$ to leading order, so the $\chi$ expansion of the prefactor of $\mathbb{P}_{\rm one}$ starts at one order higher than the leading order of the direct part of $\mathbb{P}_{\rm one}$. This also means that $\mathbb{P}^{11}$ and $\mathbb{P}^{12}$ contribute to the first nonzero order, in contrast to the trident case. Thus, the exchange term is even more important for the one-step part for double Compton scattering.  

In the trident case we could compare our saddle-point approximations for the direct terms with previous constant-crossed field results. For double Compton scattering, on the other hand, we are not aware of any previous approximations for hard photons with which we could compare our saddle-point results. The $\chi<1$ approximation in e.g.~\cite{Morozov:1975uah} is for the probability integrated over the photon momenta, which has a different form because of the contribution from softer photons. We have, however, checked that our approximations agree with the exact expressions in~\cite{King:2014wfa} for the direct part of the one-step term, see Appendix~\ref{lcfcomparison}. 

The exponential part of the above terms can be written
\be\label{PNexpLCF}
\exp\left\{-\frac{2}{3\chi}\frac{\sum_{i=1}^Nq_i}{1-\sum_{i=1}^Nq_i}\right\} \;,
\ee
where $N=2$. Assuming again one dominant field maximum, for triple Compton scattering it follows from~\eqref{P33dir} and $r_{ij}+r_{jk}=r_{ik}$ that
$\mathbb{P}_{\rm three}\sim\eqref{PNexpLCF}$ with $N=3$. Similarly, for quadruple Compton scattering we find $\mathbb{P}_{\rm four}\sim\eqref{PNexpLCF}$ with $N=4$. This suggests a simple generalization to the emission of an arbitrary number of photons.

\subsection{Sauter pulse}

In the previous section we considered $a_0\gg1$ which allows us to consider an arbitrary pulse shape. Here we will consider a particular pulse shape, namely a Sauter pulse $a(\phi)=a_0\tanh\phi$, which allows us to obtain explicit analytical expressions also for $a_0\gtrsim1$, i.e. to go beyond the LCF approximation. The calculation is very similar to the corresponding one in~\cite{Dinu:2017uoj} for the trident case. In particular, we have a saddle point at the same values of the $\phi_i$ variables as in~\cite{Dinu:2017uoj}, independently of $q_i$. For the ``two-step'' term we find
\be\label{P22to2Sauter}
\begin{split}
\mathbb{P}^{22\to2}=&\frac{\alpha^2}{8}\left[\frac{q_1}{q_2}+\frac{q_2}{q_1}+\frac{s_1s_{\bar{1}}}{q_1q_2}\right] \\
&\frac{a_0^2\exp\left\{-\frac{r_{20}}{\chi}a_0[(1+a_0^2)\text{arccot}\,a_0-a_0]\right\}}{(1+a_0^2)\text{arccot}\,a_0[(1+a_0^2)\text{arccot}\,a_0-a_0]} \;.
\end{split}
\ee
For $a_0\gg1$ we recover~\eqref{P22to2LCFpulse} to leading order. For the ``one-step'' terms we find
\be
\mathbb{P}^{22\to1}=-\frac{2}{\pi}\text{arctan}\sqrt{1-\frac{a_0}{(1+a_0^2)\text{arccot}\,a_0}}\mathbb{P}^{22\to2}
\ee
and
\be
\mathbb{P}_{\rm ex}^{22}=-\mathbb{P}^{22\to1} \;,
\ee
while $\mathbb{P}_{\rm dir}^{11}$ and $\mathbb{P}^{12}$ are again smaller than the above terms by a factor of $\chi$. 
Notice that these expressions are very similar to the ones in~\cite{Dinu:2017uoj} for trident: the dependence on $a_0$ in the exponent is exactly the same as in~\cite{Dinu:2017uoj}, and the relation between $\mathbb{P}^{22\to1}$ and $\mathbb{P}^{22\to2}$ is also exactly the same. We also find that the (leading order) exchange term $\mathbb{P}_{\rm ex}^{22}$ is on the same order of magnitude as the (leading order) direct terms $\mathbb{P}^{22\to1}$ and $\mathbb{P}^{22\to2}$. 
Here, though, $\mathbb{P}_{\rm ex}^{22}$ is not only on the same order of magnitude, but it in fact cancels $\mathbb{P}^{22\to1}$ to leading order in $\chi$; this generalizes the $a_0\gg1$ results in the previous section to $a_0\gtrsim1$. Note also that the dependence on the momenta remains the same as in the $a_0\gg1$ limit.

\subsection{Monochromatic field}

For a monochromatic field we can again find saddle-point approximations for general $a_0\gtrsim1$. For this field there are many saddle points that contribute. We begin in this section with the simplest ones, which are the same as those we studied~\cite{Dinu:2017uoj} for the integrated trident probability,
\be
\theta_{21}=\theta_{43}=2i\,\text{arcsinh}\frac{1}{a_0} \qquad \sigma_{21}=n_1\pi \qquad \sigma_{43}=n_2\pi \;.
\ee
These already give a good approximation to the locally averaged spectrum. In the next section we include additional saddle points that give oscillations to the spectrum.   
For the two-step term we have saddle points both for $n_1=n_2$ and $n_1<n_2$, where the two photons are emitted at the same and different field maxima, respectively. For the contribution from one saddle point with $n_1=n_2$ we find
\be\label{P22to2MonoSame}
\begin{split}
\mathbb{P}_{n_1=n_2}^{22\to2}=&\frac{\alpha^2}{8}\left[\frac{q_1}{q_2}+\frac{q_2}{q_1}+\frac{s_1s_{\bar{1}}}{q_1q_2}\right] \\
&\frac{\exp\left\{\!-\frac{r_{20}}{2\chi}a_0\left[(2+a_0^2)\text{arccsch}\,a_0-\sqrt{1+a_0^2}\right]\right\}}{\sqrt{1+a_0^2}\text{arccsch}\,a_0[\sqrt{1+a_0^2}\text{arccsch}\,a_0-1]} \;.
\end{split}
\ee
We have again the same function of $a_0$ in the exponent as in the trident case~\cite{Dinu:2017uoj}. For $a_0\gg1$ we recover the LCF approximation~\eqref{P22to2LCFpulse} from~\eqref{P22to2MonoSame}. The contributions from one saddle point (with $n_1=n_2$) to the dominant one-step terms are given by
\be\label{P22to1Mono}
\mathbb{P}_{n_1=n_2}^{22\to1}=-\frac{2}{\pi}\text{arctan}\sqrt{1-\frac{1}{\sqrt{1+a_0^2}\text{arccsch}\,a_0}}\mathbb{P}_{n_1=n_2}^{22\to2}
\ee
and
\be
\mathbb{P}_{{\rm ex},n_1=n_2}^{22}=-\mathbb{P}_{n_1=n_2}^{22\to1} \;.
\ee
The relation~\eqref{P22to1Mono} is exactly the same as in the trident case~\cite{Dinu:2017uoj}, and, as for the LCF and Sauter cases, we find that the exchange term cancels the direct part of the one-step term to leading order. The other one-step terms, $\mathbb{P}_{\rm dir}^{11}$, $\mathbb{P}_{\rm dir}^{12}$ and $\mathbb{P}_{\rm ex}^{12}$, are again smaller by a factor of $\chi\ll1$, but have to be included if one is interested in the first nontrivial order of the total one-step term, since $\mathbb{P}_{\rm ex}^{22}$ cancels $\mathbb{P}^{22\to1}$ to leading order.
These expressions give the contribution from one field maximum with the shape of a sinusoidal field, and for $a_0\gtrsim1$ they are on the same order of magnitude. If we have a sinusoidal field with several equivalent field maxima, then the two-step term dominates because it also receives contributions from $n_1<n_2$ and not only $n_1=n_2$, which means that it scales quadratically in the number of oscillations compared to the linear scaling of the one-step terms.
In contrast to the trident case, here the contributions from $n_2=n_1+2n-1$ are different from the ones from $n_2=n_1+2n$, where
\be
\mathbb{P}_{n_2=n_1+2n}^{22\to2}=2\mathbb{P}_{n_1=n_2}^{22\to2}
\ee 
and
\be
\begin{split}
\mathbb{P}&_{n_2=n_1+2n}^{22\to2}-\mathbb{P}_{n_2=n_1+2n-1}^{22\to2}=\frac{\alpha^2}{4}s_2\left[\frac{1}{s_1^2}+\frac{1}{s_{\bar{1}}^2}\right] \\
&\frac{\exp\left\{\!-\frac{r_{20}}{2\chi}a_0\left[(2+a_0^2)\text{arccsch}\,a_0-\sqrt{1+a_0^2}\right]\right\}}{\sqrt{1+a_0^2}\text{arccsch}\,a_0[\sqrt{1+a_0^2}\text{arccsch}\,a_0-1]} \;.
\end{split}
\ee
This difference has the same $a_0$ dependence but a different dependence on the momenta in the prefactor. This difference is due to the $({\bf w}_2-{\bf w}_1)\!\cdot\!({\bf w}_4-{\bf w}_3)$ term in~\eqref{P22dir}.

For $a_0\gg1$ we recover the LCF results. For $a_0\ll1$ the exponent goes as
\be
e^{-\frac{r_{20}}{2\chi}a_0\left[(2+a_0^2)\text{arccsch}\,a_0-\sqrt{1+a_0^2}\right]}\sim a_0^{r_{20}/b_0} \;,
\ee
which is the expected perturbative scaling: Momentum conservation at $\mathcal{O}(a_0^N)$,
\be
(p+Nk)_\mu=(p'+l_1+l_2)_\mu \;,
\ee   
implies
\be\label{N0def}
\begin{split}
N=\frac{1}{2b_0}&\left\{r_{20}+\frac{(l_1-q_1p)_\LCperp^2}{q_1}+\frac{(l_2-q_2p)_\LCperp^2}{q_2} \right. \\
&\hspace{0.5cm}\left.+\frac{(p'-s_2p)_\LCperp^2}{s_2}\right\}\geq\frac{r_{20}}{2b_0}=:N_0 \;.
\end{split}
\ee
Thus, the exponent scales as $a_0^{2N_0}$, where $N_0$ is the minimum number of photons from the background field that need to be absorbed in order to emit two photons with longitudinal momenta $q_1$ and $q_2$.

For a Sauter pulse the exponent scales as
\be
e^{-\frac{r_{20}}{\chi}a_0[(1+a_0^2)\text{arccot}\,a_0-a_0]}\sim e^{-\frac{\pi r_{20}}{2b_0}} \;.
\ee
Since the Sauter pulse has a wide Fourier transform with only exponential decay (which is slow in this context), this scaling agrees with the absorption of a single photon from the background field with (Fourier) frequency $N_0k_0$ (cf.~\cite{Dinu:2017uoj,Torgrimsson:2017pzs}).

\subsection{General antisymmetric potential}

Both the Sauter pulse and the sinusoidal field considered in the previous two sections fall in the class of fields that have antisymmetric potentials, $a(-\phi)=-a(\phi)$. In this section we derive the probability for such fields, assuming for simplicity one dominant field maximum and linear polarization but without choosing a specific field shape. Let $a(\phi)=a_0f(\phi)$. We have a saddle point at 
\be
\theta=2i z \qquad z=-if^{-1}\left(\frac{i}{a_0}\right)>0 \;,
\ee
where $f^{-1}$ is the inverse of $f$, and, as before, $\phi=\varphi=\eta=0$. We can still perform the integrals with the saddle-point method and the results are quite simple,
\be
\begin{split}
\mathbb{P}^{22\to2}=&\frac{\alpha^2}{8}\left[\frac{q_1}{q_2}+\frac{q_2}{q_1}+\frac{s_1s_{\bar{1}}}{q_1q_2}\right] \\
&\frac{\exp\left\{\!-\frac{r_{20}}{\chi}a_0z\left[1+a_0^2\langle f^2\rangle\right]\right\}}{za_0f'(iz)[za_0f'(iz)-1]} \;,
\end{split}
\ee
where
\be
\langle f^2\rangle=\frac{1}{2iz}\int_{-iz}^{iz}\!\ud u\; f^2(u) \;,
\ee
and for the one-step terms we find
\be
\mathbb{P}^{22\to1}=-\frac{2}{\pi}\text{arctan}\sqrt{1-\frac{1}{a_0 z f'(iz)}}\;\mathbb{P}^{22\to2}
\ee
and
\be
\mathbb{P}_{\rm ex}^{22}=-\mathbb{P}^{22\to1} \;.
\ee
In deriving these expressions we have assumed that $f'(iz)>0$ and $0<1-\frac{1}{a_0 z f'(iz)}<1$, which we will justify below. Note that $\mathbb{P}_{\rm ex}^{22}$ cancels $\mathbb{P}^{22\to1}$ to leading order independently on the field shape. 

\begin{figure}
\includegraphics[width=0.9\linewidth]{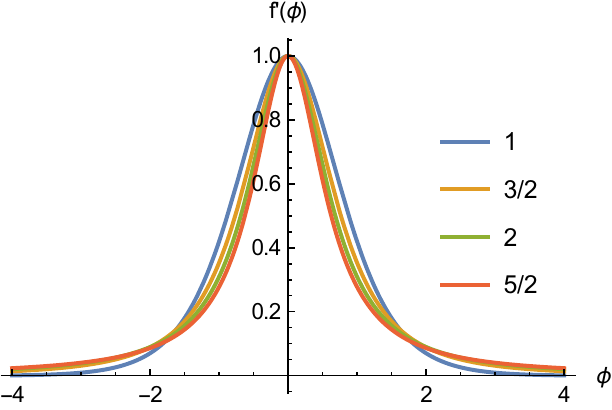}
\caption{This figure illustrates four examples from the class of fields defined by~\eqref{implicitFieldsDer} with $c=1,3/2,2,5/2$.}
\label{dfbplot-fig}
\end{figure}

To make these expressions more explicit, we consider the class of fields defined implicitly via~\cite{Gies:2015hia}
\be\label{implicitFieldsDer}
f'(\phi)=[1-f^2(\phi)]^c \;,
\ee 
where each $c$ characterizes a different field shape, see Fig~\ref{dfbplot-fig}. 
For example, $c=1/2$ and $c=1$ give us the sinusoidal field (or rather one peak of it) and  the Sauter pulse, respectively. For general $c$ the field $f(\phi)$ is given implicitly in terms of a hypergeometric function by
\be\label{phiFromf}
\phi=f{}_2F_1\left(\frac{1}{2},c,\frac{3}{2},f^2\right) \;.
\ee 
For this class of fields we find simple explicit expressions for the probability, using for the exponent
\be\label{hyperSimpExp}
\begin{split}
a_0z[1+&a_0^2\langle f^2\rangle]= \\
&{}_2F_1\left[\frac{1}{2},c,\frac{3}{2},-\frac{1}{a_0^2}\right]-\frac{1}{3}{}_2F_1\left[\frac{3}{2},c,\frac{5}{2},-\frac{1}{a_0^2}\right] \;,
\end{split}
\ee
and for the prefactor
\be\label{hyperSimpPre}
\begin{split}
za_0f'(iz)=&\left(1+\frac{1}{a_0^2}\right)^c{}_2F_1\left(\frac{1}{2},c,\frac{3}{2},-\frac{1}{a_0^2}\right)
\\
=&{}_2F_1\left(1,c,\frac{3}{2},\frac{1}{1+a_0^2}\right)
 \;.
\end{split} 
\ee
It is easy to check that $0<1-\frac{1}{a_0 z f'(iz)}<1$ for general $a_0$ and $c$.
Now everything is explicitly expressed in terms of $a_0$ and $c$, which in turn only enter in the arguments of ${}_2F_1$. For $c=1/2$ and $c=1$ we recover the results in the previous two sections for a monochromatic field and a Sauter pulse, and for arbitrary $c$ we recover for $a_0\gg1$ the LCF results above by expanding in $1/a_0$ and using the relation $c=\zeta/2$. The hypergeometric functions also simplify more generally for $c=j/2$ where $j$ is an integer. For example, for $c=3/2$, which corresponds to $f'(\phi)=(1+\phi^2)^{-3/2}$, we find a particularly simple prefactor
\be
\begin{split}
\mathbb{P}&^{22\to2}=\frac{\alpha^2}{8}\left[\frac{q_1}{q_2}+\frac{q_2}{q_1}+\frac{s_1s_{\bar{1}}}{q_1q_2}\right] \\
&\frac{a_0^4}{1+a_0^2}\exp\left\{\!-\frac{r_{20}}{\chi}a_0\left[\sqrt{1+a_0^2}-a_0^2\text{arccsch}\,a_0\right]\right\}
\end{split}
\ee
and
\be
\mathbb{P}^{22\to1}=-\frac{2}{\pi}\text{arccot}\sqrt{1+a_0^2}\; \mathbb{P}^{22\to2} \;,
\ee
while for $c=5/2$ we find a simple exponent
\be
\begin{split}
\mathbb{P}^{22\to2}=&\frac{\alpha^2}{8}\left[\frac{q_1}{q_2}+\frac{q_2}{q_1}+\frac{s_1s_{\bar{1}}}{q_1q_2}\right] \\
&\frac{9a_0^8\exp\left\{\!-\frac{2r_{20}}{3\chi}\frac{a_0}{\sqrt{1+a_0^2}}\right\}}{4+20a_0^2+31a_0^4+15a_0^6}
\end{split}
\ee
and
\be
\mathbb{P}^{22\to1}=-\frac{2}{\pi}\text{arctan}\sqrt{\frac{2+5a_0^2}{2+5a_0^2+3a_0^4}}\; \mathbb{P}^{22\to2} \;.
\ee

The prefactors above have been derived under the assumption that $a_0$ is not too small. The exponents, on the other hand, have the expected perturbative limit for $a_0\ll1$: For $c>1/2$ the exponent becomes independent of the field strength,
\be\label{PperturbGammaGamma}
a_0\ll1: \qquad \mathbb{P}\sim\exp\left(-\frac{r_{20}}{b_0}\frac{\sqrt{\pi}}{2}\frac{\Gamma(c-1/2)}{\Gamma(c)}\right) \;.
\ee
In the perturbative regime the minimum energy that needs to be absorbed is $N_0\omega$, where $N_0$ is given by~\eqref{N0def}. For a monochromatic field, $N_0$ photons have to be absorbed. For $c>1/2$, on the other hand, the Fourier transform $a(\omega_f)$ has a slow, exponential decay, which (since $|a(\omega)|^{2N_0}/|a(N_0\omega)|^2\sim a_0^{2(N-1)}\ll1$) means that the process occurs already at first order, with the absorption of a single photon with $\omega_f=N_0\omega$. 
At $\omega_f\gg\omega$, the exponential behavior of the Fourier transform is governed by the singularity $\phi_s$ closest to the real axis, i.e. $a(\omega_f)\sim e^{-|\omega_f\phi_s/\omega|}$. We find from the $|f|\to\infty$ limit of~\eqref{phiFromf} a singularity at
\be
\phi_s=i\frac{\sqrt{\pi}}{2}\frac{\Gamma(c-1/2)}{\Gamma(c)} \;.
\ee  
At $\omega_f=N_0\omega$ this implies $|a(\omega_f)|^2\sim\eqref{PperturbGammaGamma}$, so~\eqref{PperturbGammaGamma} agrees with what one can expect to find in the perturbative limit.

\subsection{Single Compton scattering}

While the results in the previous section are for double Compton scattering, it should be clear that the same method can be used to derive similar expressions for other plane-wave processes, like nonlinear Breit-Wheeler or trident pair production. In this subsection we simply give the corresponding result for single Compton scattering. The saddle-point approximation is obtained e.g. from~\eqref{glueAve} in the same way as for the above expressions for double Compton scattering, and we find
\be\label{PCsaddleAntiSym}
\mathbb{P}_{\rm C}(q)=\frac{\alpha}{2r_{10}}\frac{(\kappa_{10}-1)\exp\left\{\!-\frac{r_{10}}{\chi}a_0z\left[1+a_0^2\langle f^2\rangle\right]\right\}}{za_0f'(iz)\sqrt{1-\frac{1}{za_0f'(iz)}}} \;,
\ee
where $s_1$ now corresponds to the final electron. 
For the class of fields defined by~\eqref{implicitFieldsDer} we can again obtain explicit expressions using~\eqref{hyperSimpExp} and~\eqref{hyperSimpPre}. 

In~\eqref{PCsaddleAntiSym} we have integrated over all $\phi$ variables. In order to compare with the literature for the LCF regime we need to leave one $\phi$ integral. We find for $a_0\gg1$
\be
\mathbb{P}_{\rm C}(q)=\int\frac{\ud\sigma}{b_0}\frac{\alpha}{2\sqrt{\pi}}(\kappa_{10}-1)\sqrt{\frac{\chi}{r_{10}}}\exp\left\{-\frac{2r_{10}}{3\chi}\right\} \;,
\ee	
which for high-energy photons with $1-q\ll1$ agrees perfectly with Eq.~(19) in~\cite{Blackburn:2017dpn}.

\section{Saddle-point approximation for interference effects}\label{saddlePointInterferenceSection}

In this section we study fields with many oscillations and with several saddle points that lead to oscillations in the spectrum. We choose the following field
\be\label{aGaussSin}
a(\phi)=a_0\sin\phi\; e^{-(\phi/\mathcal{T})^2} \;.
\ee
Since the exponential part of the integrand for the $N$-step part of the $N$-photon emission probability is a simple generalization of the $N=1$ case, we focus here on single Compton scattering.
See~\cite{Mackenroth:2012rb,NarozhnyFofanov95,Mackenroth:2010jk,Mackenroth:2010jr,SeiptLaserPhys13,Meuren:2015mra,Seipt:2015rda,Nousch:2015pja} for other semi-classical/saddle-point approximations, in particular~\cite{NarozhnyFofanov95,Mackenroth:2010jk,Mackenroth:2010jr,SeiptLaserPhys13,Seipt:2015rda} for single Compton scattering and~\cite{Mackenroth:2012rb} for double Compton scattering. Note though that we consider different quantities here. 

\begin{figure}
\includegraphics[width=0.9\linewidth]{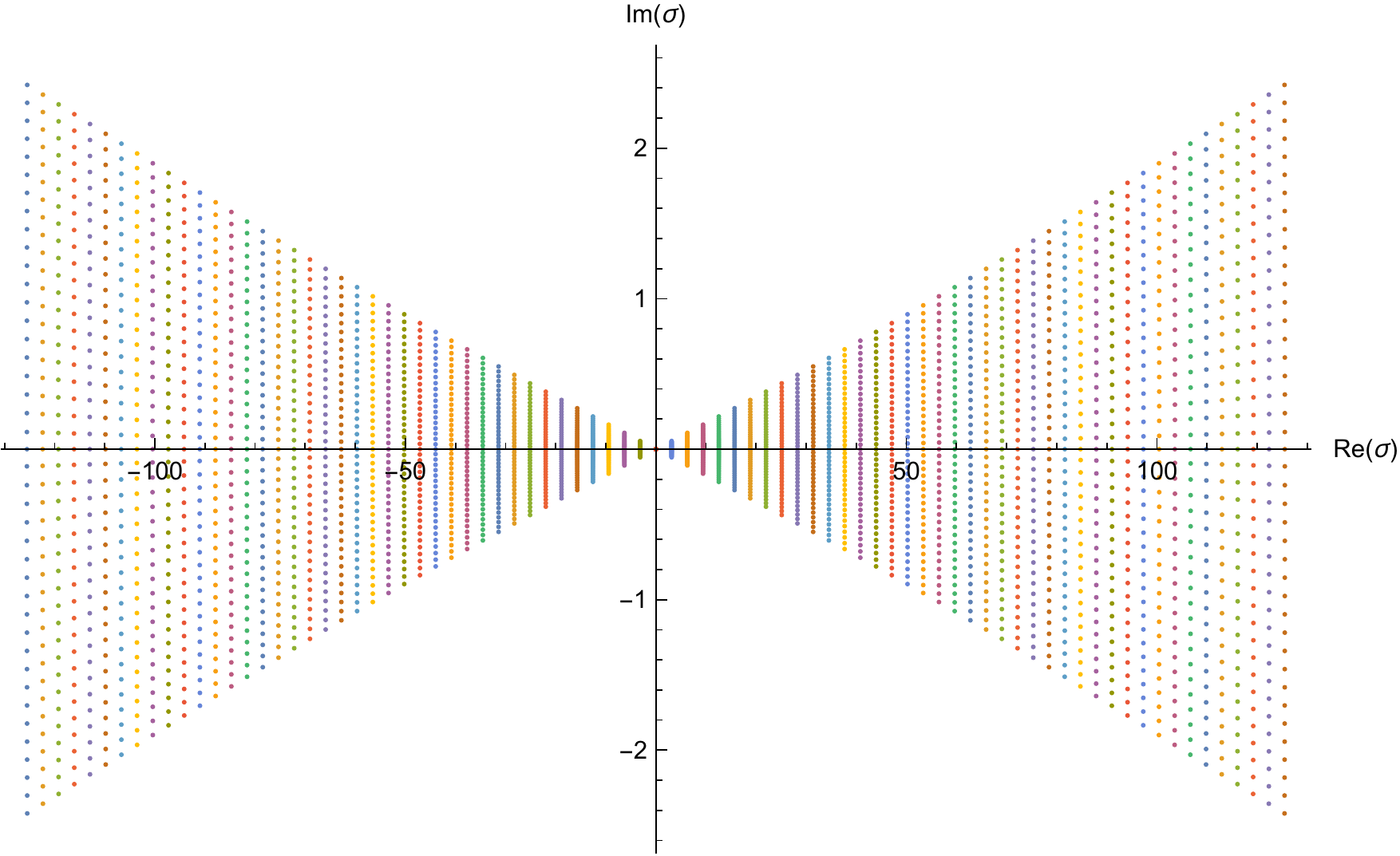} \\
\includegraphics[width=0.9\linewidth]{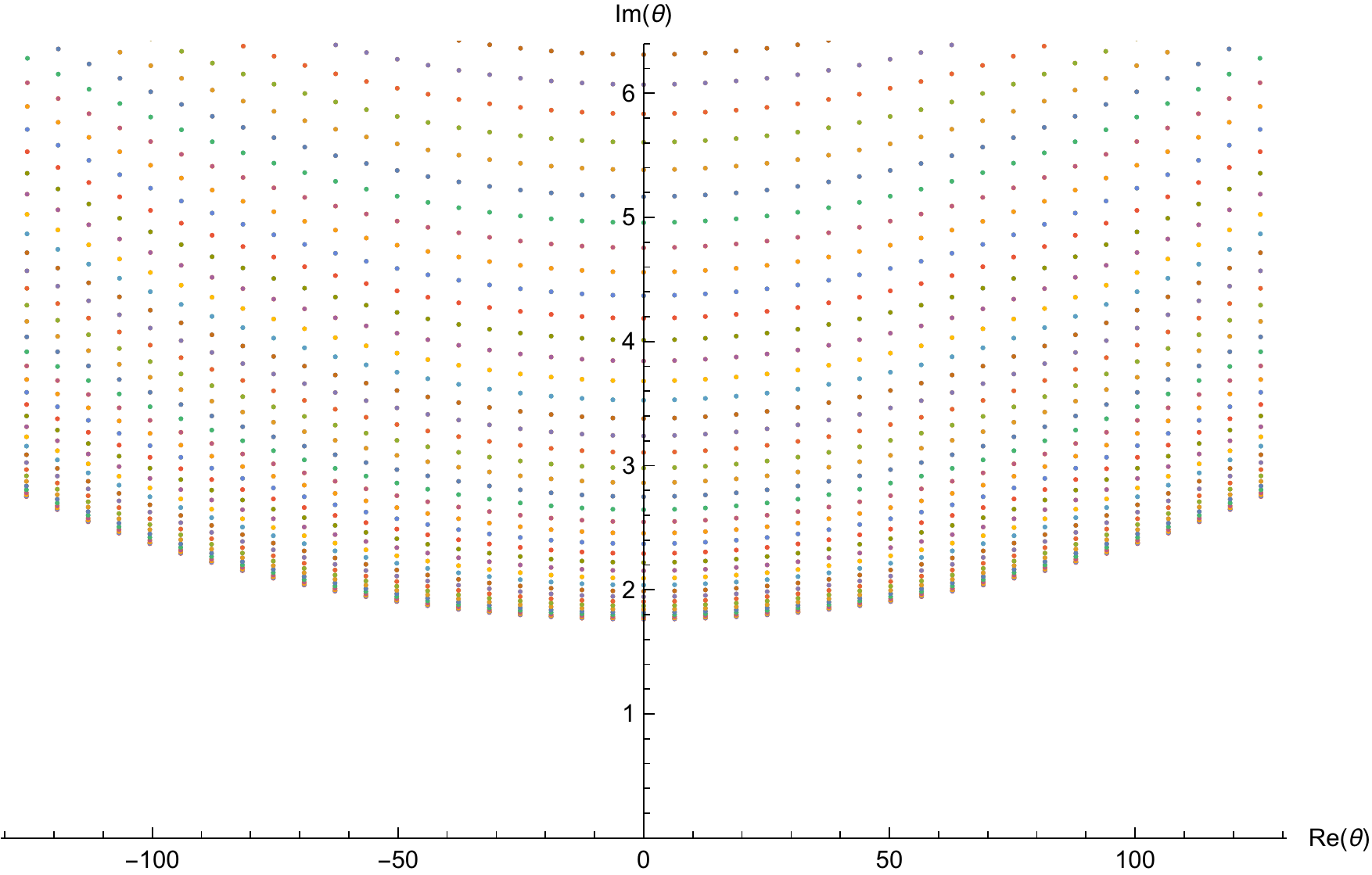}
\caption{Some of the saddle points for a pulsed field with $a_0=1$ and $\mathcal{T}=80$. For comparison, note that for a monochromatic field, $\mathcal{T}=\infty$, the saddle points are given by~\eqref{monoSaddles1}, i.e. $\text{Im }\sigma=0$ and $\text{Im }\theta\approx1.76$.}
\label{saddlePoints-fig}
\end{figure}

The saddle points for~\eqref{glueAve} are determined by
\be
\frac{\partial\Theta_{ij}}{\partial\sigma_{ij}}=\frac{\partial\Theta_{ij}}{\partial\theta_{ij}}=0 \;,
\ee 
where, again, $\theta_{ij}=\phi_i-\phi_j$ and $\sigma_{ij}=(\phi_i+\phi_j)/2$.
Note that these equations only depend on the field parameters, $a_0$ and $\mathcal{T}$ in our case, but not on the momenta $b_0$ or $q_i$. To obtain the saddle points for finite $\mathcal{T}$, we first find the saddle points for a monochromatic field ($\mathcal{T}=\infty$) and then use them as starting points for a numerical root-finding of the corresponding saddle points for finite $\mathcal{T}$. Depending on how large/small $\mathcal{T}$ is, one may find it useful to obtain the saddle points by first considering a sequence of $\mathcal{T}$ values between $\mathcal{T}=\infty$ and the desired value, and/or by starting with a simple $a_0$ value and gradually change to a more difficult one, cf. the numerical continuation in~\cite{Schneider:2018huk}. The saddle-point equations can be expressed in terms of the ``prefactor functions'' $\Delta$~\eqref{DeltaDefinition} as
\be
0=\frac{\partial\Theta_{21}}{\partial\sigma_{21}}=\Delta_{21}^2-\Delta_{12}^2
\ee
and
\be
0=\frac{\partial\Theta_{21}}{\partial\theta_{21}}=1+\frac{1}{2}\left(\Delta_{21}^2+\Delta_{12}^2\right) \;,
\ee
which imply that all saddle points, for any field shape, are determined by $\Delta=\pm i$. The saddle points are therefore necessarily complex. For the monochromatic field we find saddle points at
\be\label{monoSaddles1}
\{\sigma,\theta\}=\left\{n\pi,2i\text{arcsinh}\left[\frac{1}{a_0}\right]+2m\pi\right\} \;,
\ee 
where $n,m=0,\pm1,\pm2,\dots$. 
We also have saddle points at
\be\label{monoSaddles2}
\{\sigma,\theta\}=\left\{\left(n-\frac{1}{2}\right)\pi,\eta_m\right\} \;,
\ee
where $\eta_m$ can be found numerically by using $2i\text{arcsinh}\left[\frac{1}{a_0}\right]+(2m-1)\pi$ as starting points.
In Fig.~\ref{saddlePoints-fig} we show saddle points for a pulsed field, which are obtained numerically with the ones in~\eqref{monoSaddles1} as starting points. 
For the first set of saddle points~\eqref{monoSaddles1} we find
\be
\Delta_{21}=-\Delta_{12}=i(-1)^{n+m} 
\ee
and for the second set~\eqref{monoSaddles2}
\be
\Delta_{21}=\Delta_{12}=-i(-1)^{n+m} \;. 
\ee
Note that these values of $\Delta$ do not change as we decrease the pulse length from $\mathcal{T}=\infty$ to a finite $\mathcal{T}$. 

Let now $\delta\sigma=\sigma-\sigma_{\rm saddle}$ and $\delta\theta=\theta-\theta_{\rm saddle}$.
The quadratic fluctuation of $\Theta$ around any point can be expressed in terms of $\Delta$ and the derivative of the field $f'$, but at the saddle points we can simplify using $\Delta=\pm i$. To leading order we can put $\delta\sigma,\delta\theta\to0$ in the pre-exponential part of the integrand. Having expanded $\Theta$ to second order in $\delta\sigma$ and $\delta\theta$, we now have simple Gaussian integrals for each $n$ and $m$ which we perform analytically, i.e. we have for each saddle point
\be
\int\ud\delta\sigma\ud\delta\theta\exp\left\{-c_1\delta\sigma^2-c_2\delta\theta^2-c_3\delta\sigma\delta\theta\right\} \;,
\ee
where the coefficients $c_i$ are in general complex and obtained by finding the saddle points numerically.

For a monochromatic field we find with~\eqref{monoSaddles1} an exponential part given by (cf.~\eqref{P22to2MonoSame})
\be
\begin{split}
	&\exp\left\{\frac{ir_1}{2b_0}\Theta_{21}\right\}(n,m)=\exp\left\{\frac{ir_1a_0}{\chi}\left(1+\frac{a_0^2}{2}\right)m\pi\right\} \\
	&\exp\left\{-\frac{r_1a_0}{2\chi}\left[(2+a_0^2)\text{arcsinh}\frac{1}{a_0}-\sqrt{1+a_0^2}\right]\right\} \;.
\end{split}
\ee 
From this we see that the saddle points with $m\ne0$ lead to oscillations in the spectrum around the $m=0$ result studied in the previous section. We also see that the frequencies of these oscillations increase with decreasing $\chi$ or increasing $a_0$. Since this saddle-point approximation is good for small $\chi$, these oscillations can be relatively fast and hence contribute less after integrating over the momenta. 

\begin{figure}
\includegraphics[width=\linewidth]{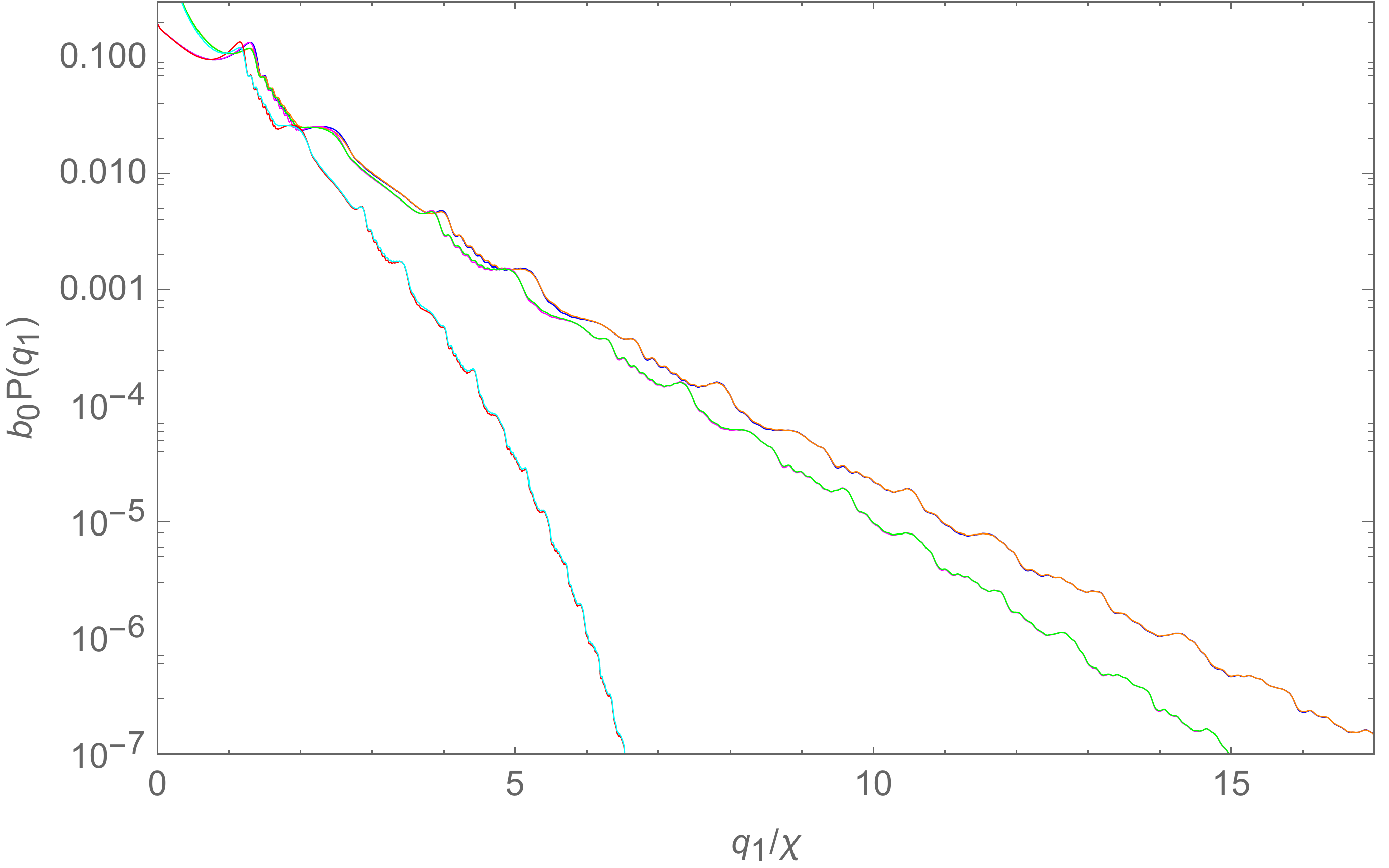} \\
\includegraphics[width=\linewidth]{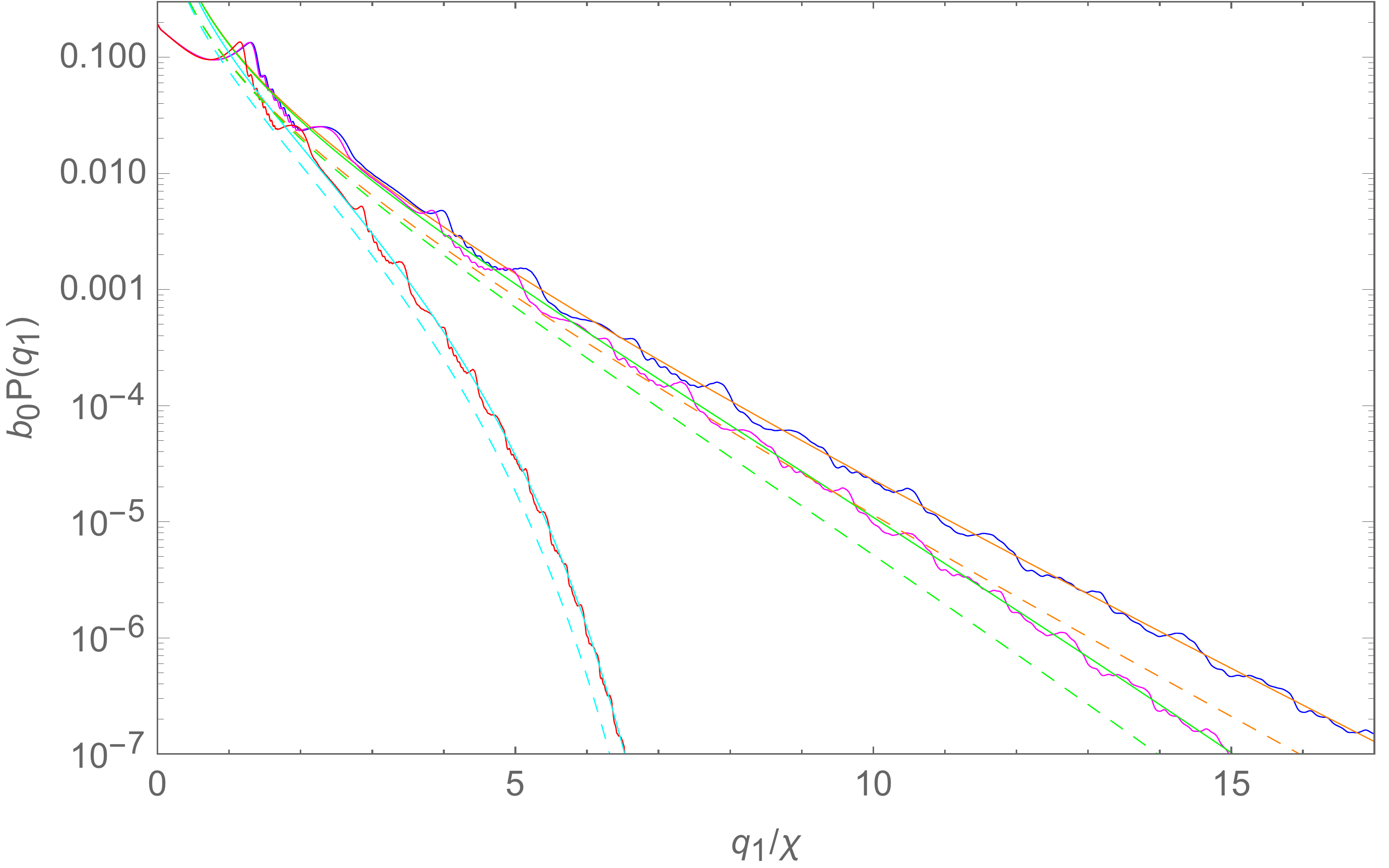} \\
\includegraphics[width=\linewidth]{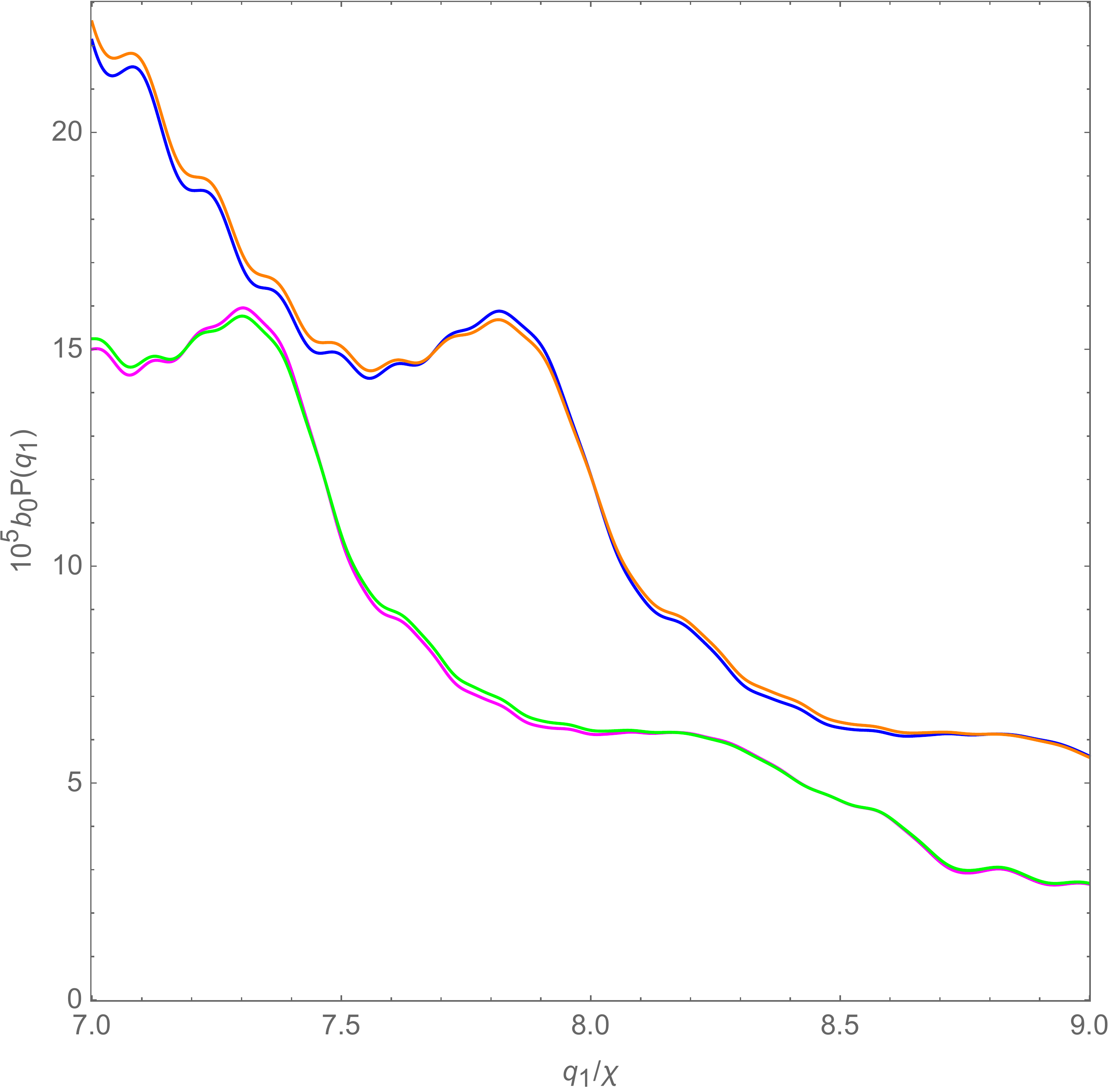}
\caption{The spectrum for single nonlinear Compton scattering for $\mathcal{T}=80$, $a_0=1$, $\chi=0.001$ (blue and orange curves), $\chi=0.01$ (magenta and green curves) and $\chi=0.1$ (red and cyan curves). The blue, magenta and red curves show the exact result and the orange, green and cyan curves are obtained with the saddle-point approximation. In the first plot we have included both sets of saddle points (the ones from~\eqref{monoSaddles1} and~\eqref{monoSaddles2}), but in the second plot only the $m=0$ saddles from the first set. The dashed lines in the second plot show the corresponding LCF approximation.}
\label{ComptonSpecApproxVsExacta01}
\end{figure}
\begin{figure}
\includegraphics[width=0.99\linewidth]{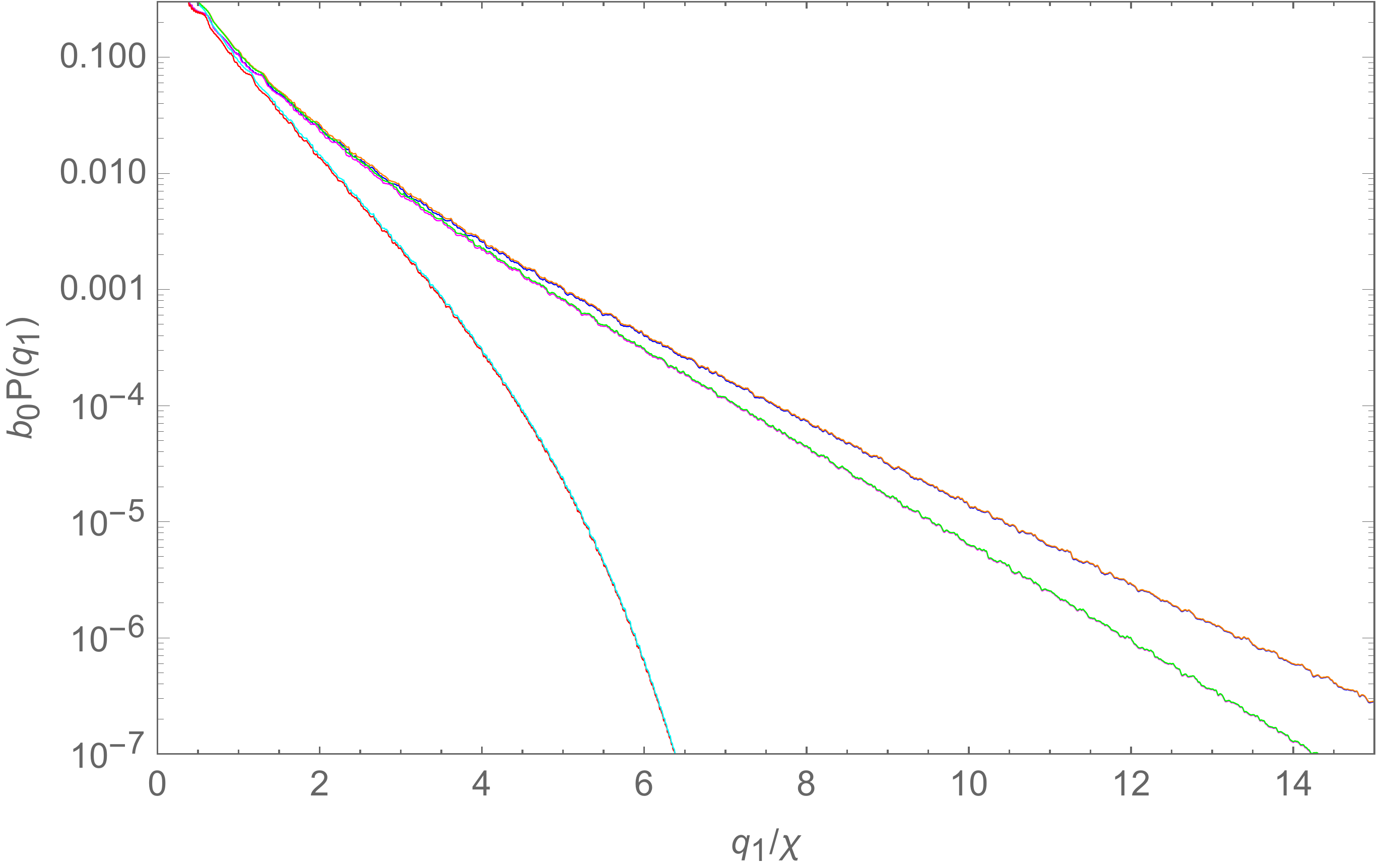} \\
\includegraphics[width=0.99\linewidth]{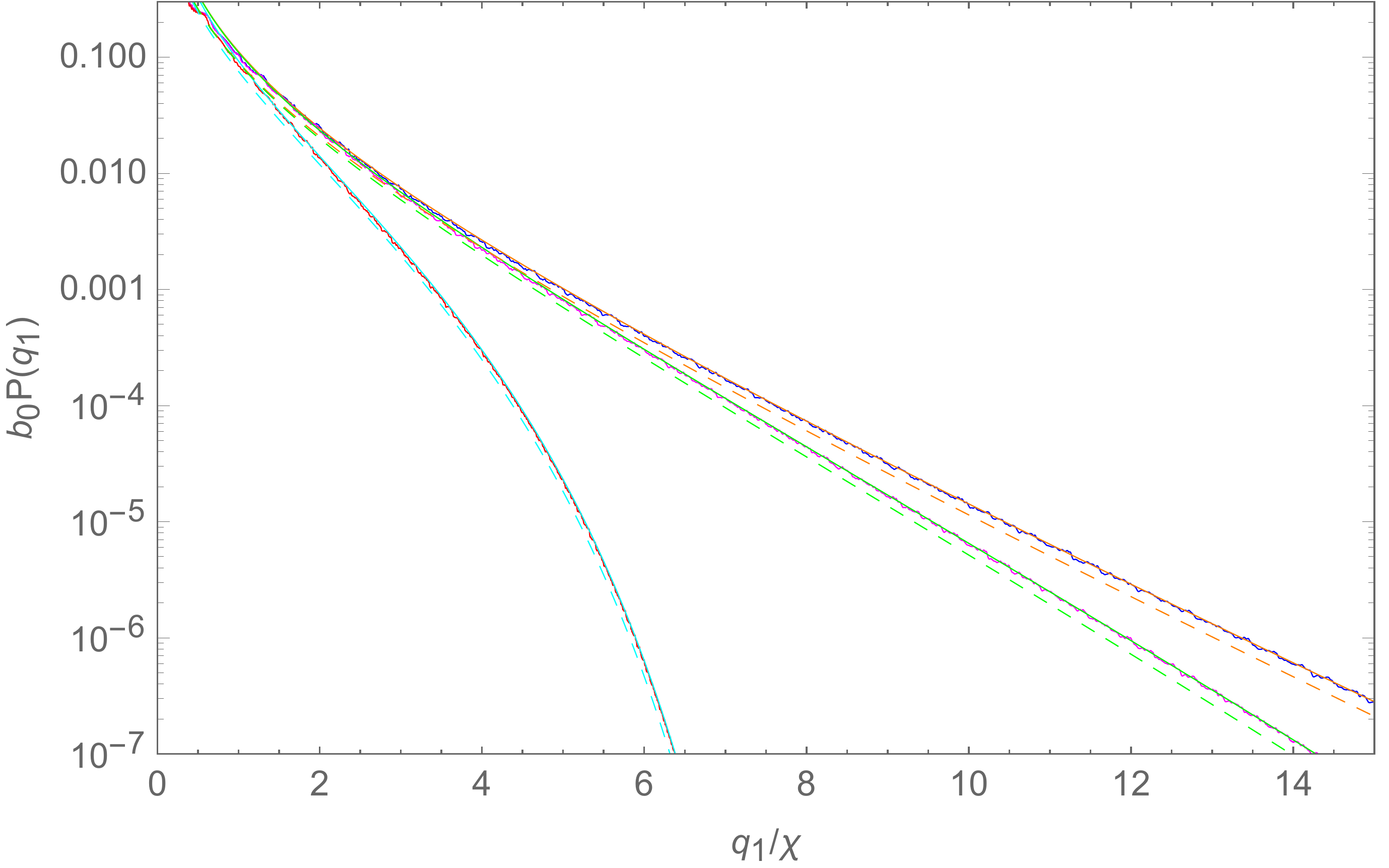} \\
\includegraphics[width=0.99\linewidth]{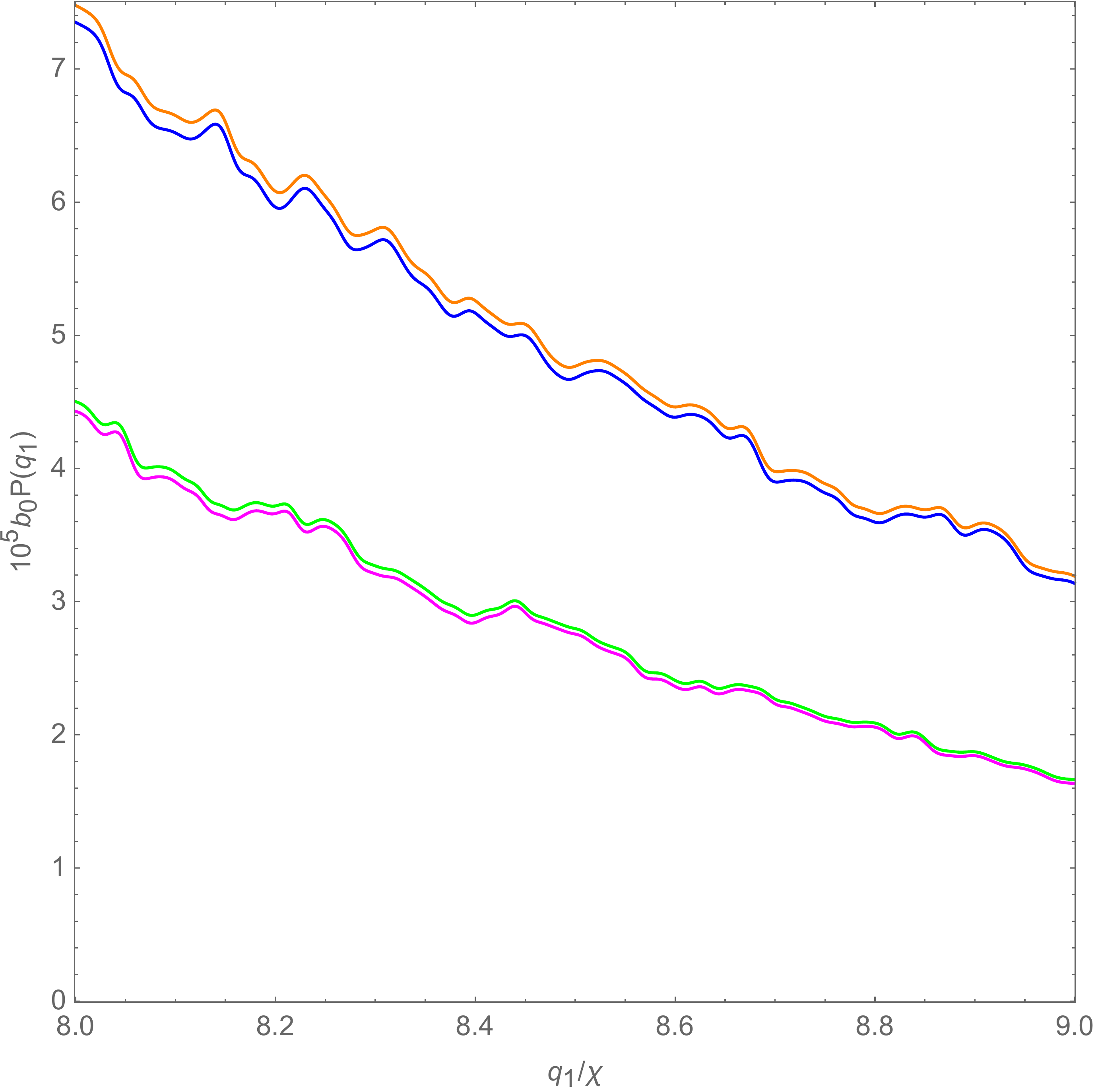}
\caption{Same as Fig.~\ref{ComptonSpecApproxVsExacta01} but with $a_0=2$.}
\label{ComptonSpecApproxVsExacta02}
\end{figure}
\begin{figure}
\includegraphics[width=\linewidth]{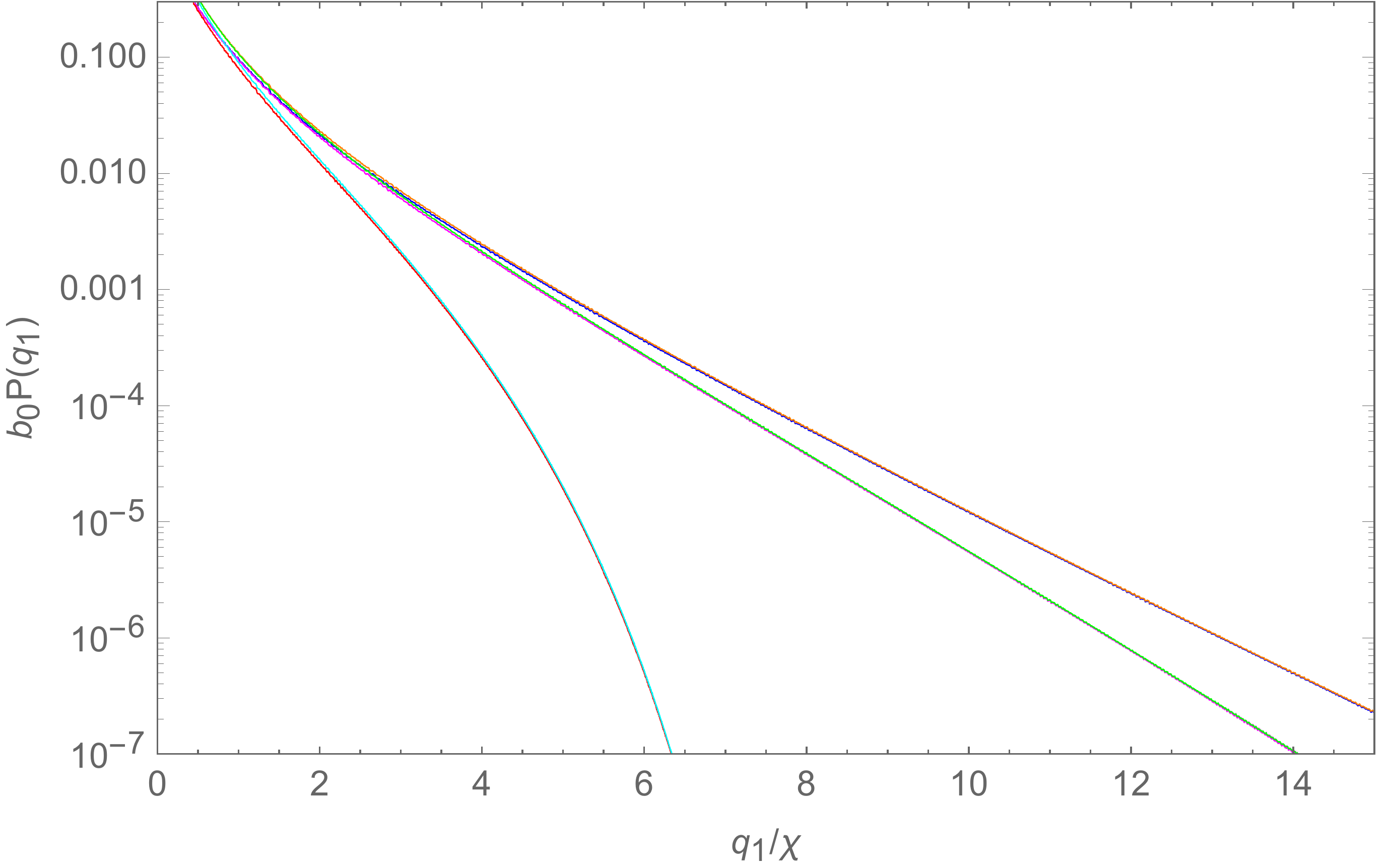} \\
\includegraphics[width=\linewidth]{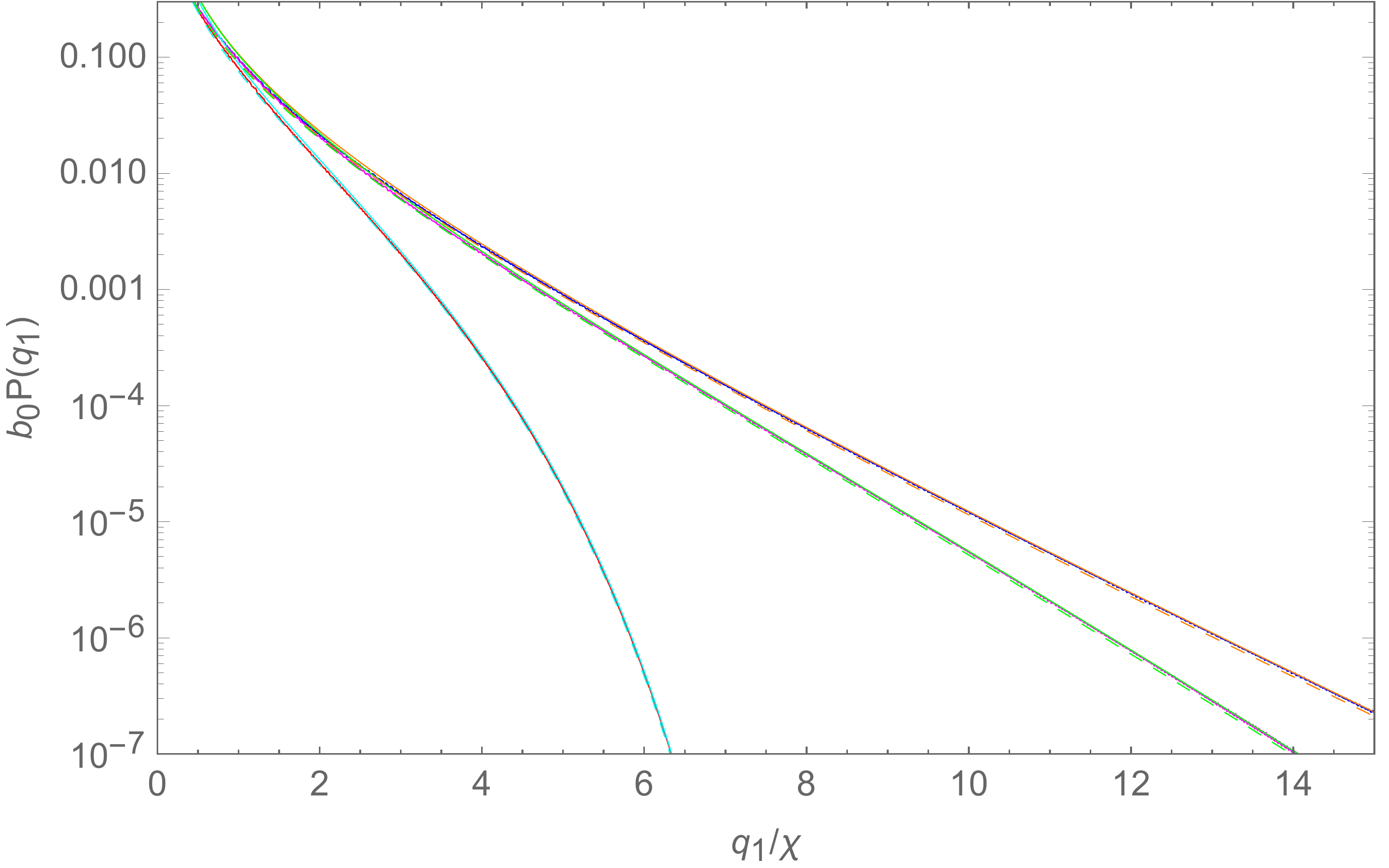} \\
\includegraphics[width=\linewidth]{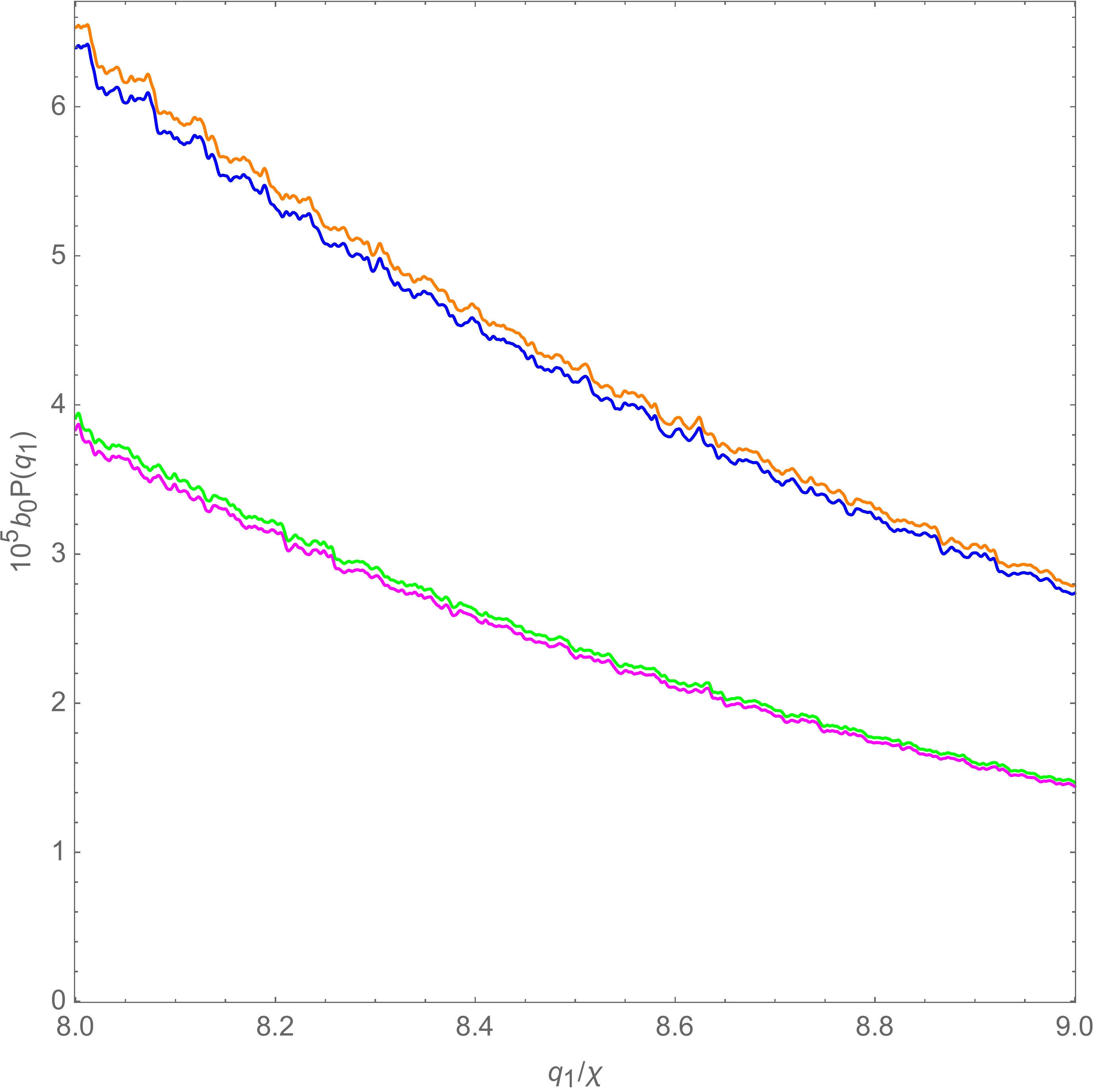}
\caption{Same as Fig.~\ref{ComptonSpecApproxVsExacta01} but with $a_0=4$.}
\label{ComptonSpecApproxVsExacta04}
\end{figure}

In Fig.~\ref{ComptonSpecApproxVsExacta01}, \ref{ComptonSpecApproxVsExacta02} and~\ref{ComptonSpecApproxVsExacta04} we compare this approximation with the results obtained by an exact numerical integration. How many saddle points one needs to include depends of course on several parameters. To obtain these results we have summed over the saddle points with $|n|\leq40$ and $|m|\leq20$.    
These plots show that the saddle-point approximation is remarkably good. It can in fact be difficult to see that there are actually two different curves in the large $q_1$ part.
Note that at $a_0=1$ the LCF approximation is not good, not even for an average where the oscillations are neglected. Our non-LCF saddle-point approximation, on the other hand, gives a very good approximation of even the nontrivial oscillations.
From these plots we see that the oscillations in the spectrum become smaller and faster as $a_0$ increases. 
Fig.~\ref{ComptonSpecApproxVsExacta02} shows that already at $a_0=2$ the oscillations are quite small on a log scale. However, by zooming in one can see that our approximation is capable of correctly describing even very fine details in the spectrum.  
In these figures we also plot the saddle-point approximation obtained by only including the $m=0$ saddles from~\eqref{monoSaddles1}. This gives a good approximation of a locally averaged spectrum. While the LCF approximation becomes more accurate for increasing $a_0$, for $a_0=2$ our approximation, even just the simpler one, is still much better.   
In Fig.~\ref{ComptonSpecApproxVsExacta04} we see that for $a_0=4$ the oscillations are so small that it might be difficult to see them without zooming in, and in this case the LCF approximation is quite good.

Although there are no IR divergences in single Compton for this field shape~\cite{Dinu:2012tj,Ilderton:2012qe}, the probability can become larger than one even for some of these non-extreme parameter values. That this can happen is well known~\cite{DiPiazza:2010mv,Dinu:2013hsd}.

\section{Double Compton scattering LCF}\label{DCnumericalLCF}

\begin{figure*}
\includegraphics[width=\linewidth,trim={2cm 1cm 3.5cm 1.1cm},clip]{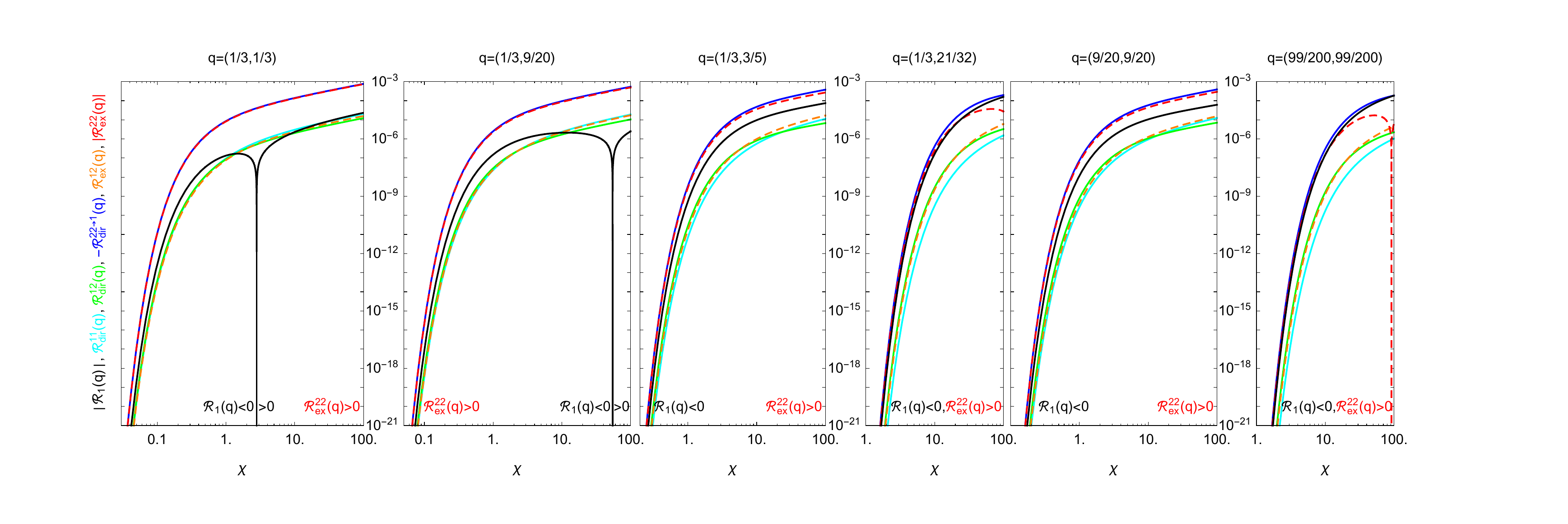}
\includegraphics[width=\linewidth,trim={0.9cm 0.5cm 3.2cm 0.5cm},clip]{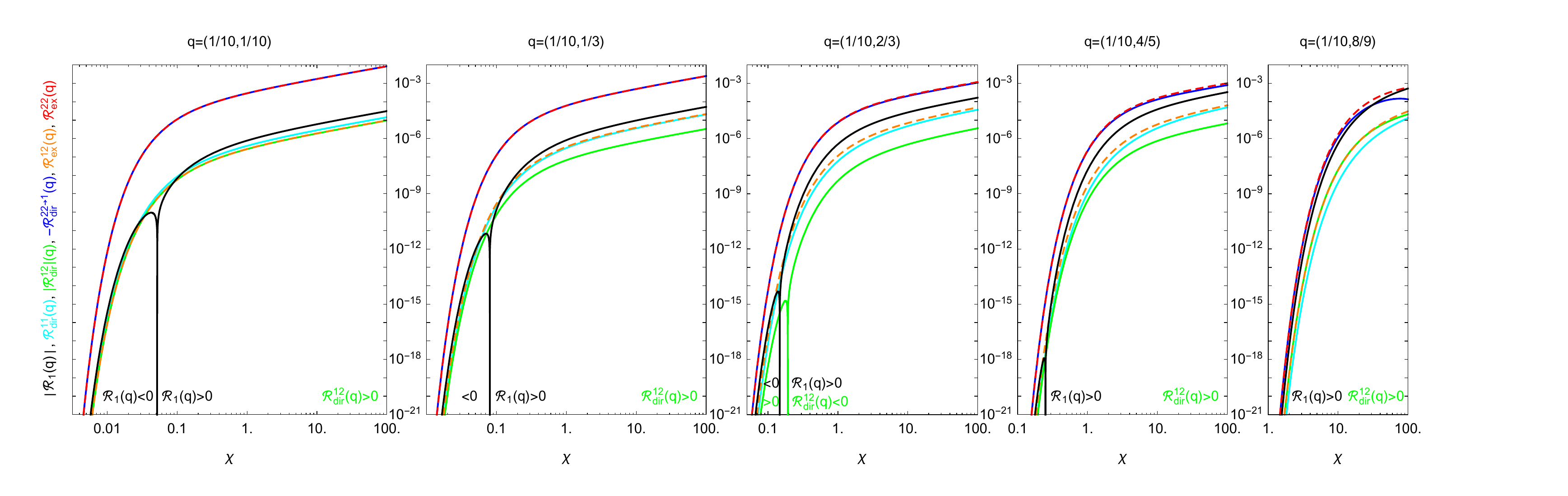}
\includegraphics[width=\linewidth,trim={5mm 0.5cm 3.3cm 0.5cm},clip]{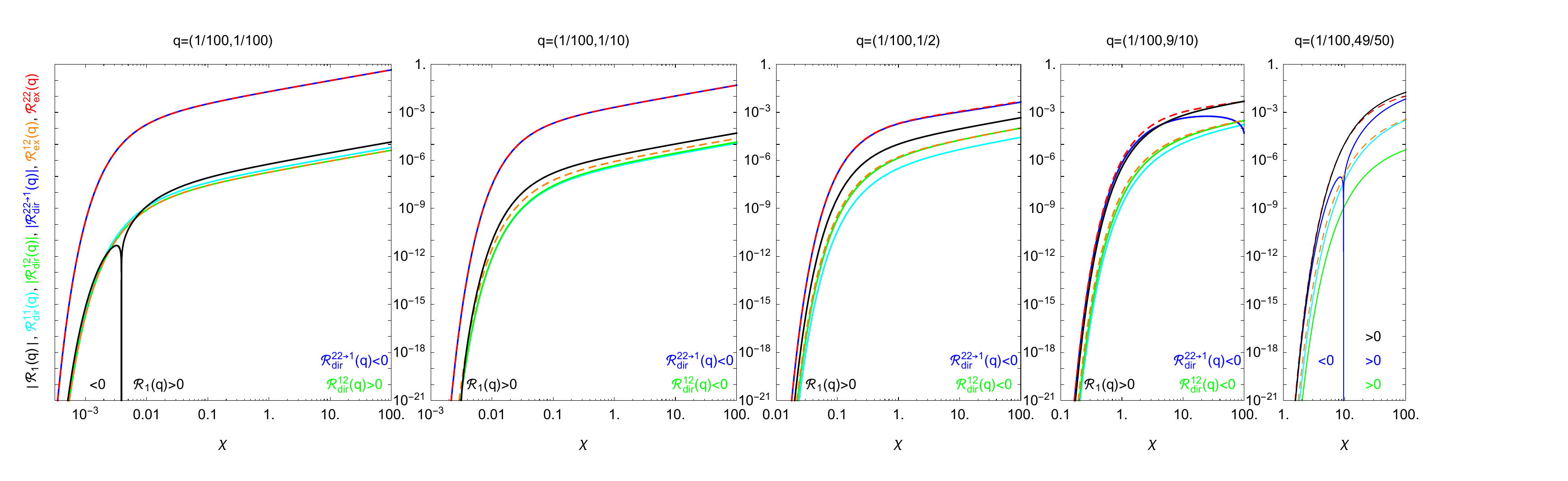}
\caption{The one-step term as a function of $\chi$ for different values of $q_i$. The blue, green and cyan solid curves show the direct terms $\mathcal{R}_{\rm dir}^{22\to1}$, $\mathcal{R}_{\rm dir}^{12}$ and $\mathcal{R}_{\rm dir}^{11}$, respectively. The red and orange dashed curves show the exchange terms $\mathcal{R}_{\rm ex}^{22}$ and $\mathcal{R}_{\rm ex}^{12}$, respectively. The black curves show the total one-step term.}
\label{oneStepLCFfig1}
\end{figure*}
\begin{figure}
\includegraphics[width=0.9\linewidth]{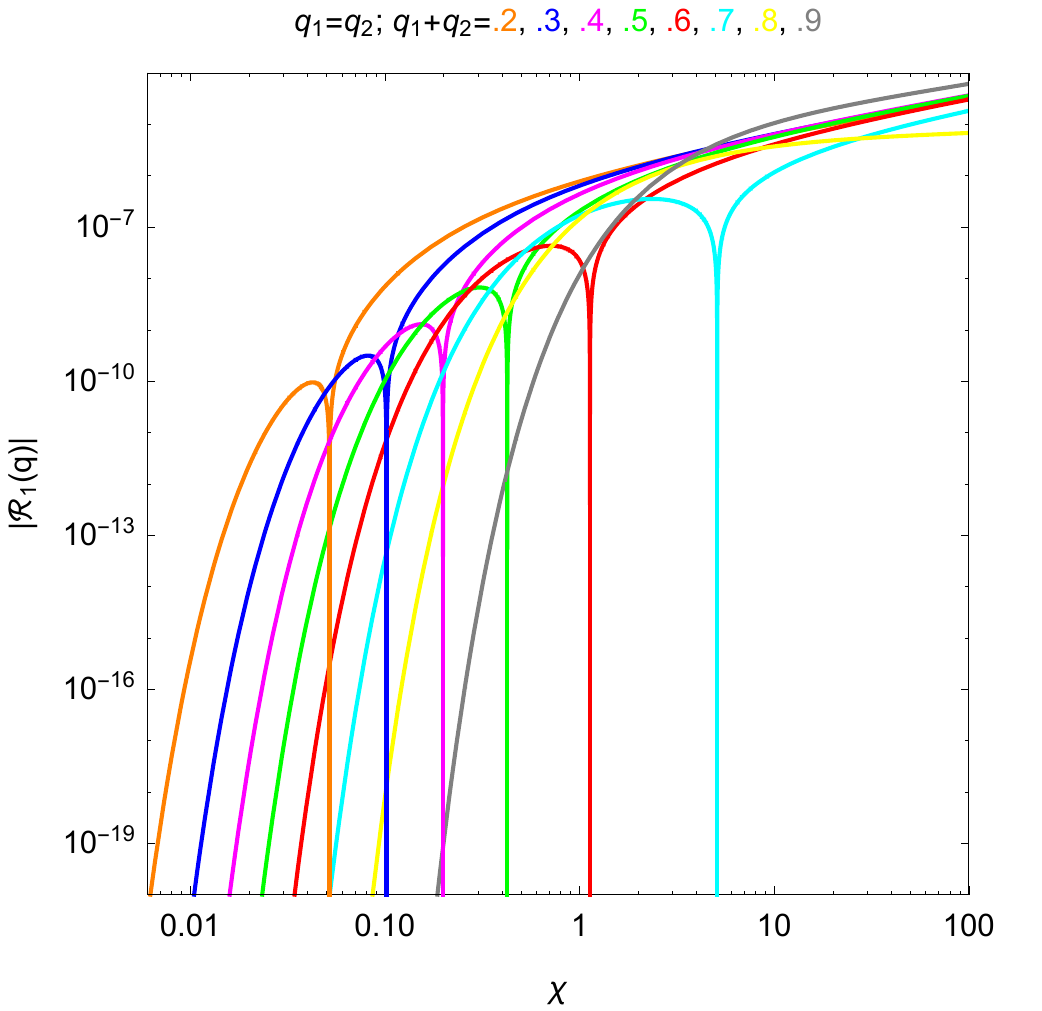}
\caption{The total one-step term as a function of $\chi$ for different values of $q_1=q_2$.}
\label{oneStepLCFfig2}
\end{figure}
\begin{figure}
\includegraphics[width=\linewidth]{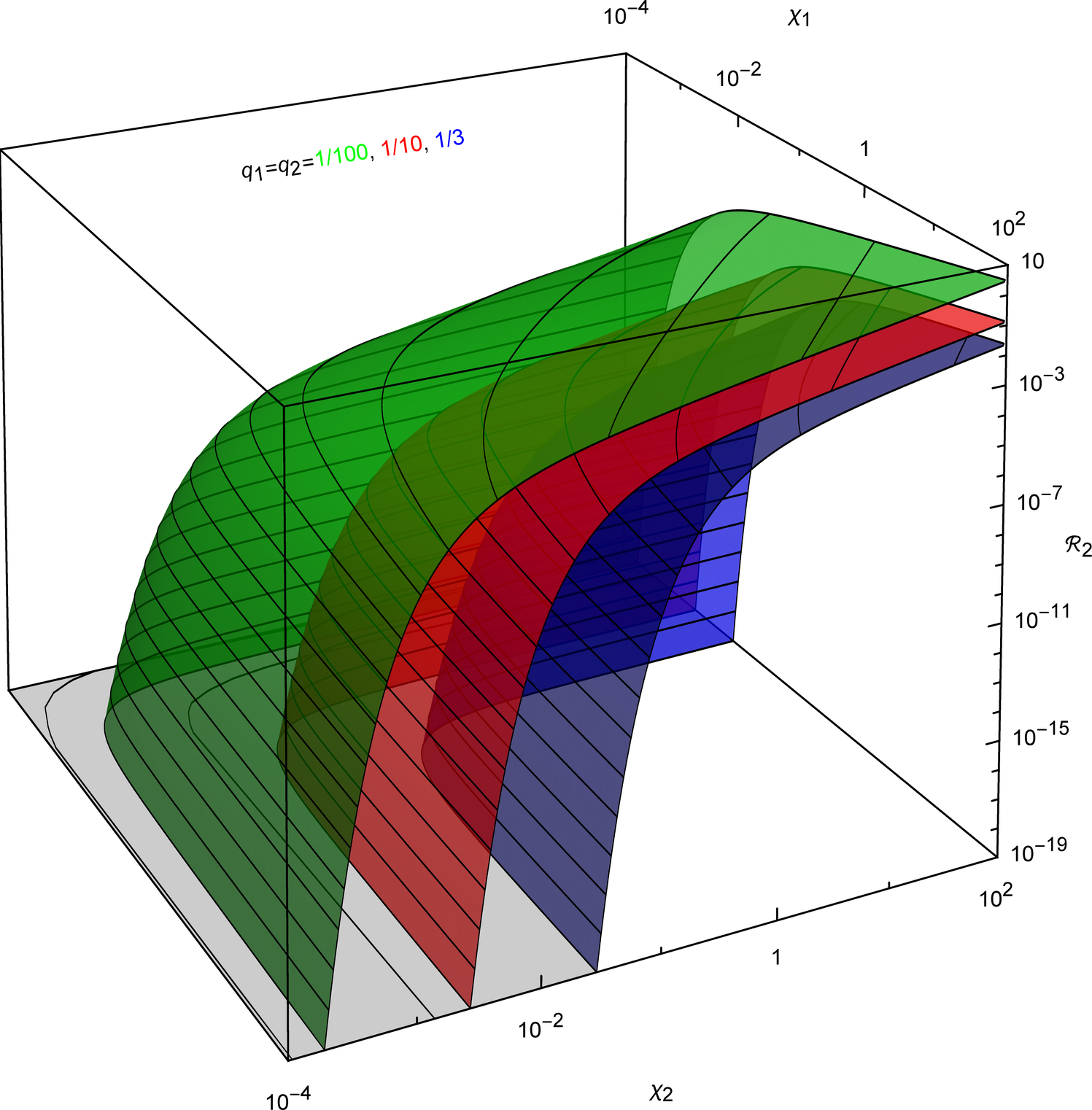}
\caption{The two-step term as a function of $\chi_1$ and $\chi_2$, the two locally constant values of $\chi$ at the two steps.}
\label{twoStepLCFfig1}
\end{figure}
\begin{figure}
\includegraphics[width=0.9\linewidth]{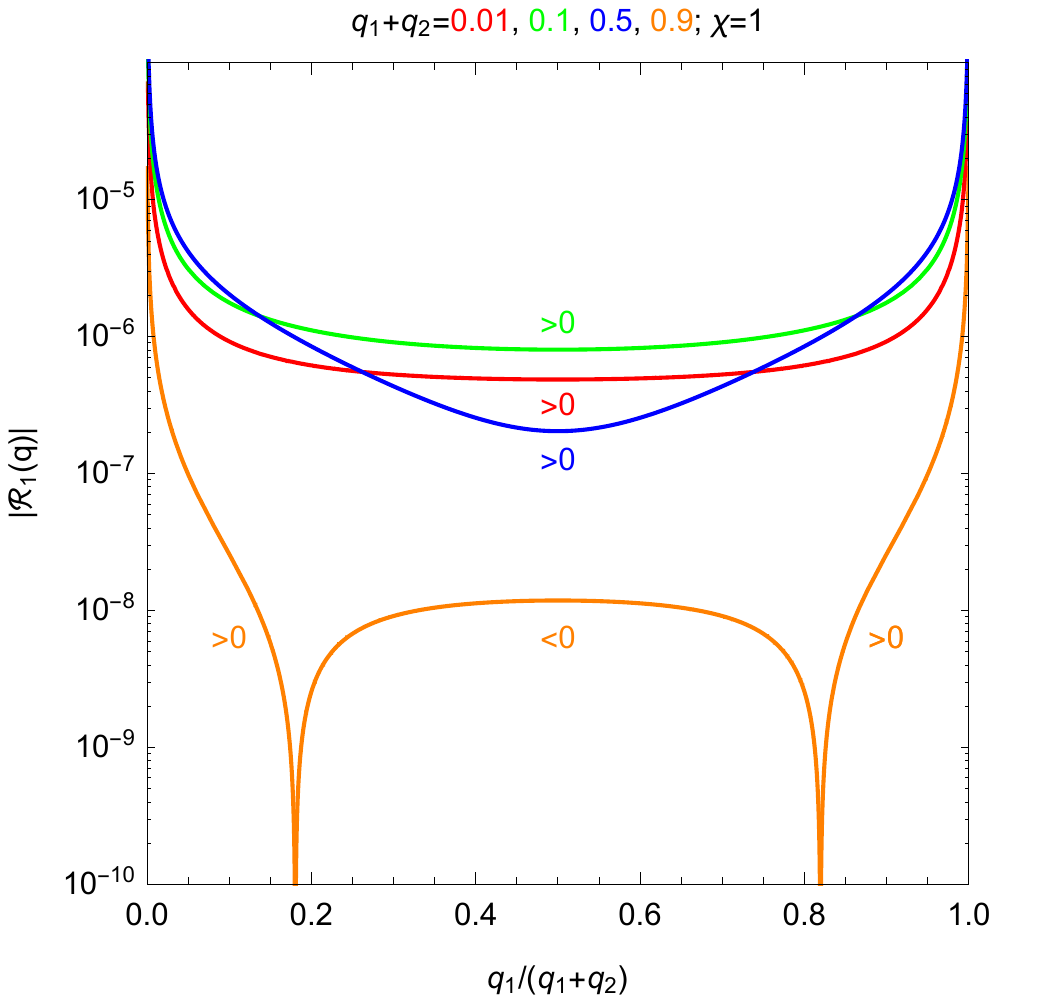}
\includegraphics[width=0.9\linewidth]{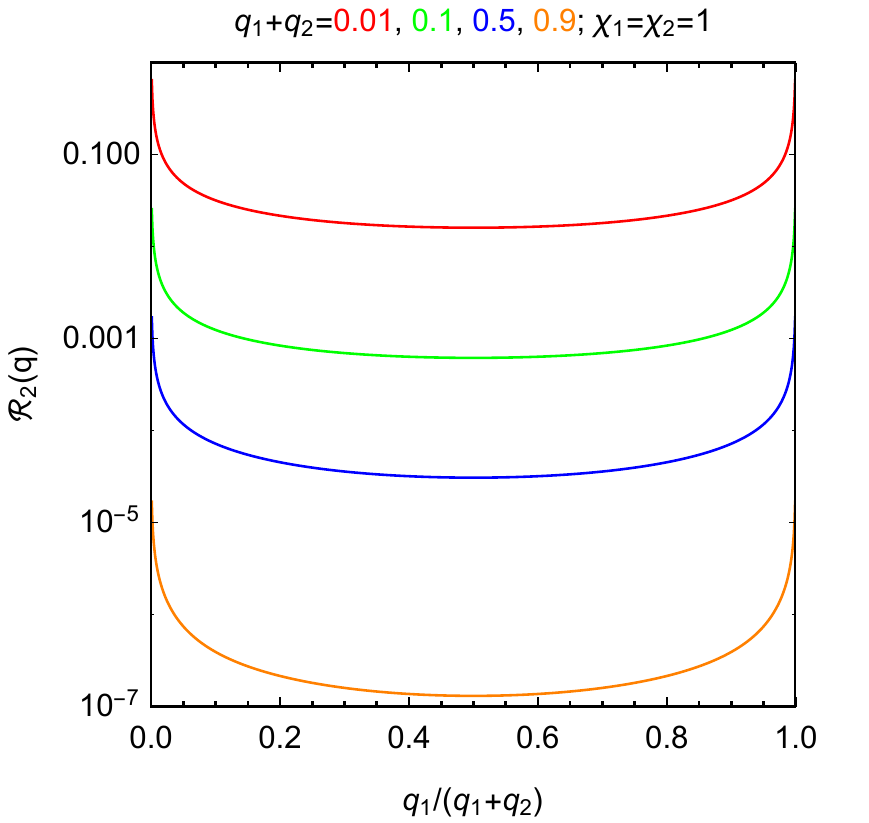}
\caption{The dependence on the longitudinal momentum of one photon $q_1$ for fixed $q_1+q_2$. The LCF approximation breaks down as $q_1\to0$ or $q_2\to0$. This happens further out for larger $a_0$.}
\label{qsymfig}
\end{figure}
\begin{figure}
\includegraphics[width=0.9\linewidth]{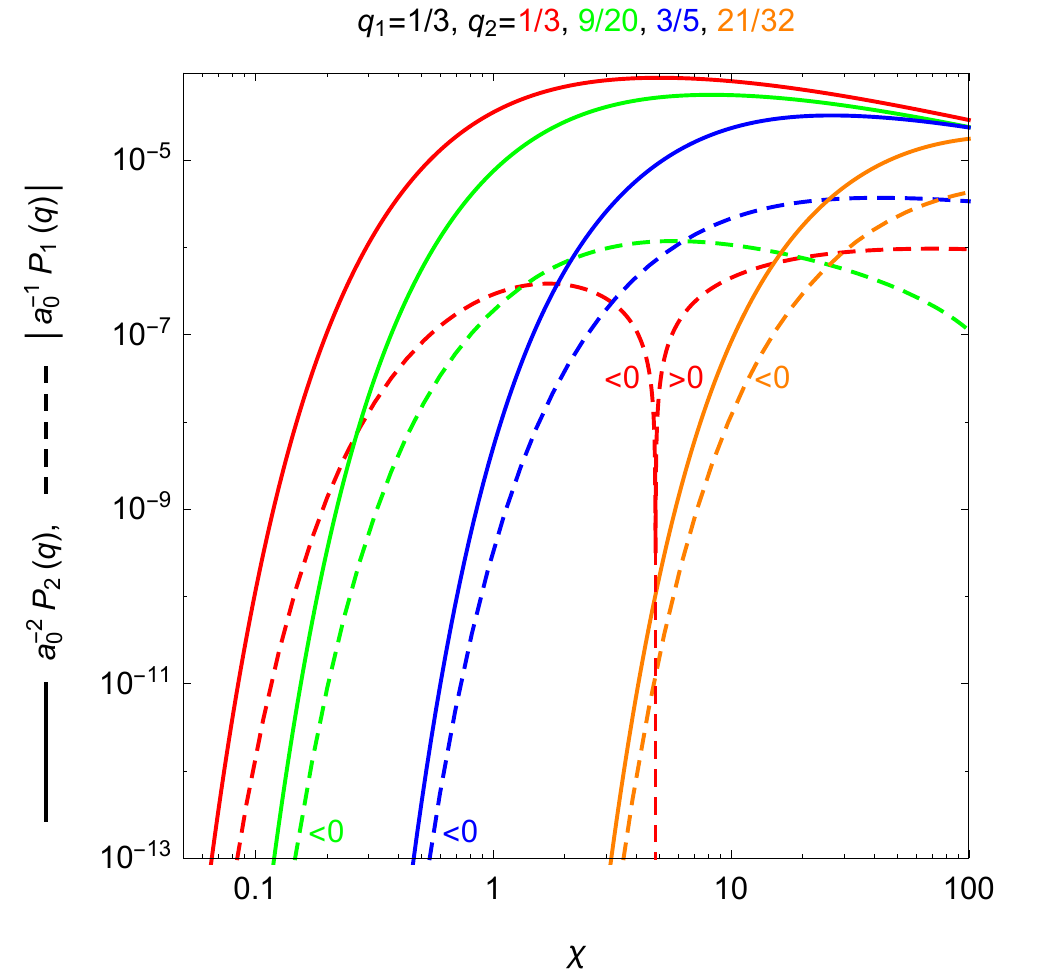}
\includegraphics[width=0.9\linewidth]{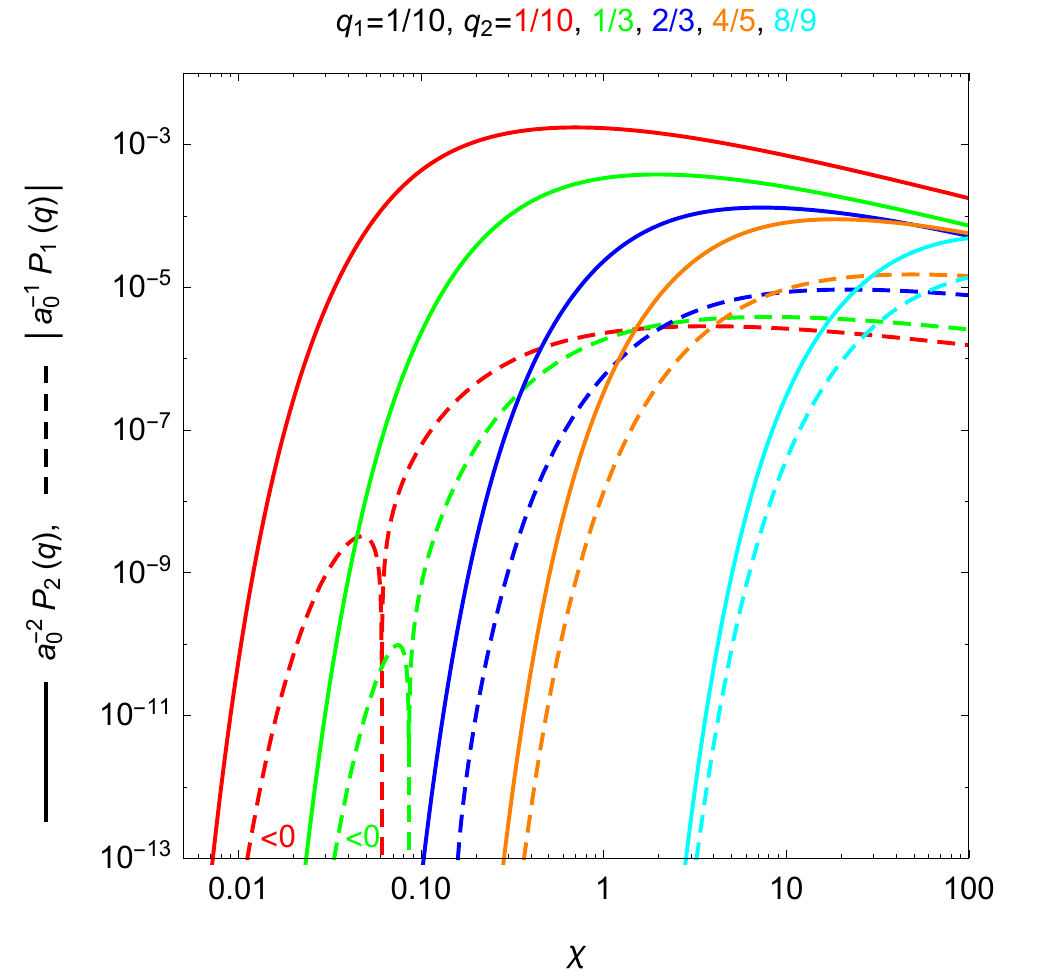}
\includegraphics[width=0.9\linewidth]{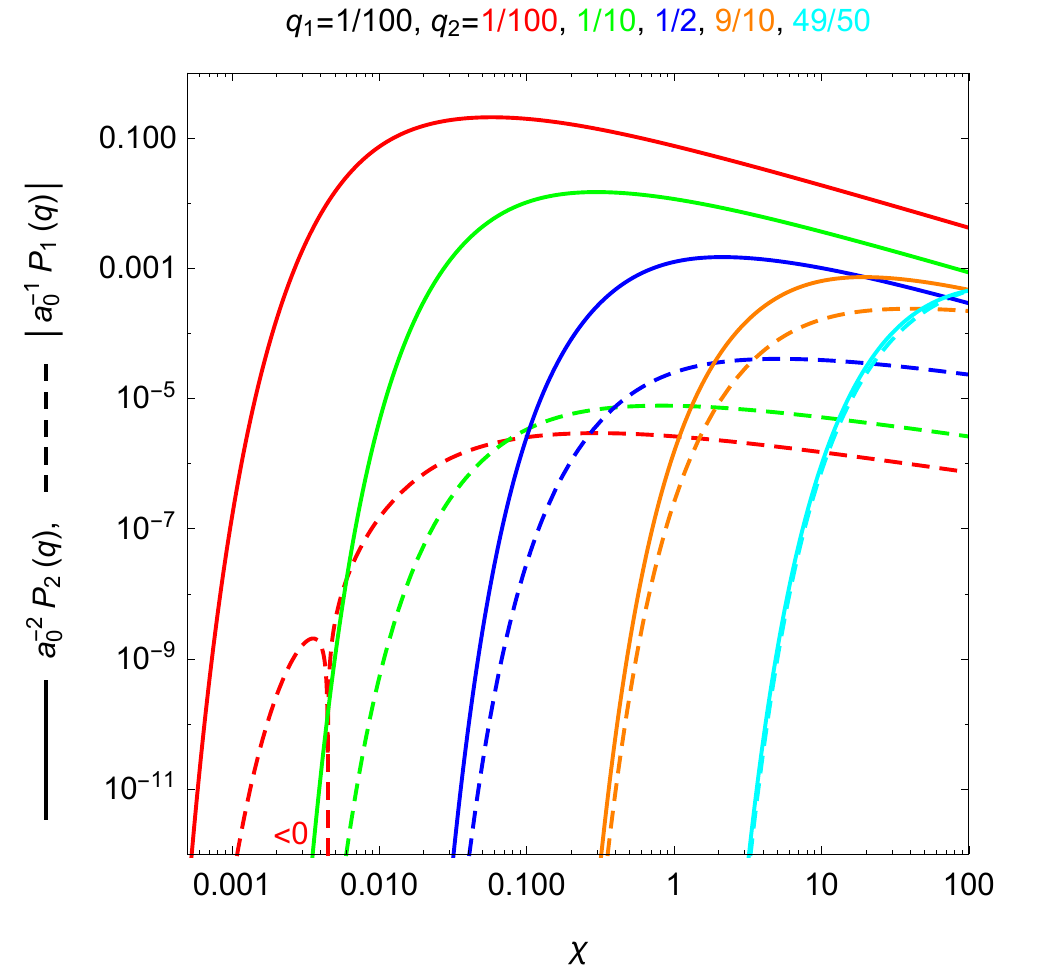}
\caption{The two-step vs. the one-step term in the LCF approximation for a Gaussian pulse~\eqref{aGaussSin} with $\mathcal{T}=\pi$.}
\label{oneVsTwoFig}
\end{figure}

We now return to double Compton scattering in the LCF regime. In the previous sections we showed that for $\chi\ll1$ the exchange term is on the same order of magnitude and even cancels the direct part of the one-step term to leading order. Here we study what happens at larger $\chi$.
We need to keep $q_i$ sufficiently large as it is known that the LCF approximation is not good for softer photons~\cite{DiPiazza:2017raw}. We expect that both $q_1$ and $q_2$ have to be considerably larger than $b_0/a_0^2$~\cite{DiPiazza:2017raw,Dinu:2015aci}. The LCF approximation only depends nontrivially on $a_0$ and $b_0$ via $\chi$. For example, to plot $\mathbb{P}_{\rm one}/a_0$ and $\mathbb{P}_{\rm two}/a_0^2$ as functions of $\chi$ we do not have to choose a value of $a_0$. So, for the lowest value of $q_i$ and the highest value of $\chi$ which we consider there should be a sufficiently large $a_0$ to justify the LCF approximation.    
In Fig.~\ref{oneStepLCFfig1} and~\ref{oneStepLCFfig2} we show the one-step term as a function of $\chi$ for different values of $q_i$. 
What is actually shown in these figures is the corresponding ``rate'' $\mathcal{R}$ defined by
\be
\mathbb{P}_{\rm one}(q)=:\int\frac{\ud\phi}{b_0}\mathcal{R}_1(\chi(\phi),q) \;,
\ee
where $\phi=(\phi_1+\phi_2+\phi_3+\phi_4)/4$, in which $\phi_4=\phi_2$ for $\mathbb{P}_{12}$ and $\phi_4=\phi_2$, $\phi_3=\phi_1$ for $\mathbb{P}_{11}$. 
As in the trident case, we find that the one-step term can be both positive and negative depending on $\chi$. 

Fig.~\ref{oneStepLCFfig1} shows that the direct and exchange parts of the one-step term can for smaller $q_1+q_2$ continue to be close to each other also for large $\chi$. 
The fact that $\mathbb{P}_{\rm dir}^{22\to1}$ and $\mathbb{P}_{\rm ex}^{22}$ almost cancel each other means that the other one-step contributions, $\mathbb{P}_{\rm dir}^{11}$, $\mathbb{P}_{\rm dir}^{12}$ and $\mathbb{P}_{\rm ex}^{12}$, are more important than in the trident case~\cite{Dinu:2017uoj}. So, even though $\mathbb{P}_{\rm dir}^{22\to1}$ and $\mathbb{P}_{\rm ex}^{22}$ are much larger than $\mathbb{P}_{\rm dir}^{11}$, $\mathbb{P}_{\rm dir}^{12}$ and $\mathbb{P}_{\rm ex}^{12}$, the size of the total one-step term is closer to the latter rather than the former.

However, we also see that $\mathbb{P}_{\rm dir}^{22\to1}$ and $\mathbb{P}_{\rm ex}^{22}$ are no longer close in magnitude for $1-q_1-q_2\ll1$, where the electron loses most of its initial longitudinal momentum to the emitted photons, and larger $\chi$. The last plot in the first row of Fig.~\ref{oneStepLCFfig1} shows one example with $1-q_1-q_2\ll1$ and $q_1=q_2$ where $\mathbb{P}_{\rm dir}^{22\to1}$ dominates and where $\mathbb{P}_{\rm ex}^{22}$ changes sign at large $\chi$. The second and third rows show examples with $1-q_1-q_2\ll1$ and $q_1\ll q_2$ or $q_1\gg q_2$ where the one-step term is instead dominated by the exchange term $\mathbb{P}_{\rm ex}^{22}$ and where $\mathbb{P}_{\rm dir}^{22\to1}$ can change sign.

We have made a comparison between our numerical results and our saddle-point approximation similar to the one in Appendix~C in~\cite{Dinu:2017uoj} for the trident case. For sufficiently small $\chi$ we again find that each of the first couple of orders give a better agreement. However, here we find that the coefficients in the series in $\chi$ increase quite fast. For example, at $q_1=q_2=1/3$ we find
\be
\begin{split}
\mathbb{P}_{\rm one}\approx&-\frac{4181\alpha^2 a_0\sqrt{\chi}}{2520\sqrt{2}\pi^{3/2}}\exp\left(-\frac{4}{3\chi}\right) \\
&(1-11.0\chi+130.5\chi^2-1847.5\chi^3+\dots) \;.
\end{split}
\ee    
Given that the saddle-point approximation can lead to asymptotic series, this growth of the coefficients should not be too surprising, but it does mean that the higher orders are less useful than in the trident case. They only provide an improvement for quite small $\chi$, but there the probability is very small because of the exponential suppression. This is a bit unfortunate if one wants an approximation for the total one-step term, because one needs at least the next-to-leading order of $\mathbb{P}_{\rm dir}^{22\to1}$ and $\mathbb{P}_{\rm ex}^{22}$ since they cancel each other to leading order.         

On the other hand, this cancellation also means that neglecting the total one-step compared to the two-step term should be a better approximation\footnote{The one-step terms give of course only the next-to-leading order term in the LCF $1/a_0\ll1$ expansion, and so are already suppressed by a factor of $1/a_0$ with respect to the two-step term.} here than in the trident case.  
The two-step term is shown in Fig.~\ref{twoStepLCFfig1} in terms of the following ``rate''
\be
\mathbb{P}_{\rm two}(q)=:\int\frac{\ud\sigma_{43}\ud\sigma_{21}}{b_0^2}\mathcal{R}_2(\chi(\sigma_{43}),\chi(\sigma_{21}),q) \;.
\ee
Fig.~\ref{qsymfig} shows that, given a fixed total emitted longitudinal momentum $q_1+q_2$, most of it is given to one of the two photons, $q_1\ll q_2$ or $q_2\ll q_1$. This is expected since the probability is in general larger for softer photons.    

However, to directly compare the one-step and the two-step terms we need to integrate over $\sigma$ for some pulse shape. In Fig.~\ref{oneVsTwoFig} we compare $\mathbb{P}_{\rm one}$ and $\mathbb{P}_{\rm two}$ for a Gaussian pulse~\eqref{aGaussSin}. We find that $\mathbb{P}_{\rm two}$ can be much larger than $\mathbb{P}_{\rm one}$ even for a very short pulse and even before taking the $a_0$-scaling into account, which gives a further increase due to $\mathbb{P}_{\rm two}\sim a_0^2$ and $\mathbb{P}_{\rm one}\sim a_0$. 
We find that for larger $q_1+q_2$ the relative difference between $\mathbb{P}_{\rm two}/a_0^2$ and $\mathbb{P}_{\rm one}/a_0$ decreases. However, the exponential suppression~\eqref{PNexpLCF} for $1-q_1-q_2\ll1$ means that we then need larger $\chi$ to have a significant probability, which is presumably more likely to be achieved by increasing $a_0$ rather than $b_0$, and a larger $a_0$ again favors $\mathbb{P}_{\rm two}$.
Further, the pulse length in Fig.~\ref{oneVsTwoFig} is probably about 30 times shorter than what one can expect in a typical experiment, so for a more realistic pulse the one-step term will be even less important. So, even in cases without near cancellation between $\mathbb{P}_{\rm dir}^{22\to1}$ and $\mathbb{P}_{\rm ex}^{22}$ the one-step term will probably not be very important.       
However, the one-step term could be important for a short pulse with $a_0\sim1$. For such fields we can of course not use LCF and so we leave that for future studies.

\section{Conclusions}

In this paper we have studied double nonlinear Compton scattering. By using the same approach as in our previous paper on trident pair production~\cite{Dinu:2017uoj}, we have showed that many of the results are very similar, which allows us to use the same methods. We have focused on the emission of ``hard'' photons which makes things more similar to the trident process than if we had included soft photons, we can for example obtain saddle-point approximations for $\chi<1$ that are similar to the ones we obtained in~\cite{Dinu:2017uoj}. Focusing on hard photons is also motivated by the fact that they can be more interesting/useful e.g. for subsequent pair production. The saddle-point method has not only allowed us to find simple analytical expressions for simple field shapes, we have also considered a more nontrivial, pulsed oscillating field. We then have to obtain the saddle points numerically, but by comparing with the exact numerical result for single Compton scattering we find a very good agreement, even for small and fast oscillations in the spectrum. Since the saddle-point approach is much faster it can therefore be a useful method for studying this as well as similar processes. Indeed, since the exponential part of the integrand is very similar for double and higher-order Compton scattering, one can also apply this method to those processes, for which an exact numerical integration would take a long or too long time. We have also made preliminary calculations for trident and found that the same saddle-point method can also be used to study oscillations in the momentum spectrum there.

The two-step part of the probability is related to two one-photon emissions. By studying this relation in detail for arbitrary polarization we have discovered a new gluing approach, i.e. a method for obtaining the dominant part for sufficiently long laser pulses. Gluing (spin-averaged) LCF probabilities is an important part of PIC simulations, where using LCF results is motivated by considering $a_0\gg1$. Our new gluing approach takes the spin of the intermediate electron into account and gives the dominant contribution for arbitrary field polarization and for $a_0\gtrsim1$. 
For $a_0\gg1$ and linear polarization our gluing approach reduces to the one in~\cite{Morozov:1975uah,King:2014wfa}.
So, this goes beyond the usual gluing approach. We have checked that our approach gives the correct results for triple and quadruple Compton scattering. To the best of our knowledge, these processes have not been studied in this regime before. In this paper we have only presented this gluing approach for intermediate electrons. Our preliminary results for trident suggest that we will be able to generalize our gluing approach to processes with intermediate photons. 
More work is needed to more precisely delineate the region of parameter space where corrections to our new gluing estimates can be neglected. One can expect that the smaller $a_0$ is or the larger $b_0$ is the longer the pulse has to be, and softer photons may also make corrections more important. To answer these questions we plan to perform a detailed numerical study for both trident and double Compton scattering.

\acknowledgments
We thank Tom Blackburn, Antonino Di Piazza, Anthony Hartin and Anton Ilderton for useful discussions.
V.~Dinu is supported by the CNCS-UEFISCDI project PN-III-P4-ID-PCE-2016-0792, and G.~Torgrimsson is supported by the Alexander von Humboldt foundation.

\appendix
\section{How to obtain the amplitude}\label{basicsSection}

In this appendix we provide the basic ingredients needed to calculate the amplitude using either the standard covariant or the lightfront quantization approach.

\subsection{Lightfront quantization}

As is standard in this field, the background is treated exactly by using Volkov solutions and the Furry picture. The amplitude can be obtained either with the standard covariant approach or with the lightfront quantization formalism~\cite{Brodsky:1997de,Heinzl:2000ht}, which naturally accommodates plane-wave background fields~\cite{Neville:1971uc,Ilderton:2013dba,Dinu:2013gaa}. The lightfront Hamiltonian governs the evolution in $x^\LCp$, and has three terms, 
\be\label{totV}
H_{\rm int}=\frac{1}{2}\int\!\ud\bar{x}\; ejA+\frac{e^2}{2}j_\LCm\frac{1}{(i\partial_\LCm)^2}j_\LCm+e^2\bar{\Psi}\fsl{A}\frac{\gamma^\LCp}{4i\partial_\LCm}\fsl{A}\Psi \;,
\ee
where $j^\mu=\bar{\Psi}\gamma^\mu\Psi$.
The first term is similar to the usual Hamiltonian, while the other terms are referred to as ``instantaneous''~\cite{Brodsky:1997de,Zhao:2013cma}. The first instantaneous term contributes to trident~\cite{Dinu:2017uoj}, while the second instantaneous term contributes to the two-photon emission considered here. The photon and fermion fields are expressed in terms of mode expansions with only on-shell momenta.    
The photon field is given by
\be
A_\mu(x)=\int\!\ud\tilde{l}\; a_\mu e^{-ilx}+a^\dagger_\mu e^{ilx} \;,
\ee
where the mode operators obey
\be
[a_\mu(l),a_\nu^\dagger(l')]=-2l_\LCm\bar{\delta}(l-l')L_{\mu\nu} \;,
\ee
with
\be
L_{\mu\nu}=g_{\mu\nu}-\frac{k_\mu l_\nu+l_\mu k_\nu}{kl}  \;.
\ee
The fermion field is given by
\be
\Psi(x)=\int\!\ud\tilde{p}\; Kub\varphi+\bar{K}vd^\dagger\varphi(-p) \;,
\ee
where the background enters via the Volkov solution~\cite{Wolkow:1935zz},
\be
\varphi=\exp\left\{-i\left(px+\int^{kx}\frac{2ap-a^2}{2kp}\right)\right\} 
\qquad
K=1+\frac{\slashed{k}\slashed{a}}{2kp} \;,
\ee
and where $\bar{K}=1-\slashed{k}\slashed{a}/(2kp)$. 

As in~\cite{Dinu:2017uoj}, we use $M_2$ to denote the term in the amplitude that comes from two vertices of the non-instantaneous part of the Hamiltonian,
\be
\begin{split}
	\frac{1}{k_\LCp}\bar{\delta}(p'+l_2+l_1-p)M_2:=&\\
	-\langle 0|b(p')\epsilon_1 a(l_1)\epsilon_2a(l_2)\int&\!\ud x_2^\LCp\ud x_1^\LCp\theta(x_2^\LCp-x_1^\LCp) \\
	H_{\rm int}^{(1)}(x_2^\LCp)&H_{\rm int}^{(1)}(x_1^\LCp)b^\dagger(p)|0\rangle \;,
\end{split}
\ee
and $M_1$ to denote the term coming from the instantaneous part of the Hamiltonian,
\be
\begin{split}
	&\frac{1}{k_\LCp}\bar{\delta}(p'+l_2+l_1-p)M_1:= \\
	&\bra{0}b(p')\epsilon_1 a(l_1)\epsilon_2a(l_2)(-i)\int\!\ud x^\LCp H_{\rm int}^{(2)}(x^\LCp)b^\dagger(p)\ket{0} \;.
\end{split}
\ee
After some straightforward calculation we obtain 
\be\label{M122}
\begin{split}
M_2^{12}=-\frac{\pi\alpha}{kp_1}&\int\ud\phi_2\underbrace{\bar{u}\bar{K}\bar{\varphi}}_{p'}\slashed{\epsilon}_2e^{il_2x_2}\underbrace{K\varphi}_{p_1}(\slashed{p}_1+1) \\
&\int^{\phi_2}\!\ud\phi_1\underbrace{\bar{K}\bar{\varphi}}_{p_1}\slashed{\epsilon}_1e^{il_1x_1}\underbrace{Ku\varphi}_{p} \;,
\end{split}
\ee
where $\bar{p}_1=\bar{p}-\bar{l}_1$
is the momentum of the intermediate electron, and
\be\label{M121}
M_1^{12}=-\frac{i\pi\alpha}{kp_1}\underset{p'}{\bar{u}}\slashed{\epsilon}_2\slashed{k}\slashed{\epsilon}_1\underset{p}{u}\int\ud\phi e^{i(l_2+l_1)x}\underset{p'}{\bar{\varphi}}\underset{p}{\varphi} \;,
\ee
where $p_1$ is the same as in $M_2$.

\subsection{Covariant approach}

Next we show how the results from the previous section can be obtained with the standard, covariant approach. In the covariant approach the amplitude is given by
\be\label{covariantStart}
\begin{split}
	&\frac{1}{k_\LCp}\bar{\delta}(p'+l_2+l_1-p)M^{12}= \\
	&(-ie)^2\int\ud^4x_1\ud^4x_2\underset{p'}{\bar{\psi}(x_2)}\slashed{\epsilon}_2e^{il_2x_2} 
	S(x_2,x_1)\slashed{\epsilon}_1e^{il_1x_1}\underset{p}{\psi}(x_1) \;,
\end{split}
\ee
where $\psi=Ku\varphi$ includes the spin factor of the Volkov solution and the fermion propagator is given by
\be
S(x,y)=i\int\frac{\ud^4P}{(2\pi)^4}K\varphi(x)\frac{1}{\slashed{P}-m+i\epsilon}\bar{K}\bar{\varphi}(y) \;.
\ee 
As in~\cite{Seipt:2012tn}, we perform the $q_\LCp$ integral by first separating the propagator into two terms using
\be\label{fermionPropSplit}
\frac{1}{\slashed{P}-m+i\epsilon}=\frac{1}{4P_\LCm}\left(\gamma^\LCp+\frac{\slashed{P}_{\rm on}+m}{P_\LCp-P_\LCp^{\rm on}+i\epsilon\text{sign}(P_\LCm)}\right) \;,
\ee
where $P_\LCp^{\rm on}=(m^2+P_\LCperp^2)/(4P_\LCm)$. The (lightfront) spatial coordinate integrals in~\eqref{covariantStart} give delta functions implying $\bar{P}=\bar{p}-\bar{l}_1=\bar{p}'+\bar{l}_2$, which means $P_\LCm>0$. Upon performing the $P_\LCp$ integral, the two terms in~\eqref{fermionPropSplit} give terms with $\delta(x_2^\LCp-x_1^\LCp)$ and $\theta(x_2^\LCp-x_1^\LCp)$, respectively. We find that the term with $\delta(x_2^\LCp-x_1^\LCp)$ is exactly equal to the term (\eqref{M121}) that comes from the instantaneous part of the lightfront Hamiltonian, and the term with $\theta(x_2^\LCp-x_1^\LCp)$ is exactly equal to the term (\eqref{M122}) that comes from two vertices with the non-instantaneous part of the Hamiltonian.

\section{Gauge invariance}

In this work we have used the lightfront gauge for the emitted photons, where $k\epsilon=0$ in addition to $l\epsilon(l)=0$, or in terms of the components $\epsilon_\LCm=0$ and $\epsilon_\LCp=l_\LCperp\epsilon_\LCperp/(2l_\LCm)$. The probability is of course gauge invariant, but that does not necessarily mean that each contribution to the probability will be separately gauge invariant. Indeed, it is well known, e.g. from QED without a background field, that individual diagrams are in general not gauge invariant. To check gauge invariance we replace $\slashed{\epsilon}_1$ in~\eqref{covariantStart} with $\slashed{l}_1$. Let us first study the following spinor part
\be
\frac{\slashed{P}+1}{P^2-1+i\epsilon}\bar{K}(P,\phi_1)\slashed{l}_1K(p,\phi_1)u \;,
\ee
where $\bar{P}=\bar{p}-\bar{l}_1$, but $P_\LCp$ is still an integration variable. We write
\be\label{sepsll1}
\begin{split}
\slashed{l}_1=\slashed{p}-\slashed{P}+c\slashed{k}=&[\slashed{\pi}(p,\phi_1)-1]-[\slashed{\pi}(P,\phi_1)-1] \\
&+[c-V(p,\phi_1)+V(P,\phi_1)]\slashed{k} \;,
\end{split}
\ee
where $c=(l_1-p+P)_\LCp/k_\LCp$ and $V(p,\phi)=(2ap-a^2)/(2kp)$. The $\phi_1$ dependent part of the exponent is given by
\be
\exp\left\{i\int_0^\phi\ud\phi[c-V(p,\phi)+V(P,\phi)]\right\} \;, 	
\ee
so the last term in~\eqref{sepsll1} is a total derivative (note that $\bar{K}\slashed{k}K=\slashed{k}$) and vanishes upon integrating over $\phi_1$. From $\slashed{\pi}K=K\slashed{p}$ we find
\be
[\slashed{\pi}(p,\phi_1)-1]K(p,\phi_1)u=K(p,\phi_1)(\slashed{p}-1)u=0 \;.
\ee
From $\bar{K}\slashed{\pi}=\slashed{p}\bar{K}$ we find
\be\label{Pminus}
\frac{\slashed{P}+1}{P^2-1+i\epsilon}\bar{K}(P,\phi_1)(-[\slashed{\pi}(P,\phi_1)-1])=-\bar{K}(P,\phi_1) \;.
\ee
The $P_\LCp$ integral gives a delta function and we find 
\be
M^{12}_{\slashed{\epsilon}_1\to\slashed{l}_1}=\frac{ie^2}{2}\int\ud\phi\;\bar{u}'\bar{K}_{p'}\slashed{\epsilon}_2K_pue^{i(l_2+l_1)x}\underset{p'}{\bar{\varphi}}\underset{p}{\varphi} \;.
\ee
This is in general nonzero, so we also have to take the exchange term $M^{21}$ into account. 

For $M^{21}$ we begin with
\be
\bar{u}'\bar{K}(p',\phi_2)\slashed{l}_1K(P,\phi_2)\frac{\slashed{P}+1}{P^2-1+i\epsilon} \;,
\ee
where $\bar{P}=\bar{p}'+\bar{l}_1$. So, this time we write
\be\label{sepsll1ex}
\begin{split}
\slashed{l}_1=&[\slashed{\pi}(P,\phi_2)-1]-[\slashed{\pi}(p',\phi_2)-1] \\
&+[c+V(p',\phi_2)-V(P,\phi_2)]\slashed{k} \;,
\end{split}
\ee
where $c=(p'+l_1-P)_\LCp/k_\LCp$. The last two terms in~\eqref{sepsll1ex} vanish as before, and
\be\label{Pplus}
[\slashed{\pi}(P,\phi_2)-1]K(P,\phi_2)\frac{\slashed{P}+1}{P^2-1+i\epsilon}=+K(P,\phi_2) \;.
\ee
The rest of the calculation is the same as for $M^{12}$, except that~\eqref{Pplus} has opposite sign compared to~\eqref{Pminus}, and hence
\be
M_{\slashed{\epsilon}_1\to\slashed{l}_1}=M^{12}_{\slashed{\epsilon}_1\to\slashed{l}_1}+M^{21}_{\slashed{\epsilon}_1\to\slashed{l}_1}=0 \;.
\ee 
Thus, although $M^{12}$ and $M^{21}$ might not be separately gauge invariant, the total amplitude $M=M^{12}+M^{21}$ is.

Note that in the trident case the terms corresponding to $M^{12}$ and $M^{21}$ are separately gauge invariant, which makes the direct-exchange separation of the one-step term gauge invariant. In the double Compton case, in general one has to consider both the direct and the exchange parts of the one-step term together to have a gauge-invariant result. Note though that the separation between the two-step and the total one-step term should be gauge invariant because the two-step term can be obtained by gluing together two first-order processes.

\section{Exact comparison}

In this section we provide a nontrivial consistency check demonstrating that our results agree with the previous literature. We have already shown that our two-step term agrees with previous results in the LCF limit. To check also the one-step term we will compare the LCF limit of our results with the results for the direct part of the one-step term as obtained with the approach in~\cite{King:2013osa,King:2014wfa}.

\subsection{Previous constant-crossed approach}

In this subsection we follow closely the approach described in detail in~\cite{King:2013osa}. We first Fourier transform the lightfront-time dependencies of the vertices,
\be
\underset{P}{\bar{K}\bar{\varphi}}\slashed{\epsilon}_1e^{il_1x_1}\underset{p}{K\varphi}=\int\frac{\ud r_1}{2\pi}e^{i(P+l_1-p-r_1k)x_1}\Gamma(r_1)
\ee
and
\be
\underset{p'}{\bar{K}\bar{\varphi}}\slashed{\epsilon}_2e^{il_2x_2}\underset{P}{K\varphi}=\int\frac{\ud r_2}{2\pi}e^{i(p'+l_2-P-r_2k)x_1}\Delta(r_2) \;.
\ee
The spatial as well as the lightfront-time integrals now give delta functions
\be
\begin{split}
	\int\ud x_1^4\ud x_2^4\to&(2\pi)^4\delta^4(p'+l_1+l_2-p-[r_1+r_2]k) \\
	&(2\pi)^4\delta^4(P-[p-l_1+r_1k]) \;.
\end{split}
\ee
We use the second delta function to perform the $P_\mu$ integral and the $k_\LCp$ component of the first delta function to perform the $r_2$ integral. In the electron propagator we have $P^2-1=2kP(r_1-r_1^*)$, where $r_1^*=pl_1/kP$. It is therefore natural change variable from $r_1=:r+r_1^*$ to $r$.
The result we want to compare with is for the probability integrated over the transverse momenta. For these integrals we make the following change of variables $l_{1\LCperp}=q_1(L_{1\LCperp}+p_\LCperp)$ and $l_{2\LCperp}=\frac{q_2}{s_1}(L_{2\LCperp}+s_1p_\LCperp-q_1L_{1\LCperp})$, where $L_{1\LCperp}$ and $L_{2\LCperp}$ are the new integration variables. In terms of these variables we have $r_1^*=\frac{a_0r_{10}}{2\chi}(1+L_{1\LCperp}^2)$, and $r_2=\frac{a_0r_{21}}{2\chi}(1+L_{2\LCperp}^2)-r$.
For a constant field the Fourier transforms $\Gamma$ and $\Delta$ can be expressed in terms of the Airy function ${\rm Ai}(c)$ and ${\rm Ai}'(c)$. The argument of these Airy functions is 
\be
c_1=\frac{1}{2^{2/3}}\left(z_1+\frac{2r}{a_0\sqrt{z_1}}+z_1L_{1\LCperp}^2\right)
\ee
for $\Gamma$ and 
\be
c_2=\frac{1}{2^{2/3}}\left(z_2-\frac{2r}{a_0\sqrt{z_2}}+z_2L_{2\LCperp}^2\right)
\ee
for $\Delta$, where 
\be
z_1=\left(\frac{r_{10}}{\chi}\right)^\frac{2}{3} \qquad z_2=\left(\frac{r_{21}}{\chi}\right)^\frac{2}{3} \;.
\ee
The momentum integrals parallel to the field, i.e. $L_{11}$ and $L_{21}$, can be interpreted in terms of lightfront-time volume factors by changing variables to $\sigma=(L_{21}+L_{11})/(2a_0)$ and $\varphi=(L_{21}-L_{11})/a_0$. The integral over $\sigma$ gives the overall volume factor, while the other gives
\be
\begin{split}
	\int\ud&\varphi\left|\int\frac{\ud r_1}{2\pi}\frac{e^{i\varphi r}F(r)}{r+i\epsilon}\right|^2=\int\ud\varphi\theta(-\varphi)|F(0)|^2 \\
	&+\frac{1}{2\pi}\int_{-\infty}^{\infty}\frac{\ud r}{r^2}(|F(r)|^2-|F(0)|^2) \;,
\end{split}
\ee  
where the $\varphi$ integral in the first term gives an additional volume factor, so the first term is the two-step part and the $r$ integral term gives the one-step part.
The integrals over the momentum components perpendicular to the field, i.e. $L_{12}$ and $L_{22}$, take the following form
\be
\int\ud x\; x^{2n}\{{\rm Ai}^2,{\rm Ai}{\rm Ai}',{\rm Ai}'^2\}(c+x^2) \;,
\ee
which can be performed as in the appendix of~\cite{King:2013osa}.
We can now express the two-step as
\be
\mathbb{P}_2=\frac{\Delta\phi^2}{2}\mathcal{A}(0)
\ee
and the direct part of the one-step as
\be
\mathbb{P}_1^{\rm dir}=\Delta\phi\frac{1}{2\pi}\int_{-\infty}^\infty\frac{\ud r}{r^2}[\mathcal{A}(r)-\mathcal{A}(0)] \;,
\ee
where $\mathcal{A}$ is given below.

At this point one has to decide what to do with the photon polarization sum. Let us first check the standard replacement
\be\label{epsepstog}
\sum_{\rm pol.}\epsilon_\mu\epsilon_\nu\to-g_{\mu\nu} \;.
\ee  
This gives
\be\label{mathcalAfromAiAi}
\begin{split}
\mathcal{A}=\frac{\alpha^2a_0^2}{2s_1^2\chi^2}\bigg\{&\frac{q_1q_2}{s_1^2\sqrt{z_1z_2}}\hat{C}^{\cdot\cdot}{\rm Ai}(z_{1r}){\rm Ai}(z_{2r}) \\
&+\frac{\kappa_{10}\kappa_{21}}{z_1z_2}{\rm Ai}'(z_{1r}){\rm Ai}'(z_{2r}) \\
&+\frac{\kappa_{10}}{z_1}\hat{C}^{\prime1}{\rm Ai}'(z_{1r}){\rm Ai}_1(z_{2r}) \\
&+\frac{\kappa_{21}}{z_2}\hat{C}^{1\prime}{\rm Ai}_1(z_{1r}){\rm Ai}'(z_{2r}) \\
&+\hat{C}^{11}{\rm Ai}_1(z_{1r}){\rm Ai}_1(z_{2r})\bigg\}+(q_1\leftrightarrow q_2) \;,
\end{split}
\ee
where ${\rm Ai}_1(z)=\int_z^\infty\ud y{\rm Ai}(y)$, the arguments of the Airy functions are given by
\be
z_{1r}=z_1+\frac{2R}{\chi\sqrt{z_1}} \qquad z_{2r}=z_2-\frac{2R}{\chi\sqrt{z_2}} \;.
\ee
$R=r\chi/a_0$, and where the coefficients $\hat{C}$ are given by (the $g$-subscript indicates~\eqref{epsepstog}) 
\be
\hat{C}^{\cdot\cdot}_g=1-\frac{s_1(1+s_1)(s_1+s_2)}{q_1q_2}R
\ee
\be
\hat{C}^{\prime1}_g=1-q_2R \qquad \hat{C}^{1\prime}_g=1+q_1R
\ee
\be
\hat{C}^{11}_g=1+\frac{s_1^2-q_1q_2}{s_1}R+(s_2-q_1q_2)R^2 \;.
\ee
This agrees\footnote{There seem to be some trivial typos in~\cite{King:2014wfa}: The signs can be independently checked by comparing the two-step part with what one finds by gluing together the results in~\cite{King:2014wfa} for single Compton scattering. The sign errors would otherwise lead to a negative probability. There also seems to be a missing factor of $2$ in the one-step term coming from rescaling $r\to a_0r/2$.} with the result in~\cite{King:2014wfa}.
	
However, to compare with our results for the direct part we should calculate the corresponding quantity in the lightfront gauge $k\epsilon=0$, so that we can be sure that we are comparing exactly the same quantities. In this gauge we have
\be
\sum_{\rm pol.}\epsilon_\mu\epsilon_\nu(l)=-\left(g_{\mu\nu}-\frac{k_\mu l_\nu+l_\mu k_\nu}{kl}\right) \;,
\ee
where the $l_\mu$ terms vanish when applied to a gauge invariant term.
We then find that $\mathcal{A}$ is given by~\eqref{mathcalAfromAiAi}, but this time the coefficients are given by
\be
\hat{C}^{\cdot\cdot}_{\rm LF}=1-\frac{s_1s_2}{q_1q_2}R
\ee
\be
\hat{C}^{\prime1}_{\rm LF}=1-\frac{\kappa_{21}}{r_{21}}R \qquad \hat{C}^{1\prime}_{\rm LF}=1+\frac{\kappa_{10}}{r_{10}}R
\ee
\be
\hat{C}^{11}_{\rm LF}=1+\left[\frac{\kappa_{10}}{r_{10}}-\frac{\kappa_{21}}{r_{21}}-\frac{q_1q_2}{s_1}\right]R +\left[s_2-\frac{\kappa_{10}}{r_{10}}\frac{\kappa_{21}}{r_{21}}\right]R^2 \;.
\ee
By comparing the coefficients $\hat{C}_g$ and $\hat{C}_{\rm LF}$ we see that $\mathcal{A}_g$ and $\mathcal{A}_{\rm LF}$ agree for $R=0$, which they should because $\mathcal{A}(0)$ gives the two-step term. However, for $R\ne0$ they do not have to be the same because of gauge dependence. In any case, we can directly compare~\eqref{mathcalAfromAiAi} for $\mathcal{A}_{\rm LF}$ with the LCF approximation of our exact results.

\subsection{Our approach}

The starting point in this subsection is our exact expressions for $\mathbb{P}_1^{\rm dir}=\mathbb{P}^{22\to1}+\mathbb{P}^{12}_{\rm dir}+\mathbb{P}^{11}_{\rm dir}$, where $\mathbb{P}^{22\to1}$, $\mathbb{P}^{12}_{\rm dir}$ and $\mathbb{P}^{11}_{\rm dir}$ are given by~\eqref{thetaSeparate}, \eqref{P22dir}, \eqref{P12dir} and~\eqref{P11dir}. These expressions are valid for any inhomogeneous field. The locally-constant-field approximation is obtained by expanding in $1/a_0\ll1$ as explained in~\cite{Dinu:2017uoj}.  
For $\mathbb{P}^{22\to1}$ we use the following integration variables, $\sigma=(\sigma_{43}+\sigma_{21})/2$ and $\varphi=\sigma_{43}-\sigma_{21}$, where $\sigma_{ij}=(\phi_i+\phi_j)/2$, and $\theta_{43}$, $\theta_{21}$. In the constant field limit the $\varphi$ integral is trivial and gives
\be
\int\ud\varphi\left[-\theta(\varphi)\theta\left(\frac{|\eta|}{2}-\varphi\right)\right]=-\frac{|\eta|}{2}
\ee  
where $\eta=\theta_{43}-\theta_{21}$. To factorize the $\theta_{21}$ and $\theta_{43}$ integrals we write
\be
-\frac{|\eta|}{2}=\frac{1}{2\pi}\int\ud r\frac{e^{-i\eta r}-1}{r^2} \;.
\ee
For $\mathbb{P}^{12}_{\rm dir}$ we change variables from $\phi_1$ and $\phi_3$ to $\theta_{21}$ and $\theta_{23}$. To factorize these integrals we note that we can replace $\theta(-\eta)\to-{\rm sign}(\eta)/2$, where $\eta=\theta_{23}-\theta_{21}$, and
\be
-\frac{1}{2}{\rm sign}(\eta)=-\frac{i}{2\pi}\mathcal{P}\int\ud r\frac{e^{-i\eta r}}{r} \;,
\ee
where $\mathcal{P}$ is an instruction to take the principal value.
Finally for $\mathbb{P}^{11}_{\rm dir}$ we write
\be
\begin{split}
	&\int\frac{\ud\theta_{21}}{\theta_{21}^2}e^{\frac{ir_{20}}{2b_0}\Theta_{21}}= \\
	&\int\frac{\ud\theta_{21}\ud\theta_{43}}{\theta_{21}\theta_{43}}\delta(\theta_{43}-\theta_{21})e^{\frac{ir_{21}}{2b_0}\Theta_{43}+\frac{ir_{10}}{2b_0}\Theta_{21}} \;,
\end{split}
\ee 
and use the representation
\be
\delta(\eta)=\frac{1}{2\pi}\int\ud r\; e^{-i\eta r} \;.
\ee
The $\theta_{ij}$ integrals in each of these terms can now be performed and give Airy functions. We find exactly the same result as in~\eqref{mathcalAfromAiAi} with the coefficients given by $\hat{C}_{\rm LF}$. Since these two approaches are entirely different, this gives a highly nontrivial check of our methods and results. 

Note that the gauge dependence does not necessarily mean that $\mathcal{A}_g$ and $\mathcal{A}_{\rm LF}$ are completely different. Indeed, in the next section we will show that they are equal to leading order for $\chi<1$ and hard photons. \\

\subsection{Saddle-point comparison}\label{lcfcomparison}

As already noted, there are no explicit results for hard photons that we can immediately compare with, but it is possible to derive such results from e.g. the analytic expressions in~\cite{King:2014wfa}, see~\eqref{mathcalAfromAiAi}.
To obtain the longitudinal momentum spectrum for hard photons, there is only one integral to perform, namely
\be
\int\ud t\frac{\mathcal{A}(t)+\mathcal{A}(-t)-2\mathcal{A}(0)}{t^2} \;,
\ee   
where $\mathcal{A}(t)$ is given by~\eqref{mathcalAfromAiAi} and $t=2r/a_0$.
For $\chi\ll1$ and hard photons we can perform the $t$ integral with the saddle-point method. It turns out that $|\chi t|\sim\sqrt{\chi}\ll1$, so we rescale $t=\tau/\sqrt{\chi}$ and expand the integrand to leading order in $\chi$, which involves
\be
{\rm Ai}(\xi\gg1)\approx\frac{e^{-\frac{2\xi^{3/2}}{3}}}{2\sqrt{\pi}\xi^{1/4}}
\ee
and similar expansions for ${\rm Ai}'$ and ${\rm Ai}_1$.
Now we can perform the resulting, elementary $\tau$ integral
\be
\int\ud\tau\frac{e^{-c\tau^2}-1}{\tau^2}=-2\sqrt{\pi c} \;.
\ee 
There is no difference between $\mathcal{A}_g$ and $\mathcal{A}_{\rm LF}$ to leading order, they both give
\be\label{Igg1}
\mathbb{P}_1^{\rm dir}=
-\frac{\alpha^2a_0}{4\pi^\frac{3}{2}}\sqrt{\frac{r_{20}}{\chi}}
\left[\frac{q_1}{q_2}+\frac{q_2}{q_1}+\frac{s_1s_{\bar{1}}}{q_1q_2}\right]e^{-\frac{2r_{20}}{3\chi}} \;.
\ee
This agrees perfectly with our result~\eqref{dironeleadlcfsaddle} for the direct part of the one-step term,
demonstrating consistency. To obtain the corresponding expression for the two-step term $\mathbb{P}_2$ one just has to omit the $t$ integral, the result agrees with~\eqref{twosteplcfsaddle}.

We have thus demonstrated consistency between our expressions and those of~\cite{King:2014wfa}, with all the nontrivial dependencies on the various parameters. In addition, the short calculation here is very different from the derivation of~\eqref{dironeleadlcfsaddle} in the main text, providing further reassurance that our results are consistent.
Since we have checked that our analytical results agree with our numerical results, this comparison also gives us a benchmark of our numerical results.

\end{document}